\documentclass[]{aastex63}
\usepackage{amsmath}
\usepackage{amsfonts}
\usepackage{amssymb}
\usepackage{soul,color}
\usepackage{multirow}
\graphicspath{{./}{figures/}}
\definecolor{marc}{rgb}{0.4, 0.0, 0.4}
\definecolor{asif}{rgb}{0.0, 0.0, 1.0}

\begin{document}
\title{Modeling Magnetically Channeled Winds in 3D: I. Isothermal Simulations of a Magnetic O Supergiant}

\correspondingauthor{Sethupathy Subramanian}
\email{ssubrama@nd.edu}

\author[0000-0003-0017-0811]{Sethupathy Subramanian}
\affiliation{University of Notre Dame, 225 Nieuwland Science Hall, Notre Dame, IN 46556, USA}

\author[0000-0003-3309-1052]{Dinshaw S. Balsara}
\affiliation{University of Notre Dame, 225 Nieuwland Science Hall, Notre Dame, IN 46556, USA}

\author[0000-0001-7721-6713]{Asif ud-Doula}
\affiliation{Penn State Scranton, 120 Ridge View Drive, Dunmore, PA 18512, USA}

\author[0000-0003-4632-5962]{Marc Gagn\'e}
\affiliation{West Chester University, Merion Science Center 201, West Chester, PA 19383, USA}



\begin{abstract}
In this paper we present the first set of 3D magnetohydrodynamic (MHD) simulations
performed with the {\sc Riemann Geomesh} code. We study the dynamics of the
magnetically channeled winds of magnetic massive stars in full three dimensions
using a code that is uniquely suited to spherical problems. 
Specifically, we perform isothermal simulations of a smooth wind on a rotating 
star with a tilted, initially dipolar field. 
We compare the mass-loss, angular momentum loss, and magnetospheric dynamics of 
a template star (with the properties that are reminiscent of the O4 supergiant
$\zeta$~Pup) over a range of rotation rates, magnetic field strengths, and 
magnetic tilt angles.
The simulations are run up to a quasi-steady state and the results are
observed to be consistent with the existing literature, showing the episodic
centrifugal breakout events of the mass outflow, confined by the magnetic field 
loops that form the closed magnetosphere of the star. 
The catalogued results provide perspective on how angular-momentum loss varies 
for different configurations of rotation rate, magnetic field strength and 
large magnetic tilt angles. 
In agreement with previous 2D MHD studies, we find that high 
magnetic confinement reduces the overall mass-loss rate,
and  higher rotation increases the mass-loss rate.
This and future studies will be used to estimate 
the angular-momentum evolution, spin-down time, 
and mass-loss evolution of magnetic massive stars
as a function of magnetic field strength, rotation rate,
and dipole tilt.

\end{abstract}

\begin{keywords}
{Massive Stars -- Winds, Outflows -- MHD -- Simulations}
\end{keywords}


\section{Introduction}


Although fewer than 1\% of stars end their lives in core-collapse supernovae,
massive stars have had a profound influence on the chemical, ionization and 
star-formation history of the universe. The first very massive stars formed from the 
pristine H/He gas of the early universe, prior to the formation of the first galaxies. 
How massive stars evolve over cosmic time depends on understanding the role that 
binarity, mass transfer, mergers, rotation, metallicity, mass loss,
and magnetic fields play on the pre-supernova evolution of individual systems
\citep[see][for a review]{2012ARA&A..50..107L}.

The last two decades have seen the discovery of an exciting new class of stars: magnetic massive stars.
Strong, organized magnetic fields on O and early-B stars strongly influence angular momentum evolution, and impact
their late-stage evolution prior to core-collapse. {The occurrence rate of magnetic fields in
O- and early-B stars is $7-10\%$} \citep{2016MNRAS.456....2W, 2017MNRAS.465.2432G, 2017A&A...599A..66S}.
That said, the origin of the magnetic fields is not known.
Among the massive stars with detectable magnetic fields, their
magnetic geometries appear to be stable on timescales of many decades.
Single magnetic massive stars rotate more slowly than non-magnetic stars, suggesting
long-term magnetic braking \citep{2016MNRAS.456....2W}.

In the presence of a strong, large-scale magnetic field the stellar wind is channeled
toward the magnetic equator,
the magnetosphere is closed at relatively small radii. The wind pulls open the field at high magnetic 
latitude, and at large radii.
For weak fields, the field is unconfined and blown open by the wind.
\citet{Uddoula02} showed that the effect of the field on the stellar wind
depends largely on just a single {\it wind magnetic confinement parameter}, 
$\eta_\star$, which characterizes the competition between field and wind outflow energy density.
\begin{equation}\label{eq:eta}
\eta_\star \equiv \frac{{B_{\mathrm{eq}}^2 R_{\star}^2}} {\dot{M}_{B=0} v_{\infty}},    
\end{equation}
where the equatorial field strength $B_{\mathrm{eq}} = \frac{1}{2} B_\mathrm{p}$
is half the polar field strength,
$R_{\star}$ is the stellar radius, and $\dot{M}_{B=0}$ and $v_{\infty}$
are the mass-loss rate and terminal wind speed in the absence of a field. 

In an extensive series of papers,
2D MHD simulations were used to understand the dynamics of
magnetic wind channeling and the consequent hard X-ray emission seen with {\it Chandra}
\citep{Uddoula02, Owocki04c, Gagne05, 2006ApJ...640L.191U, 2008MNRAS.385...97U, 2009MNRAS.392.1022U, 2013MNRAS.428.2723U, 2014MNRAS.441.3600U}.

Specifically, \citet{2008MNRAS.385...97U} showed how magnetic channeling
depends on rotation and magnetic confinement.
We hasten to point out that all this early modeling work was done within the context of aligned rotators, where the rotation axis and the dipole axis coincide. Indeed, that is the only case for which models can be built.
\citet{Gagne05} incorporated an energy equation to model the X-rays from
$\theta^1$\,Orionis\,C, the central star of the Orion Nebula.
These early 2D models were able to explain many of the observational properties of
the few magnetic O stars that were known at the time.
More recently, \citet{Petit2013} compiled  an exhaustive list of 64
confirmed magnetic OB stars with $T_{\rm eff} > 16$\,kK, along with their physical,
rotational and magnetic properties.

Many of the observational properties of magnetic massive stars depend
on the ratio of the Alfv\'en and the Kepler corotation radii \citep{Petit2013}.
The Alfv\'en radius $R_{\mathrm{A}}$, which, at the magnetic equator,
is the approximate boundary between the closed magnetosphere and the open field,
can be expressed in terms of the magnetic confinement parameter
\citep{2008MNRAS.385...97U}:
\begin{equation}\label{eq:RA}
\frac{R_{\mathrm{A}}}{R_\star} \approx 0.3 + (\eta_\star + 0.25)^{\frac{1}{4}}
\end{equation}
The torque of the magnetic field on the wind outflow maintains
rigid body rotation within the magnetosphere.
The rotation parameter $W$ is the ratio of the rotational and orbital velocities 
at the photosphere. For near-critical rotation, the star is oblate, but for $W \leq 0.5$
we use a spherical approximation \citep{2008MNRAS.385...97U}:


\begin{equation}
W \equiv \frac{V_{\rm rot}}{V_{\rm orb}} = \frac{\Omega R_{\star}}{\sqrt{\frac{G M_{\star}}{R_{\star} } } },
\end{equation}
where $\Omega$ is the rotational angular frequency and 
$W = \Omega / \Omega_{\rm crit}$ is the critical rotation ratio.
The outward centrifugal acceleration from rigid-body rotation will exactly balance the inward
gravitational acceleration at the Kepler corotation radius.
\begin{equation}
    R_{\mathrm{K}} = \left( \frac{GM}{{\Omega}^2} \right)^{\frac{1}{3}} = W^{-\frac{2}{3}} {R_\star}
\end{equation}

Many magnetic O stars are slow rotators with approximately kG fields such that
the Alfv\'en radius is smaller than the Kepler radius
\citep{Petit2013, 2016MNRAS.456....2W, 2016AdSpR..58..680U, 2017MNRAS.465.2432G}. 
In these so-called {\em dynamical} magnetospheres (DM), wind material fed into the magnetosphere falls back
onto the photosphere within a dynamical time scale.
But for the more rapidly rotating B stars, and, notably, the O7.5\,III secondary in Plaskett's star,
$R_\mathrm{K} < R_\mathrm{A}$; wind mass fed into the magnetosphere builds up at the magnetic equator
between the Kepler and Alfv\'en radii. These are the so-called {\em centrifugal} magnetospheres (CM) \citep{Petit2013}.
Some mechanism must allow for mass and density to leak out of these
{\em centrifugal} magnetospheres: slowly via diffusion, or sporadically, via centrifugal breakout events,
wherein magnetic tension can no longer contain the built-up mass \citep{2020MNRAS.499.5379S}.
\citet{2020MNRAS.499.5366O} show that the variable H$\alpha$ profiles of centrifugal magnetosphere stars
strongly favor centrifugal breakout events.

\begin{figure}
    \centering
    \includegraphics[scale=0.60]{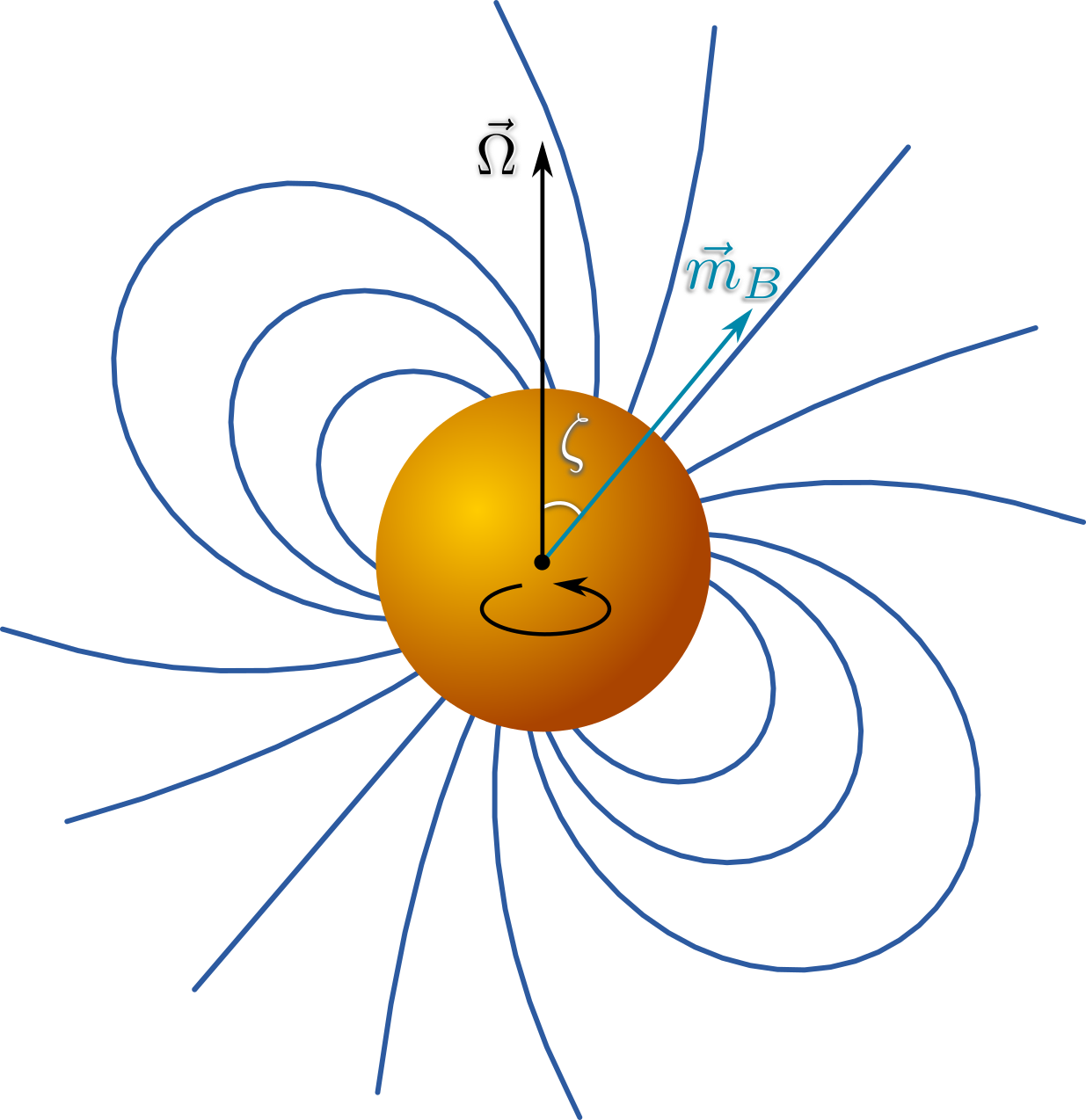}
    \caption{An oblique magnetic rotator. The photosphere is orange,
    the vertical rotation axis is shown in black, with the angular velocity
    ${\vec \Omega}$ along the $z-$axis.
    The magnetic dipole axis ${\vec m_B}$ is shown in blue
    with the initially dipolar field lines. The magnetic tilt angle is $\zeta$.}
    \label{fig:schematic}
\end{figure}

Although 2D simulations have been used to successfully model aligned rotators,
magnetically channelled stellar winds are intrinsically 3D.
The magnetic dipole axis is nearly always tilted to the rotation axis,
with typical obliquity angles between $30\degr$ and $90\degr$ \citep{2016MNRAS.456....2W}.
A schematic diagram of an oblique magnetic rotator is shown in Figure~\ref{fig:schematic}.
\citet{2019MNRAS.489.3251D} use 3D isothermal MHD simulations performed with the {\sc PLUTO} code to
predict the radio emission from the winds of oblique magnetic rotators.
Their simulations show the formation of a two-armed spiral structure.

In this work, our aim is to extend this modeling effort using the 
{\sc Riemann Geomesh} MHD code \citep{2019MNRAS.487.1283B, 2020ComAC...7....1F}.
Rather than using a conventional spherical $(r, \theta, \phi)$ grid,
{\sc Riemann Geomesh} employs a recursive partitioning of the
spherical icosahedron based triangular meshing of the sphere,
and a non-linear radial grid to map the volume of the wind, typically out to $10-20 R_\star$.
The resultant mesh maps the surface of the sphere as uniformly as possible,
which traditional Cartesian based meshing is unable to accomplish,
especially near the rotational poles.
The {\sc Riemann Geomesh} code is therefore uniquely suited to simulate
oblique magnetic rotators in a bias free fashion.

\begin{figure}
    \includegraphics[scale=0.67]{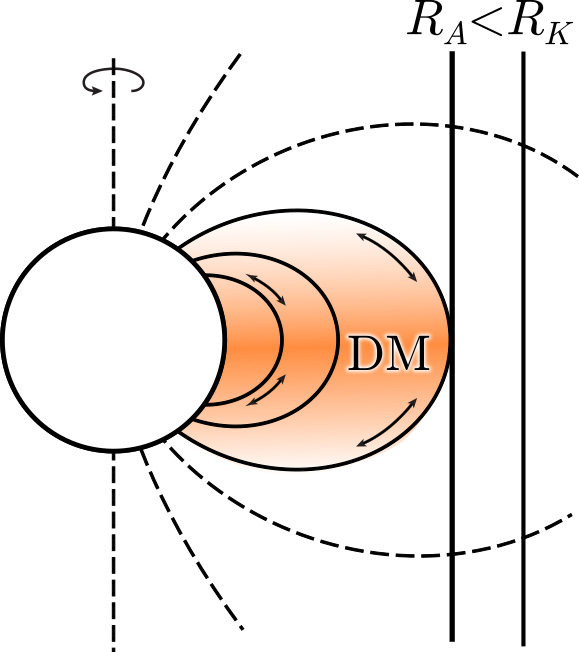}\hspace{3mm}
    \includegraphics[scale=0.67]{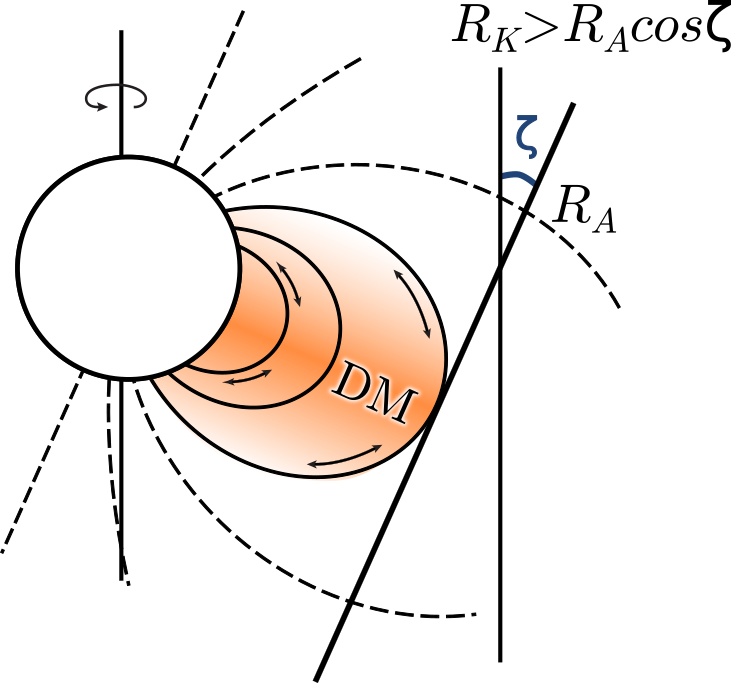}\hspace{5mm}
    \includegraphics[scale=0.67]{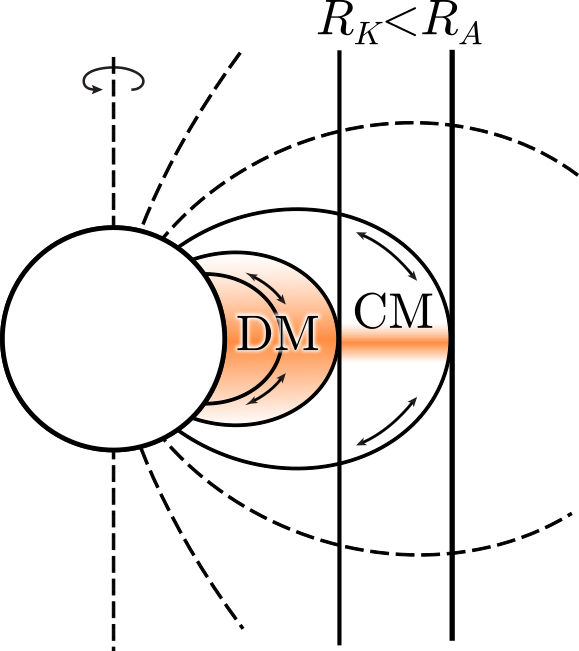}
    \includegraphics[scale=0.67]{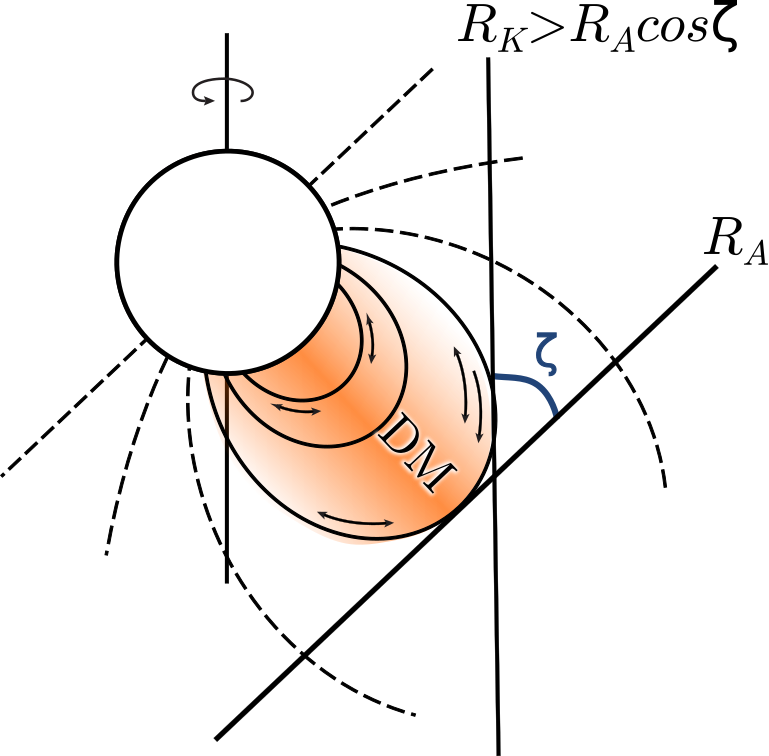}\par
    \hspace{18mm}(a)\hspace{30mm}(b)
    \hspace{36mm}(c)\hspace{38mm}(d)\par

\caption{Schematic magnetosphere diagrams. (a) An aligned dynamical magnetosphere (DM):
the Alfv\'en radius $R_{\rm A}$ is inside the Kepler corotation radius $R_{\rm K}$. Mass inside
the magnetosphere (orange) will fall back onto the photosphere.
(b) A sketch of a tilted DM with $R_{\rm A} < R_{\rm K}$.
(c) An aligned centrifugal magnetosphere (CM): 
the  Kepler corotation radius $R_{\rm K}$ is inside the Alfv\'en radius $R_{\rm A}$
(d) A conjectured sketch of a tilted magnetosphere (that would have been a CM if it were aligned). The
tilt causes $R_{\rm A}cos\zeta < R_{\rm K}$ even if we initially had $R_{\rm K} < R_{\rm A}$.
We caution that (b) and (d) are just conjectures at this point in the narrative. Figures (a) and (c) are adopted from \citet{Petit2013}.}
\label{fig:dmcm_tilt}
\end{figure}

The classification of centrifugal 
magnetosphere (CM) and dynamical magnetosphere (DM) is
based on an aligned magnetic rotator, as displayed in 
Figures {\ref{fig:dmcm_tilt}a} and {\ref{fig:dmcm_tilt}c}. 
This system of classifying magnetospheres is especially relevant to aligned 
magnetic rotators but including a tilt angle complicates this simple classification scheme,
as can be seen in Figures {\ref{fig:dmcm_tilt}b} and {\ref{fig:dmcm_tilt}d}.
Figure~{\ref{fig:dmcm_tilt}b} shows a tilted ``dynamical'' magnetosphere. Figures {\ref{fig:dmcm_tilt}b} and {\ref{fig:dmcm_tilt}d} are just conjectured sketches intended to sensitize the reader that the inclusion of three dimensionality introduces a third important parameter into our discussion; i.e. the tilt between the rotation axis and the dipole axis. A major aim of this work is to examine the dynamics of these magnetospheres in 3D
to understand the role of magnetic tilt.

In this paper, we present and analyze 3D simulations of magnetically channeled winds
around oblique magnetic rotators. The first goal of this paper
is to show that the geodesic mesh MHD code can perform accurate simulations
with large tilt angles and rotation rates.
The second goal of this paper is to examine the dependence of the quasi-steady state 
mass-loss rate with rotation, magnetic confinement, and magnetic tilt angle, and to test
the prediction that the overall mass-loss rate should decrease with the magnetic 
field, and increase with the rotation rate \citep{2009MNRAS.392.1022U} .
The third goal of this paper is to use the 3D simulations to look for and study the episodic
breakout events which are predicted to occur in centrifugal magnetospheres
\citep{2006ApJ...640L.191U,2020MNRAS.499.5379S}.
The corresponding angular momentum flux is also shown 
as a function of $\theta, \phi$ at the outer most boundary of the simulation.
The fourth goal is to estimate spindown as the star loses mass and
angular momentum over its lifetime. This is achieved with the help of mass flux 
and angular momentum flux obtained at the outer boundary of the simulation.
\citet{2008MNRAS.385...97U} analyze angular momentum evolution and spindown
for aligned magnetic rotators in 2D.

The outline of this paper is as follows: 
in section \ref{sec:method} we describe the geodesic mesh along with the 
value it provides for this work.
In section \ref{sec:model} we describe the model, the boundary conditions, and the numerical simulations.
In section \ref{sec:quasi_ss} we present the quasi-steady state mass-loss rate for 
different magnetic tilt angles and rotation rates.
In section \ref{sec:dynm} we show the episodic centrifugal breakout events in 3D, as well as 
the mass fallback close to the star.
And in section \ref{sec:jdot} we plot angular momentum flux as a function 
of $\theta, \phi$, and show regions of high angular momentum flux,
and estimate characteristic stellar spindown time.
Lastly, in section \ref{sec:summ} we present the summary and 
conclusion of this work.


\section{Mesh and Methods} \label{sec:method}

The overarching problem consists of simulating the magnetically channeled, line
driven winds around rotating massive stars. Due to the spherical
aspect of the problem, it is best to carry out the simulation on a mesh that 
is uniquely suited for this problem. 

One of the innovative aspects of this paper is that we have carried out 
the simulations using a geodesic mesh, and imposed a globally divergence-free formulation 
of the MHD equations. The other major advance in the simulation of line
driven winds is that the magnetic field is split into two components: 
one the curl free, background magnetic field ${\bf B_0}$ and the other 
being the time evolving component ${\bf B_1}$. 
This would be reflective of the fact that the primary
source of the magnetic field i.e., the star itself, lies outside the simulation domain. 
The variations in the magnetic field due to the material outflow is 
modeled with the help of the evolving field ${\bf B_1}$. 
The simulations are carried out in the rotating frame of reference of the 
star. Consequently, the simulations include the forces that arise due to 
rotation, which are centrifugal and coriolis forces. 
The complete description of the MHD equations employed in this manuscript is
provided in Appendix \ref{sec:App_MHD}.
In Subsection \ref{sec:geo_struct} we describe the structure of the geodesic 
mesh. In Subsection \ref{sec:geo_value} we explain why it is particularly useful for this problem. 


\subsection{Structure of the Geodesic Mesh}\label{sec:geo_struct}


The simulation setup is comprised of a spherical domain around the star, discretised 
with uniform triangular patches using a spherical icosahedron as the base. 
The uniformity of such a geodesic mesh results in high accuracy and consistent 
time-steps, when compared to other meshing strategies 
\cite{2019MNRAS.487.1283B, 2020ComAC...7....1F}. 
In the following subsections, because of the novelty of the approach, the 
implementation of the meshing is described. Then the value of such a uniform 
mesh in a spherical domain is explained, in context with the simulation of a 
magnetized star. 

\begin{figure}
    \includegraphics[scale=0.51]{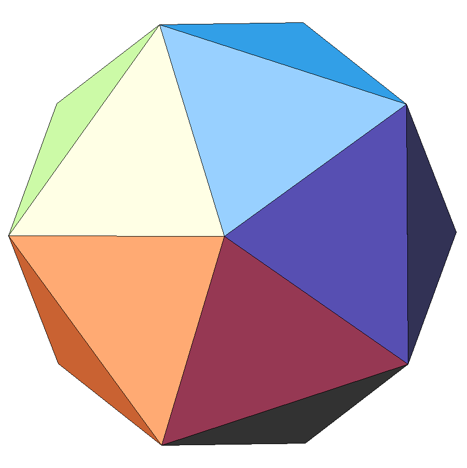}
    \includegraphics[scale=0.51]{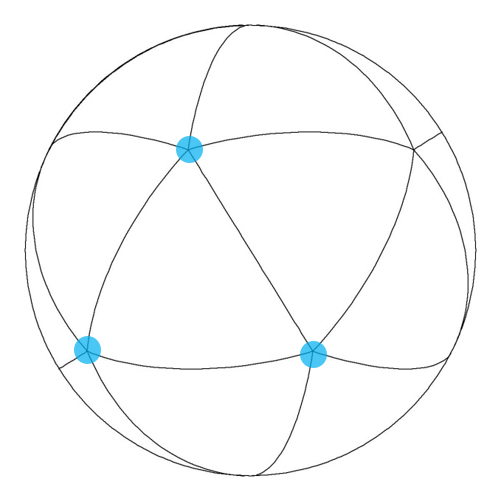}
    \includegraphics[scale=0.51]{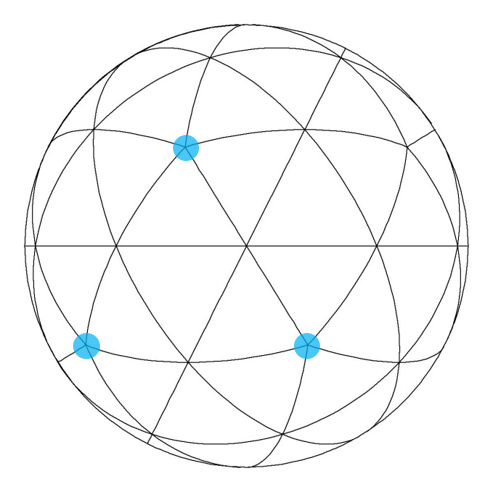}
    \includegraphics[scale=0.51]{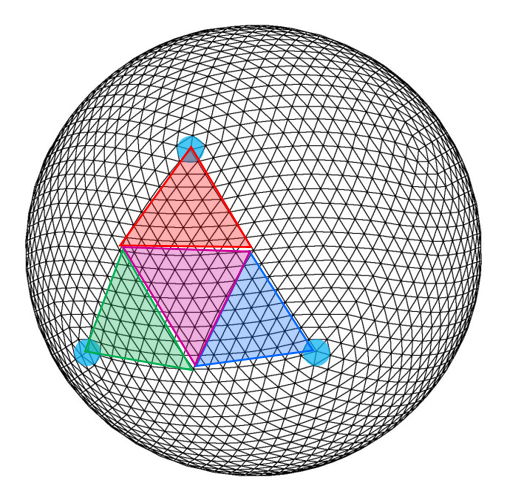}\par
    \hspace{18mm}(a)\hspace{38mm}(b)
    \hspace{38mm}(c)\hspace{38mm}(d)\par
\caption{Icosahedron based spherical mesh: (a) a regular icosahedron 
\cite{wikimedia_icosa} (b) a spherical icosahedron, (c) first subdivision of the spherical icosahedron,
and (d) fourth subdivision of the spherical icosahedron.}
\label{fig:geomesh_mesh}
\end{figure}

High accuracy MHD computations of spherical objects require a high quality 
meshing of the spherical domain.
The traditional logically Cartesian based meshing of the sphere has two severe shortcomings:
i) smaller zones at the poles lead to a shorter time-steps and 
ii) the geometric singularity causes a loss of accuracy at the poles. 
As a result, it is preferred to discretise the sphere by the structures 
emerging from the spherical icosahedron. Then recursive bisections of the 
geodesic curves can be used to obtain the required level of angular 
refinement shown in Figure~\ref{fig:geomesh_mesh}. 

The icosahedron shown in Figure~\ref{fig:geomesh_mesh}a is used to obtain a spherical 
icosahedron shown in Figure~\ref{fig:geomesh_mesh}b. The first subdivision of the 
spherical icosahedron is illustrated in Figure~\ref{fig:geomesh_mesh}c, continuing 
this subdivision three more times, we get finer discretisation of the 
icosahedron-based spherical mesh, shown in Figure~\ref{fig:geomesh_mesh}d.
The colored patches in Figure~\ref{fig:geomesh_mesh}d suggest that such a mesh structure
leads naturally to large-scale numerical parallelisation. We exploit the 
symmetry and the divisibility of the geodesic mesh to perform these simulations 
efficiently on thousands of compute cores.

The discretisation shown in Figure~\ref{fig:geomesh_mesh}b is referred to as level-0 
sectorial division. It consists of 20 great triangles that subtend an angle 
of $\pi/2 - \tan^{-1}(1/2) {\rm rad} \approx 63.4^\circ$ from the center of the sphere.
The first subdivision shown in Figure~\ref{fig:geomesh_mesh}c is referred to as level-1 
discretisation, with 80 great triangles, each subtending an average 
angle of $33.9\degr$.
Each colored triangle in Figure~\ref{fig:geomesh_mesh}d therefore represents a level-1 sector. 
In the same way, the level-4 discretisation depicted in Figure~\ref{fig:geomesh_mesh}d
subtends an angle of $\approx 4.33\degr$.
While it is difficult to display the level-5 subdivision, it yields an angular resolution of
$2.16\degr$. At this angular resolution, the simulations in this study
cover the spherical surface with $20,480$ spherical triangles.

\begin{figure}
    \centering
    \includegraphics[scale=0.6]{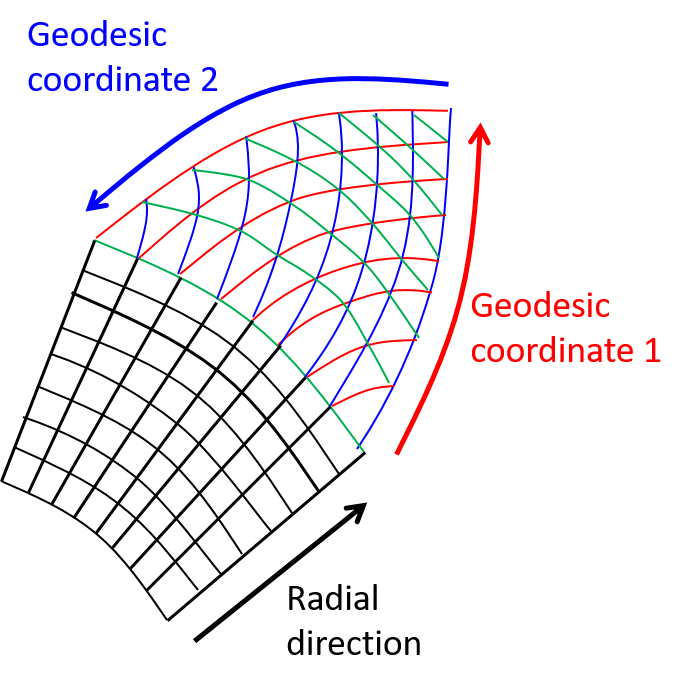}
    \caption{Discretisation along the radial direction and numbering strategy 
             for the triangulation.}
    \label{fig:geomesh_radial}
\end{figure}

Due to the fast changing nature of the CAK velocity and density profiles, it is 
essential to have a very thin radial discretisation close to the star:
$\Delta r = 1.0 \times 10^{-3} R_\star$ or slightly larger. 
This way, the thin radial zones near the photosphere initially expand geometrically,
and then expand logarithmically out to 15 stellar radii.
The 3D mesh is formed by extruding the 2D mesh with 240 radial zones.
The mesh volume is therefore divided into $240 \times 20,480 \approx 5.0\times10^6$ zones.
The geodesic coordinates 1 and 2 in Figure~\ref{fig:geomesh_radial} are shown in red and blue.
With the radial coordinate (in black) they define access to each zone on the geodesic mesh.


\subsection{Value of {\sc Geomesh}} \label{sec:geo_value}


Our geodesic-based mapping of the sphere does not have any preferred 
orientation, it is highly regular and uniform in the angular direction.
While a logically rectangular mesh in $(r, \theta, \phi)$ coordinates has 
singularities at the poles. The drawbacks of coordinate singularities include,
but are not limited to, i) difficulty in doing constrained transport at the poles 
and ii) reduction in time-step and accuracy of the solution at the poles,
\cite{2019MNRAS.487.1283B}.
It has also been reported in \cite{2019MNRAS.489.3251D} that there can be fictitious
heating of gas at the poles.
The mesh shown in Figure~\ref{fig:geomesh_mesh} is free of such singularities. 
Since such a mapping is uniform, a general 
case of magnetic rotation can be described by choosing a fixed axis for the 
rotation, and only varying the orientation of the magnetic field.
Also, the Geomesh code employs state-of-the-art Multidimensional
Riemann solvers and a WENO-AO scheme, which are specifically developed for 
unstructured meshes \citep{2015JCoPh.287..269B, 2020JCoPh.40409062B}. 
The divergence free evolution of magnetic field $\nabla \cdot {\bf B_1} = 0$
is enforced with the help of Yee-type collocation of facially averaged magnetic 
fields. This along with the edge-integrated electric fields, achieve a 
high-order accurate numerical implementation of Faraday's law; which ensures a
divergence free evolution of magnetic field \cite{2019MNRAS.487.1283B}.
Please see a recent review by \cite{2017LRCA....3....2B} 
for a detailed explanation of these computational techniques,
as they apply to astrophysical MHD simulations.

In this work, the rotation axis of the star is chosen to be 
$z$-axis and the orientation of the dipole magnetic field is setup 
with respect to the $z$-axis. The simulations are carried out for different
tilt angles ($\zeta$), such as 0, 45, 75 degrees of angular separation between the 
$z$-axis and the magnetic dipole axis. The next section provides the
explanation for modeling of the massive star wind and the boundary 
conditions used.


\section{Modeling and Boundary Conditions} \label{sec:model}

The schematic of the simulation setup is shown in Figure~\ref{fig:schematic}. The 
radius of the stellar surface $R_{\star}$ is taken as the inner boundary and 
the outer boundary $R_{\rm o}$ is chosen as $15R_{\star}$. 
The dipolar magnetic field is maintained throughout the simulation domain 
by means of the background magnetic field ${\bf B_0}$. The magnetic flux density is 
parametrized by $\eta_\star$ (see Eq. \ref{eq:eta}).
%
\subsection{ Initial Conditions } \label{ssec:init_cond}

%

Our template star is the O4\,I(n)fp supergiant $\zeta$~Puppis. Specifically, 
we use the spherical parameters derived by \citet{2019MNRAS.484.5350H} 
based on the Hipparcos distance $d=332\pm11$\,pc:
$L_\star=4.47\times10^5 L_\odot, T_{\rm eff} = 40\,000\,{\rm K}, R=13.5 R_\star, M=25 M_\odot$, and
$\dot{M} \approx 1.5\times10^{-6} M_\odot \,{\rm yr}^{-1}$.
$\Gamma_{\rm e}=0.471$ is the Eddington parameter for $\zeta$~Pup.
In the case of spherically symmetric mass loss, the CAK mass loss rate is given by,
\citep{2004EAS....13..163O},
\begin{equation}\label{eq:Mdot_cak}
    {\dot M}_{\rm CAK} = \dfrac{L_\star}{c^2} 
                     \dfrac{\alpha }{(1-\alpha)(1+\alpha)^{1/\alpha}}  \left(
                     \dfrac{ \bar{Q} \Gamma_{\rm e} }{1 - \Gamma_{\rm e}}         \right)
                     ^{(1-\alpha)/\alpha}
\end{equation}
where, $\alpha$ is the CAK exponent \citep{Castor75}, $L_\star$ is the stellar bolometric luminosity,
$\bar{Q}$ is the quality of the resonant absorption and $c$ is the speed of the light.
We initialize our model with a spherically symmetric wind with standard $\beta$-law.
The radial velocity along the radial-$r$ direction is given by,
\begin{equation}\label{eq:Vr_cak}
    v(r) = v_\infty \left( 1 - \dfrac{R_\star}{r} \right)^{\beta}
\end{equation}
where, $v_\infty$ is the terminal wind speed, and $\beta$ is the velocity exponent.
The terminal velocity speed is assumed to be $3$ times the escape velocity,
$v_\infty = 3 v_{\rm esc}$ and the $\beta = 0.8$ 
\citep{1986ApJ...311..701F, 2004EAS....13..163O}. 
The density is initialized using the mass-loss rate and the velocity,
\begin{equation}\label{eq:rho_cak}
    \rho(r) = \dfrac{ {\dot M}_{\rm CAK} }{4 \pi r^2 v(r)}
\end{equation}


\subsection{ Boundary conditions } \label{ssec:sim_bdy}


Owing to the rapidly changing velocity and density profiles of the 
CAK wind, it was essential to do a resolution study with different $\Delta r$
values close to star. For this, three different values of 
$\Delta r = 1.0 \times 10^{-3} R_\star$, 
           $3.0 \times 10^{-3} R_\star$ and 
           $5.0 \times 10^{-3} R_\star$
were chosen. The resulting density and velocity profiles from these
simulations matched each other well,
producing the expected mass loss rate ${\dot M}$.
Hence, as mentioned earlier the discretisation close to the star has
$\Delta r \approx 1.0 \times 10^{-3} R_\star$.

At the inner boundary ghost zones, following \cite{2002ApJ...576..413U}, 
we set the radial velocity by constant-slope extrapolation and fix the 
density at a value $\rho_{\rm c}$ chosen to ensure subsonic base outflow for 
the characteristic mass flux of a one-dimensional, nonmagnetic CAK model, i.e.,
$\rho_{\rm c}= \dot{M}/{4 \pi R_\star^2 a}/15$.
The evolving magnetic field component ${\bf B_1}$ (see Appendix 
\ref{sec:App_MHD}) is initialized to zero at the inner boundary. 
At the outer boundary, far from the star, standard outflow conditions 
are maintained.

In the case of line driven winds, the acceleration just above the photosphere is high.
Consequently, the density declines rapidly, while maintaining a quasi-steady mass-loss rate
\cite{2004EAS....13..163O}. In order to numerically account for the steep density gradient,
the zones near the inner boundary are very thin in the radial direction, 
increasing geometrically out to $1.5R_\star$, and increasing logarithmically at larger radii.
A heuristic factor $k_\rho$ is applied at the inner radial boundary, such that
$\rho_\ast=k_\rho \rho_{\rm sp}$, where $\rho_{\rm sp}$ is the estimated sonic-point density.

The value of $k_\rho$ is taken to be $15$ and it can be
seen in Table \ref{tab:mdot} of section \ref{sec:quasi_ss} that the 
simulated mass-loss rate matches within $5\%$ of the theoretical  
mass-loss rate, which is $1.43 \times 10^{-6} M_\odot/{\rm yr}$. 
In the next subsection CAK acceleration source terms are described. 


\subsection{Source terms } \label{ssec:cak_accel}


The basic formalism for the acceleration of line driven winds was developed by 
Castor, Abbott and Klein in 1975 (CAK) \citep{Castor75}.
The one-dimensional CAK line acceleration along the radial direction $r$ is given by,
\begin{equation}\label{eq:g_cak_1}
    g_{\rm CAK} = \frac{f_{\rm d}}{1-\alpha}
              \frac{\kappa_{\rm e} L_\star \bar{Q} } {4 \pi r^2 c}
       \left( \frac{dv/dr} { \rho c {\bar Q} \kappa_{\rm e} } \right)^\alpha
\end{equation}
where, $f_{\rm d}$ is a finite disk correction factor, which accounts for the 
solid angle of the photosphere. The remaining terms 
represent the CAK line force from a luminous point source. 
$\kappa_{\rm e}$ is the opacity due to free electrons, and $v$ is the velocity. 
There are two additional implicit acceleration terms:
i) the radial gravitational acceleration $g_{\rm grav}$ and
ii) accelerations induced by the rotating reference frame of the simulation,
namely, the Coriolis and centrifugal accelerations ${\bf g}_{\rm rot}$.
Therefore, the total acceleration is:
\begin{equation} \label{eq:g_tot}
    {\bf g}_{\rm tot} = g_{\rm CAK}{\bf{\hat r}} + g_{\rm grav} {\bf{\hat r}} + {\bf g}_{\rm rot}
\end{equation}
The expressions for the gravitational and rotational accelerations are 
provided in Appendix~\ref{sec:App_MHD}, along with the complete set 
of MHD equations. 
The simulations are performed using these initial and boundary conditions
and the source terms for a range of magnetic confinement parameters $\eta_\star$,
magnetic tilt angles $\zeta$, and critical rotation ratios $W$.
The next section describes the simulations, and the transition from initial conditions
to quasi-steady state behavior.


\section{Transition to Quasi-steady State} \label{sec:quasi_ss}


The 3D Riemann Geomesh simulations described here span a range of magnetic tilt angles
$\zeta = 0, 45\degr, 75\degr$,
two rotation rates $W = 0, 0.5$, and three magnetic confinement parameters, $\eta_\star =0, 10, 50.$
The ten cases are listed in Table~\ref{tab:mdot}, with the corresponding Kepler and Alfv\'en radii,
and the corresponding aligned-magnetosphere classification: CM for centrifugal magnetospheres, and
DM for dynamical magnetospheres. The outgoing mass-loss rates in the last column are discussed below.

\begin{table}
	\centering
	\caption{Quasi-steady-state mass-loss rates as a function of magnetic confinement
	$\eta_\star$, magnetic tilt angle $\zeta$, and critical rotation ratio $W$.
	$R_{\rm A}$ and $R_{\rm K}$ are the corresponding Alfv\'en and Kepler radii in stellar radius $R_\star$.
	CM indicates a centrifugal magnetosphere, for which $R_{\rm A} > R_{\rm K}$.
	DM indicates a dynamical magnetosphere, for which $R_{\rm A} \leq R_{\rm K}$.
	${\dot M}$ is the time-averaged mass-loss rate out of the outer boundary in solar masses per year.}
	\label{tab:mdot}
	\begin{tabular}{ccccccc}
		\hline
		$\eta_\star$ &
		$\zeta$ 
        & $W$ & $R_{\rm K}$ & $R_{\rm A}$ & CM/DM 
        & ${\dot M}$ \\
        \multicolumn{3}{c}{ } &
        \multicolumn{2}{c}{$(R_\star)$} &
         &
        ($M_\odot/{\rm yr}$) \\
		\hline
         0  &   $0\degr$  &   0   &$\infty$&  -  &   -  & 1.49 $\times 10^{-6}$  \\
         0  &   $0\degr$  &  0.5  &  1.59  &  -  &   -  & 1.70 $\times 10^{-6}$  
         \vspace{1.5mm} \\
        10  &   $0\degr$  &   0   &$\infty$& 2.1 &  DM  & 6.92 $\times 10^{-7}$  \\
        10  &   $0\degr$  &  0.5  &  1.59  & 2.1 &  CM  & 8.37 $\times 10^{-7}$  \\
        10  &  $45\degr$  &  0.5  &  1.59  & 2.1 &  CM  & 7.96 $\times 10^{-7}$  \\
        10  &  $75\degr$  &  0.5  &  1.59  & 2.1 &  CM  & 5.88 $\times 10^{-7}$  
         \vspace{1.5mm} \\
        50  &   $0\degr$  &   0   &$\infty$& 2.9 &  DM  & 4.35 $\times 10^{-7}$  \\
        50  &   $0\degr$  &  0.5  &  1.59  & 2.9 &  CM  & 4.58 $\times 10^{-7}$  \\
        50  &  $45\degr$  &  0.5  &  1.59  & 2.9 &  CM  & 5.84 $\times 10^{-7}$  \\
        50  &   $75\degr$ &  0.5  &  1.59  & 2.9 &  CM  & 4.41 $\times 10^{-7}$  \\
		\hline
	\end{tabular}
\end{table}

\begin{figure}
    \includegraphics[scale=0.58]{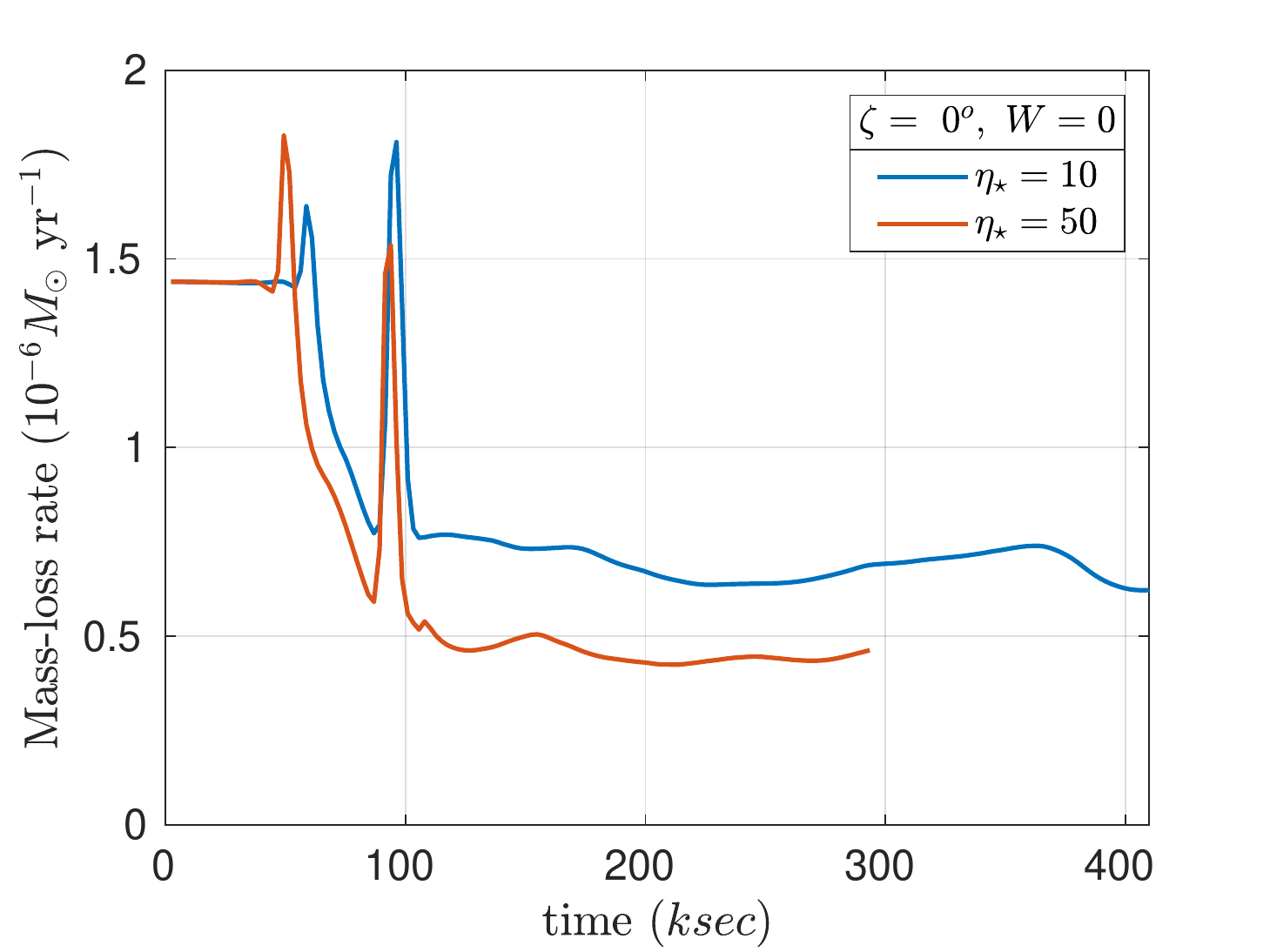}
    \includegraphics[scale=0.58]{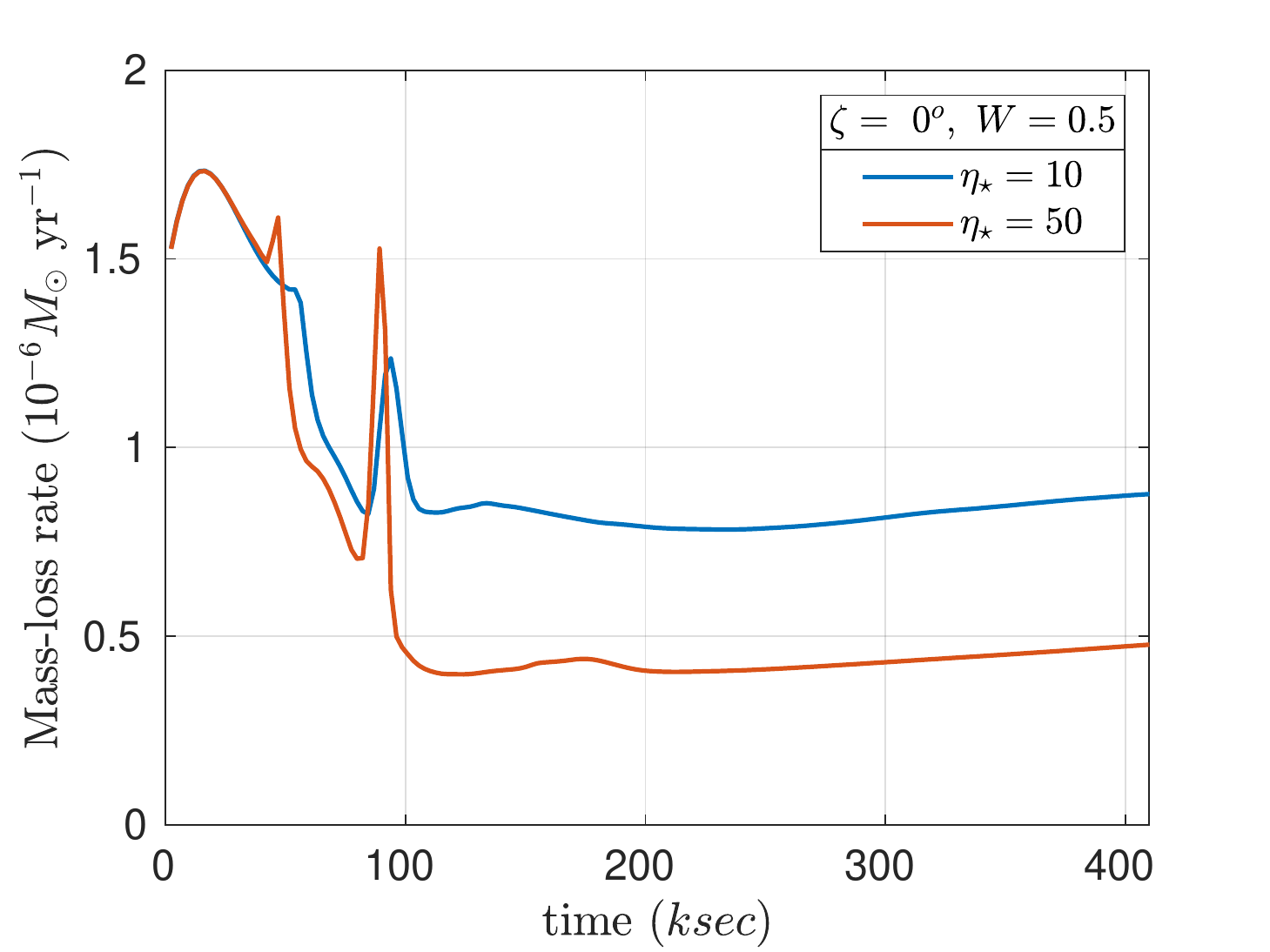}\par
    \hspace{42mm}(a)\hspace{80mm}(b) \par
    \includegraphics[scale=0.58]{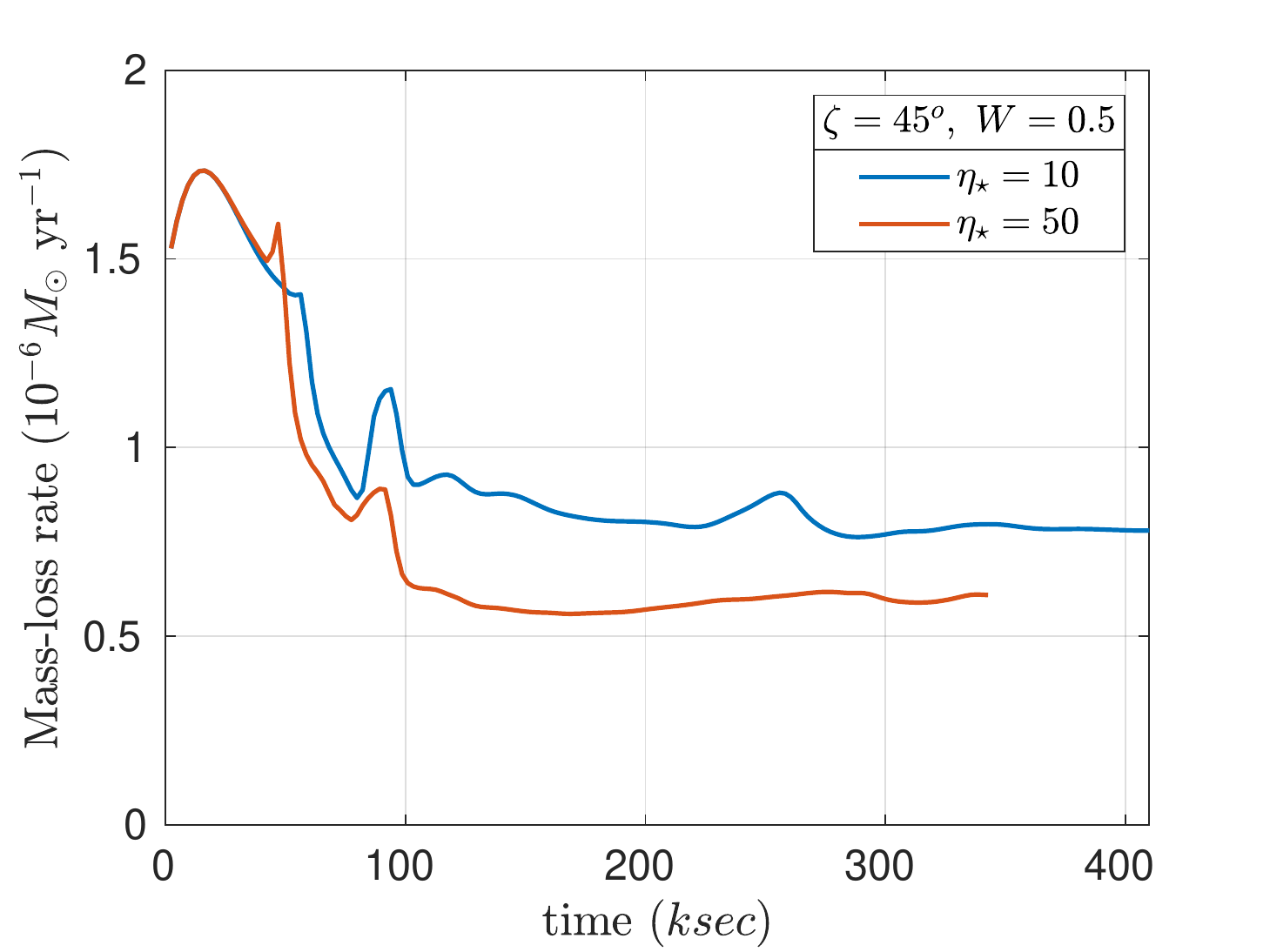}
    \includegraphics[scale=0.58]{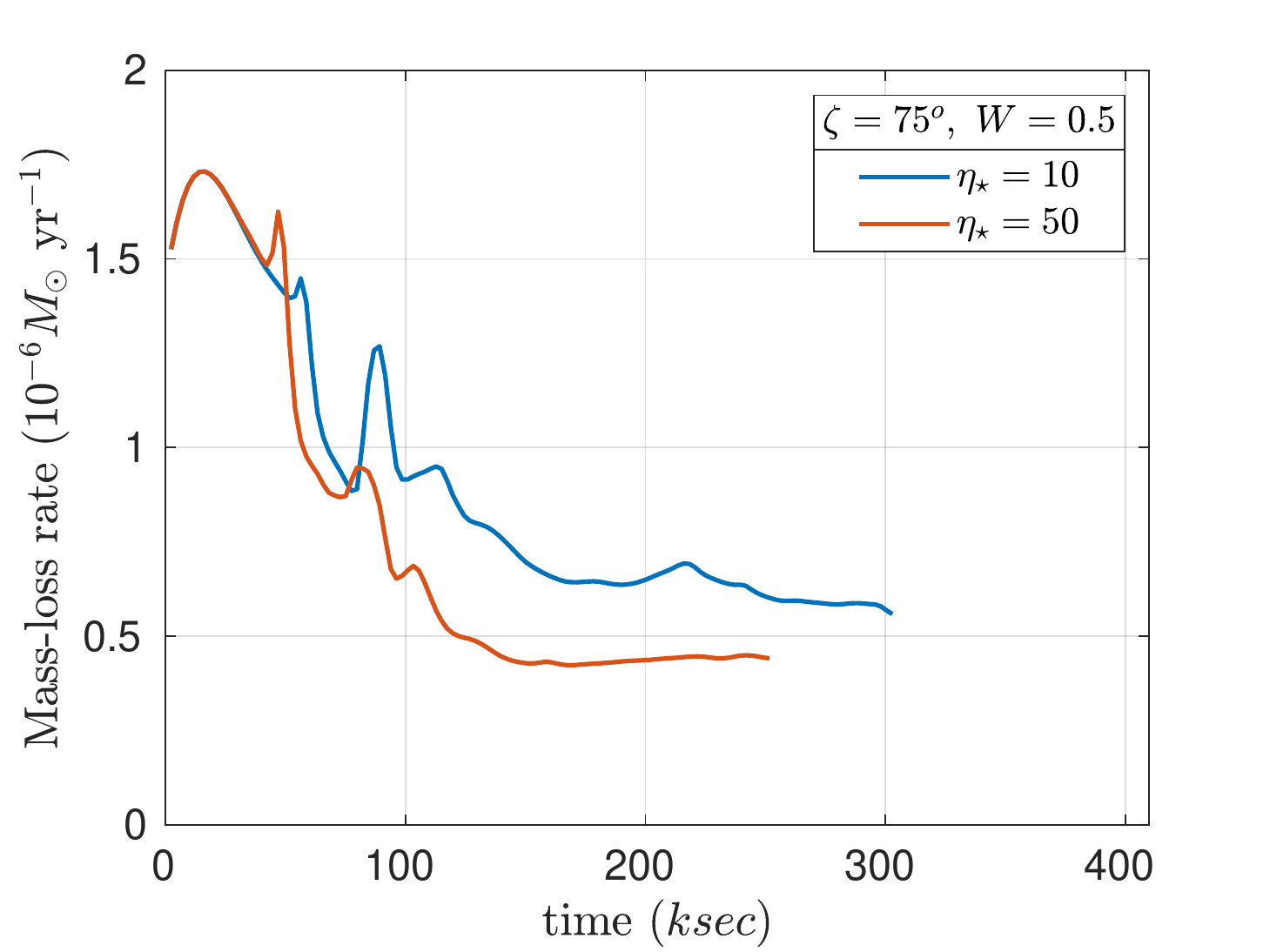}\par
    \hspace{42mm}(c)\hspace{80mm}(d) \par
\caption{Mass-loss rate at the outer boundary as a function of time showing the approach to quasi-steady state.
Shown in red (lower curve) $\eta_\star = 10$, and in blue (upper curve) $\eta_\star 50$ for:
         (a) $\zeta$ = $ 0\degr$,  $W = 0$,
         (b) $\zeta$ = $ 0\degr$, $W = 0.5$,
         (c) $\zeta$ = $45\degr$, $W = 0.5$, and
         (d) $\zeta$ = $75\degr$, $W = 0.5.$
         Quasi-steady state is achieved typically after $\approx 110$~ks.}
\label{fig:mdot}
\end{figure}
We first note that although we do not include our simulations at $W=0.25$, the behavior of the DMs
can be seen in the $W=0$ simulations. Second, we did not consider $W>0.5$ because at very high rotation
the photosphere becomes oblate \citep{2008MNRAS.385...97U}, and the spherical mesh is no longer appropriate.
Third, we note that higher values of $\eta_\star$ lead to significantly longer compute times.
High magnetic confinement simulations will be the subject of a future paper.

Figure~{\ref{fig:mdot}} shows the integrated mass loss from the outer 
boundary of the simulation as a function of time. The figures 
show the mass-loss rate reaching a quasi-steady value after $100-120$~ksec.
The mass-loss spikes seen in all simulations are the initial mass blow out, 
which occurs as the initially dense inner CAK wind exits the outer boundary. 

\begin{figure}
    \includegraphics[scale=0.58]{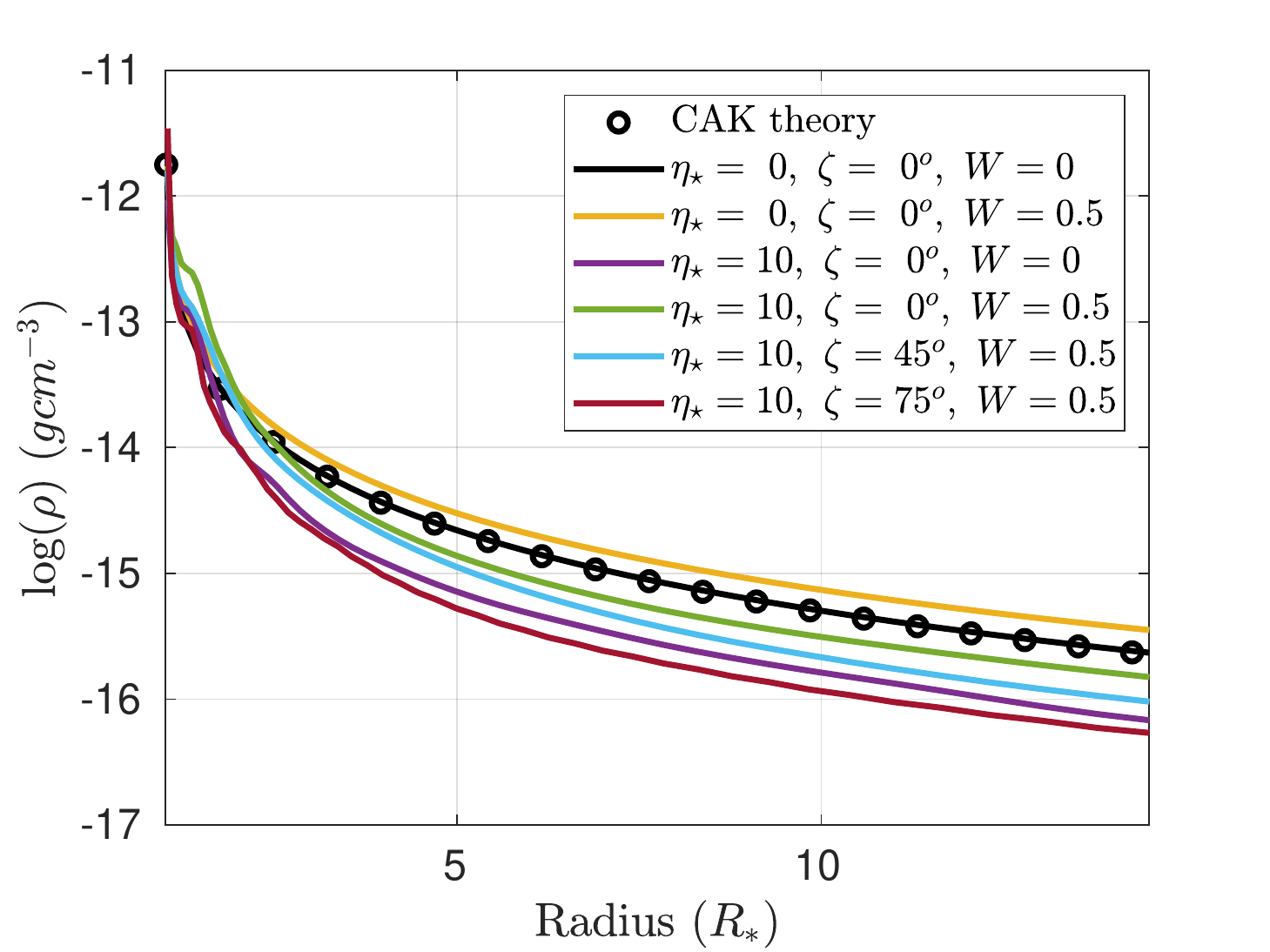}
    \includegraphics[scale=0.58]{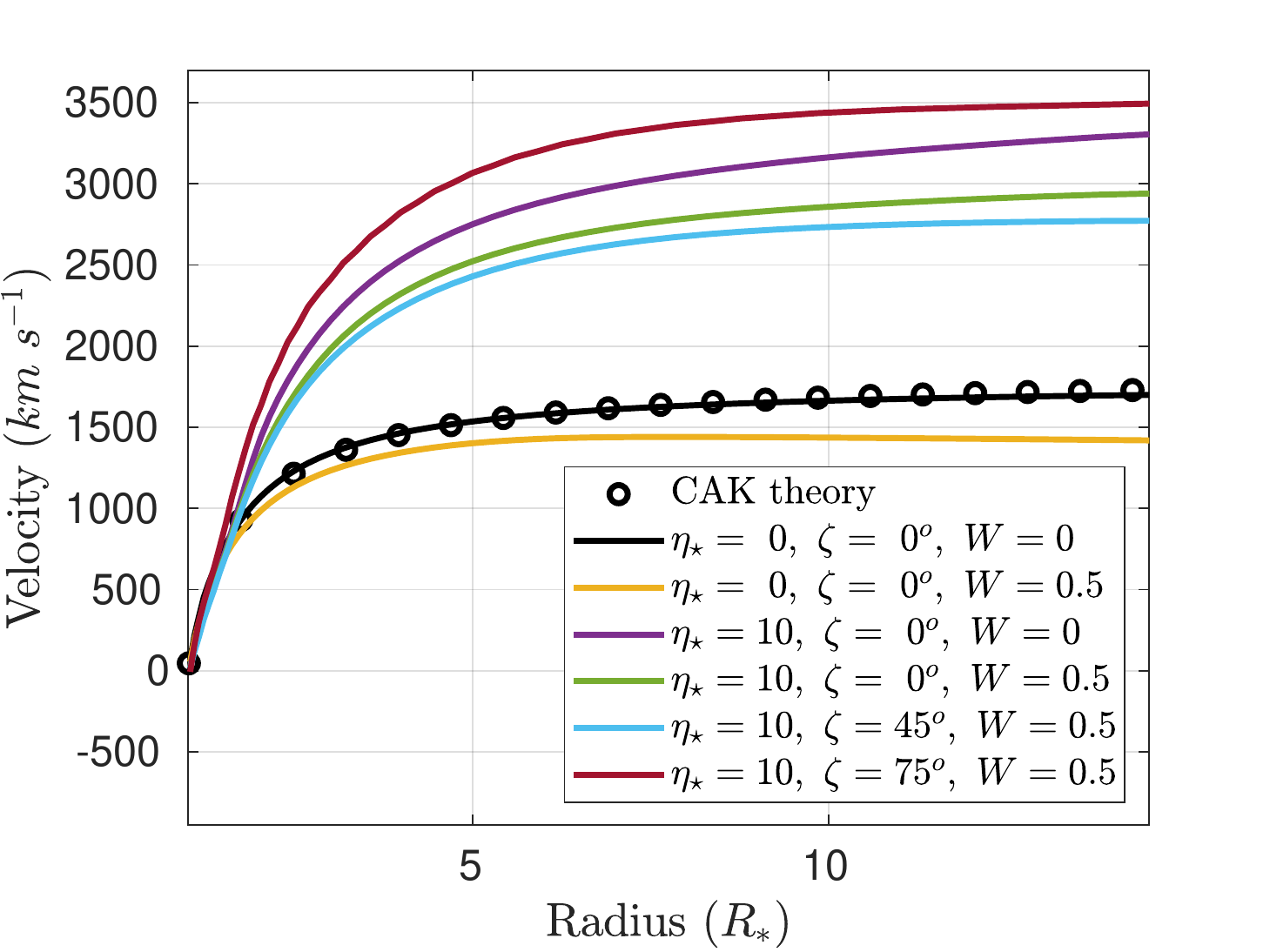}\par
    \hspace{42mm}(a)\hspace{80mm}(b) \par
    \includegraphics[scale=0.58]{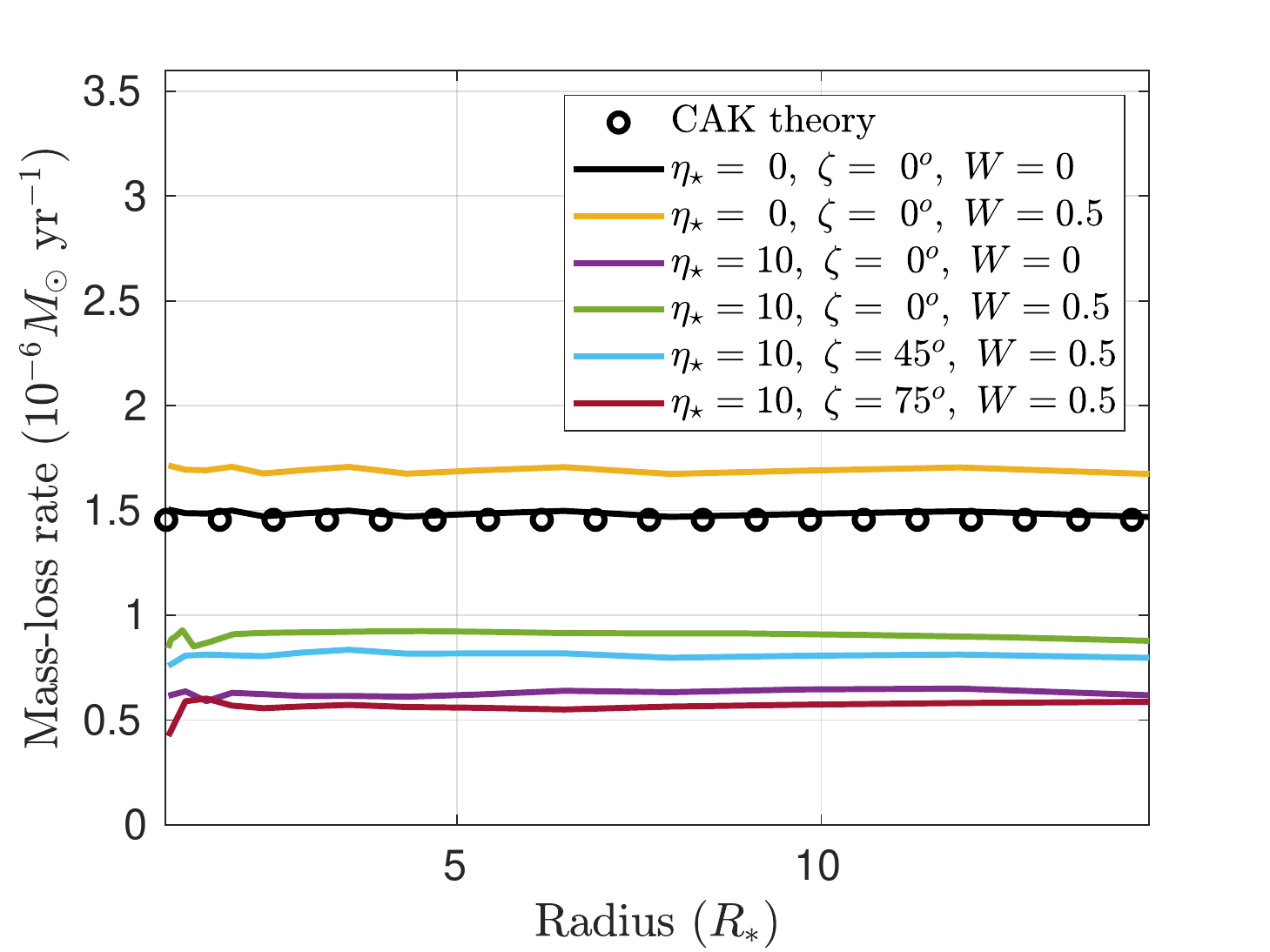}
    \includegraphics[scale=0.58]{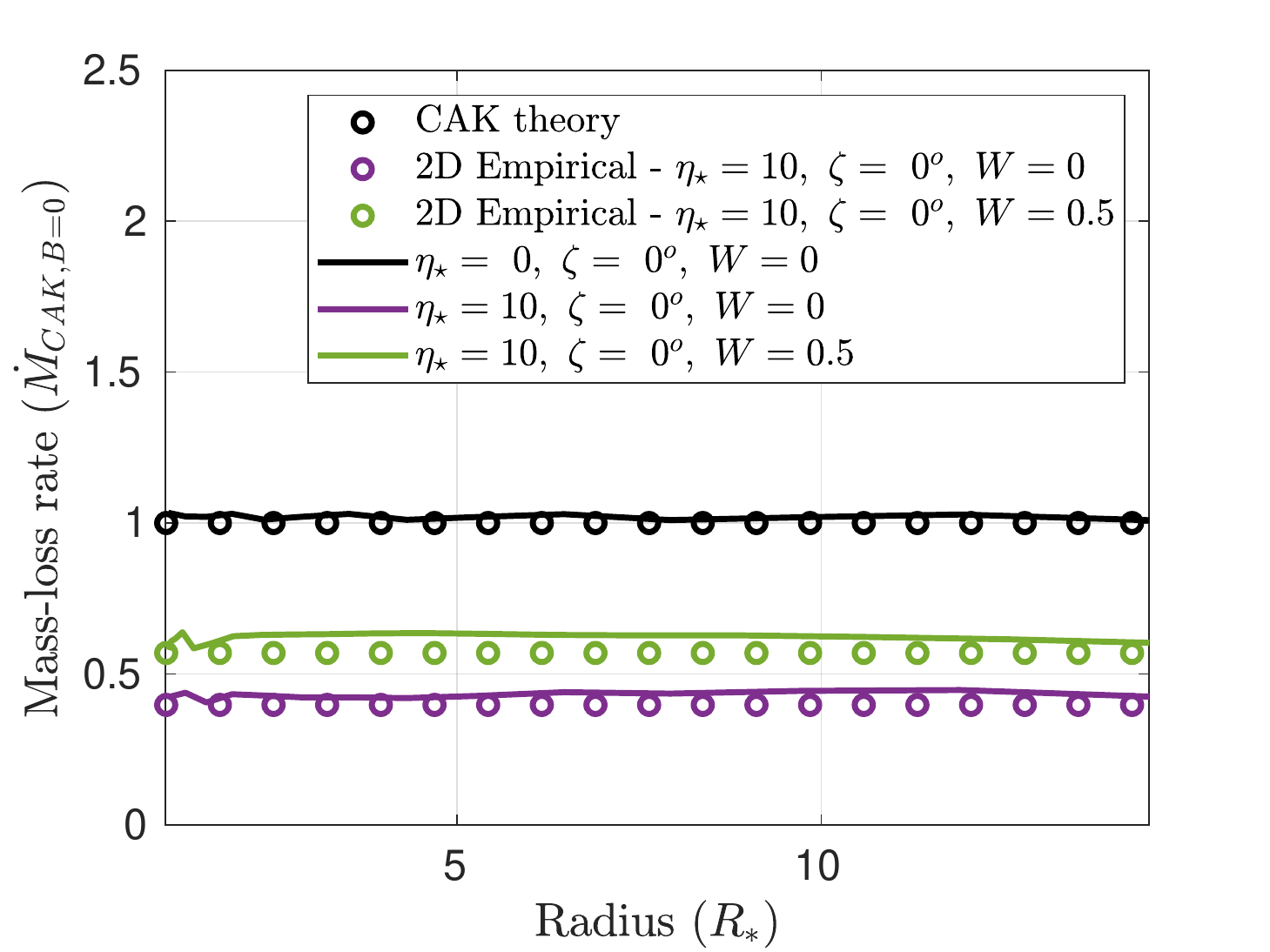}\par
    \hspace{42mm}(c)\hspace{80mm}(d) \par
\caption{Variation of (a) log density, (b) radial velocity and (c) mass-loss
         rate as a function of radial distance $R$. (d) mass-loss rate for the aligned cases
         in comparison with empirical formula derived in \cite{2008MNRAS.385...97U} The quantities are
         averaged over $\theta, \phi$ directions, for six sets
         of magnetic confinement $\eta_\star$, magnetic tilt angle $\zeta$ and critical rotation
         fraction $W$. The non-magnetic CAK mass-loss rate
         is shown with open circles.}
\label{fig:RhoVr_R}
\end{figure}

In the first simulation with no magnetic field and no rotation, a steady state is reached quickly,
and, reassuringly, the density, velocity and mass-loss profiles follow those predicted by CAK theory.
These can be seen, respectively in Figures~6a, 6b and 6c as solid black lines (simulation) 
and black open circles (CAK theory).
In the second simulation, rapid rotation is accounted for by means of centrifugal and Coriolis forces,
as described in Appendix \ref{sec:App_MHD}.
These forces act perpendicular to the $z$ rotation axis and  
tend to increase density and mass-loss rate, but decrease radial velocity,
approaching the rotational equator in the $xy$-plane.
Averaged over $\theta, \phi$, this can
be seen as the yellow solid radial profiles in Figs.~6a-c.
The $\sim 15 \%$ increase in the mass loss rate, averaged over $\theta, \phi$ at the outer boundary,
can be seen in the last column of Table~\ref{tab:mdot}, and 
is similar to the 2D simulation results obtained by \cite{2008MNRAS.385...97U}.
In the absence of a magnetic field, the increase in $\dot{M}$ is accompanied by an average
{\em decrease} in $v_\infty$.
This can be understood as a decrease in the escape velocity $v_{\rm esc}$ with increased rotation.

For the magnetic simulations with $\eta_\star = 10, 50$, $W=0, 0.5$
and $\zeta = 0\degr, 45\degr, 75\degr$ (see Table \ref{tab:mdot})
the initially dipolar field is stretched open near the magnetic poles, 
and remains closed near the magnetic equator, out to approximately the Alfv\'en radius,
channeling wind material into the closed magnetosphere, thereby reducing the mass-loss rate at the outer
boundary. The reduction in $\dot{M}$ is more pronounced for higher magnetic confinement,
consistent with prior 2D results
\citep{2008MNRAS.385...97U}.
Table~1 also shows that, in all but one case ($\eta_\star=50$, $\zeta=75\degr$),
higher rotation leads to higher $\dot{M}$.

From Table \ref{tab:mdot} for $\eta_\star = 10$, we see that as the tilt increases,
the mass-loss rate is reduced. This behaviour suggests a dot-product $\cos\zeta$ variation involving
the magnetic dipole moment ${\bf m_B}$ and the angular velocity  ${\bf \Omega}$.
That is, the moment arm of the mass outflow is reduced as tilt increases,
thereby reducing the net mass-loss rate. The same is not true, however, at higher confinement,
where the maximum $\dot{M}$ is achieved for $\zeta=45\degr$.

In fact, the phenomenon depends upon the geometry of the tilt in addition to the competition
between magnetic confinement and centrifugal forces. Therefore, a larger parameter study 
will be required to yield meaningful predictions.
The complexity of the situation can be seen for 
$\eta_\star=50$, $W=0.5$ and $\zeta=45\degr$, where the mass-loss rate is 
higher than both the $\zeta=0\degr$ and $\zeta=75\degr$ cases.
Here we resort to the simulations to explain the situation.
At high field $\eta_\star = 50$, with no tilt $\zeta=0$, and rapid rotation $W=0.5$,
much of the outflow close to the rotational equator is trapped in the magnetosphere.
Because the centrifugal acceleration is greatest in the equatorial XY plane,
$\dot{M}$ is not significantly increased by rapid rotation. This is evident in
rows 7 and 8 of Table~1: the mass-loss rates at $W=0$ and $W=0.5$ are comparable.
Examining Figure {\ref{fig:dynet5e1d45W0p50}}, with $W=0.5$, $\eta_\star=50$, and $\zeta=45\degr$,
most of the outflow near the rotational equator is along open field lines,
thereby maximizing $\dot{M}$ and $\dot{J}$, the total rate of angular momentum loss.
Figure {\ref{fig:dynet5e1d45W0p50}} will be discussed in more detail in section 5.
The total angular momentum loss is listed in Table~2 and discussed in section 6 below.

The Figure \ref{fig:RhoVr_R} shows the variation in log density, velocity and 
mass-loss rate as a function of radius at the end of each simulation.
These are averaged over the $\theta, \phi$ to illustrate their radial variation.
Outside the Alfv\'en radii, the mass-loss rate remains constant,
while the density decreases rapidly and the velocity increases, as expected from CAK.
In general, the presence of magnetic confinement $\eta_\star > 1$
increases the overall radial velocity, while the density profile
undergoes a reduction due to magnetic confinement. In the same way,
in the presence of rotation, the density profile and the mass-loss rate
increase, as the wind is flung out by rotational forces, consistent
with prior 2D simulations \citep{2008MNRAS.385...97U}. 

For the aligned rotator case, there is some well-developed theory for the mass loss case even when we have rotation and magnetic fields. We compare our simulations for aligned rotators with that theory in this paragraph. In figure {\ref{fig:RhoVr_R}}d, the mass-loss rate for aligned magnetic field
cases are presented. The mass-loss rate is normalized with respect to the
non-magnetic and no-rotation case ($\dot M_{B=0}$). The green and purple
circles in the plot are obtained from the analytic formula provided in equations 23 and 24 of
\citet{2008MNRAS.385...97U} and repeated here as:

\begin{equation*}\label{eq:djgas_dt}
    \dfrac{\dot M_B}{\dot M_{B=0}} \approx 1 - \sqrt{1 - R_*/R_c} 
\end{equation*}

\begin{equation*}\label{eq:djgas_dt}
    \dfrac{\dot M_B}{\dot M_{B=0}} \approx 1 - \sqrt{1 - R_*/R_c} + 1 - \sqrt{1 - 0.5R_*/R_K}
\end{equation*}
where, $R_c$ is the confinement radius of the magnetic closed loops. From our simulation
results the confinement radius is observed to be of the form, $R_c \approx R_* + 0.6(R_A - R_*)$. We see from figure {\ref{fig:RhoVr_R}}d that the agreement between our simulations and the theory is quite good. 
In the quasi-steady state, the simulations show 
the magnetic confinement of the wind and the episodic breakout events at 
the magnetic equator. This is detailed in section \ref{sec:dynm} below.


\section{The Dynamics of Magnetic Channeling in the Wind } \label{sec:dynm}

\begin{figure} 
    \includegraphics[scale=0.235]{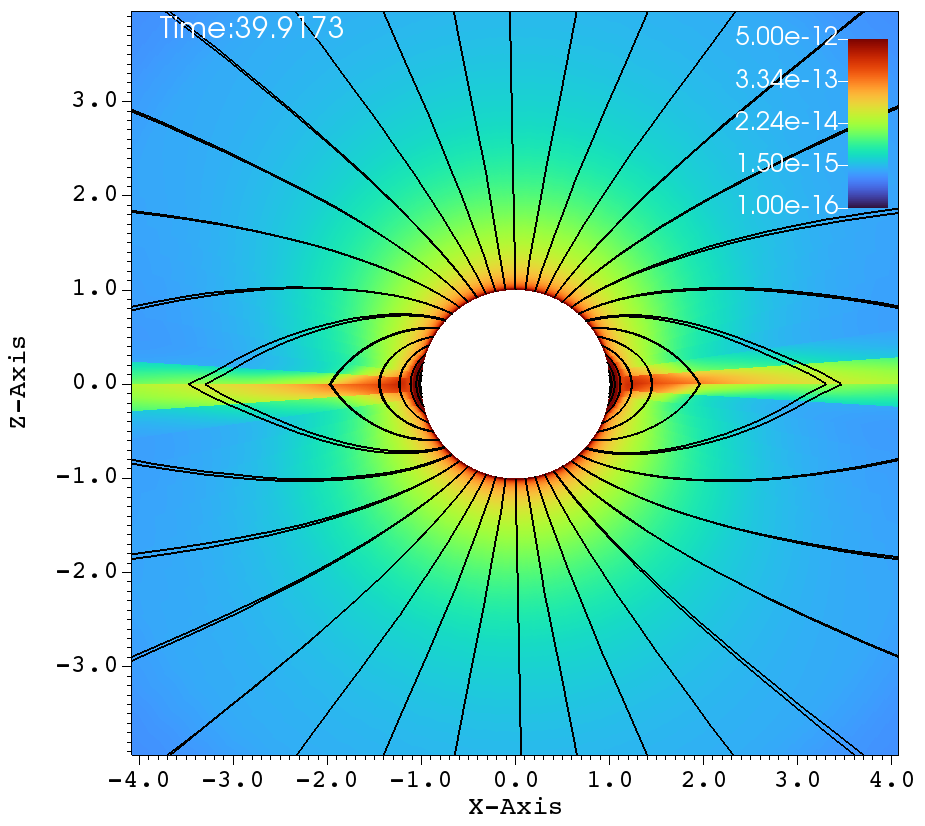}
    \includegraphics[scale=0.235]{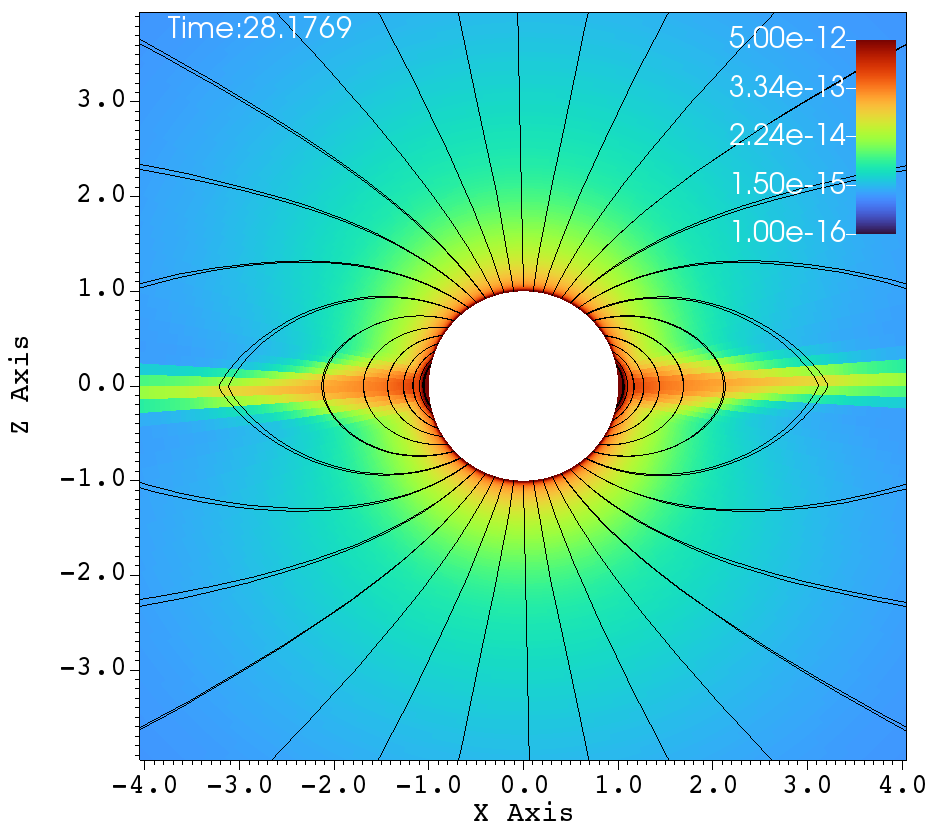}\par

    \hspace{38mm}(a)\hspace{78mm}(b) \par
    \includegraphics[scale=0.235]{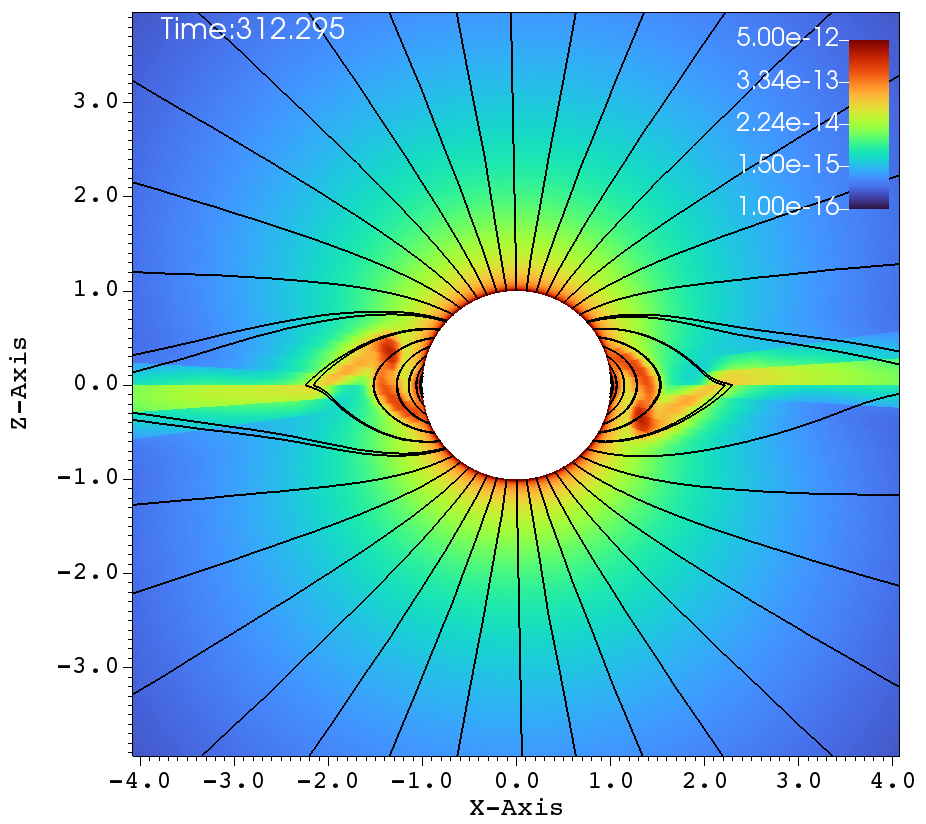}
    \includegraphics[scale=0.235]{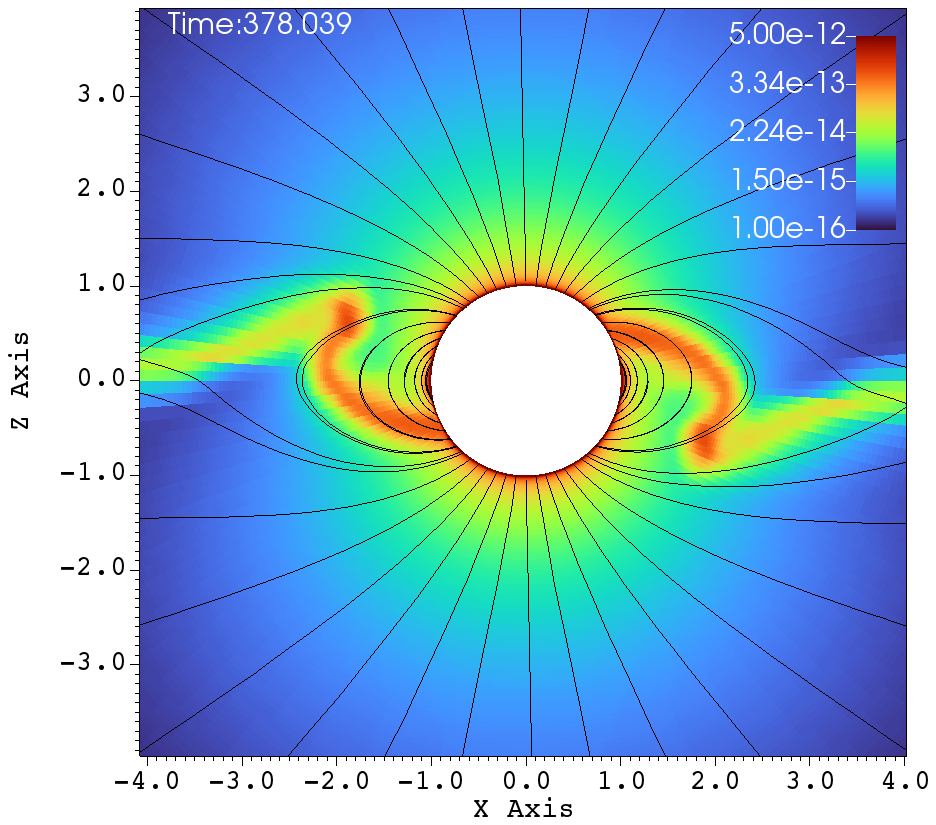}\par

    \hspace{38mm}(c)\hspace{78mm}(d) \par
    \includegraphics[scale=0.235]{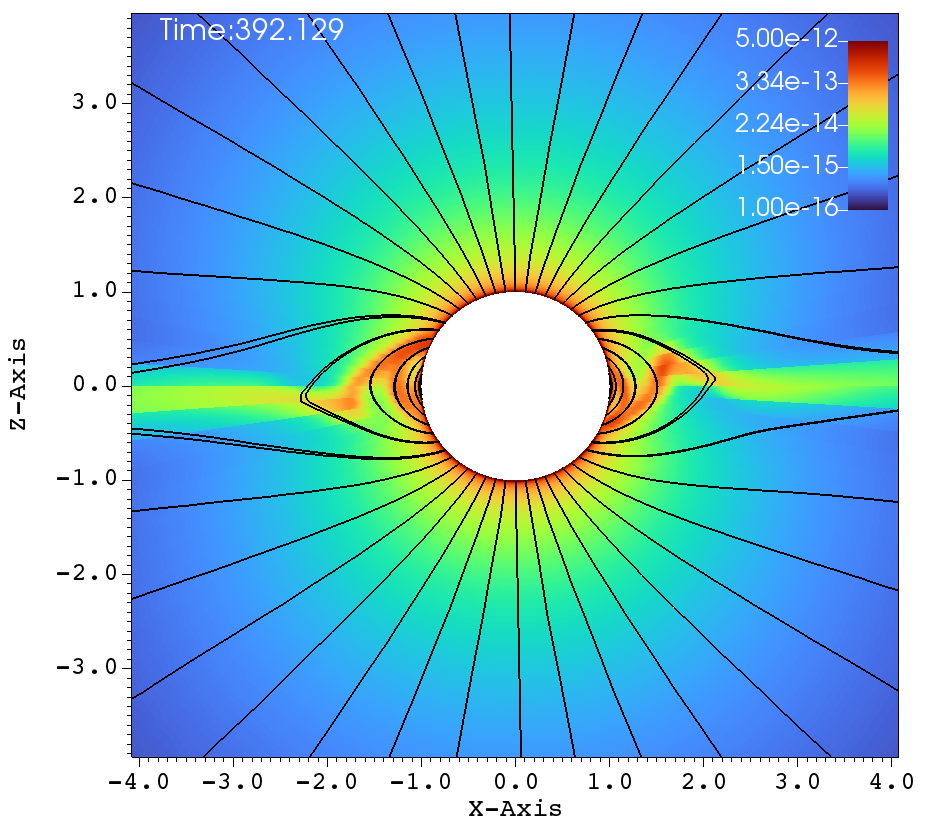}
    \includegraphics[scale=0.235]{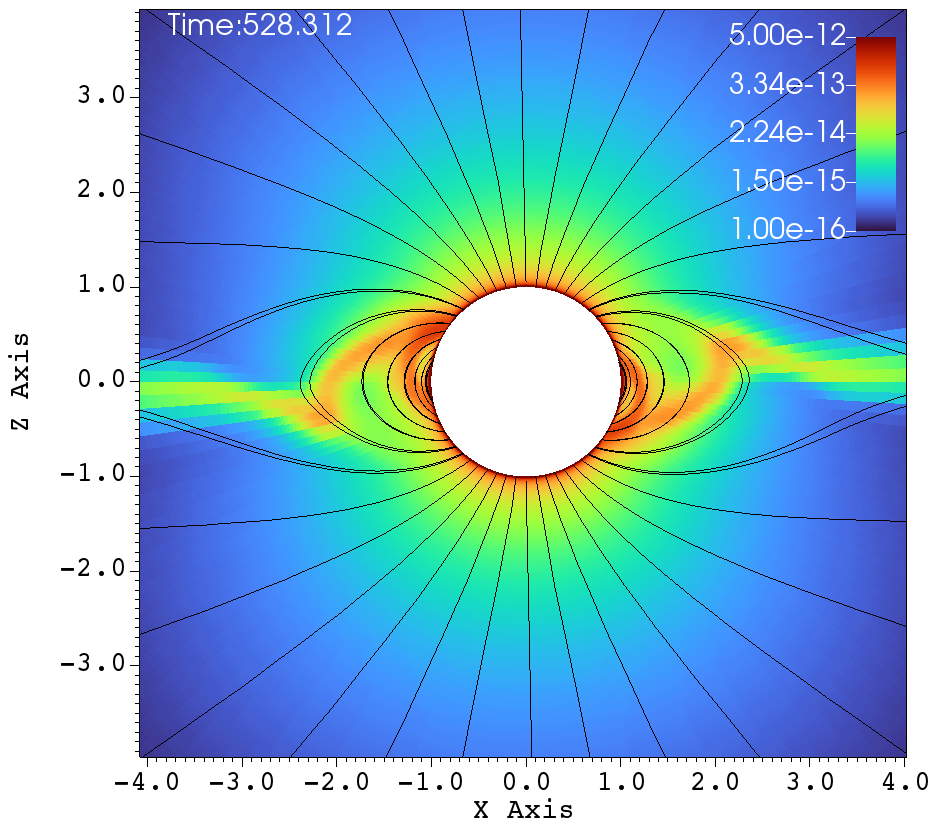}\par
    
    \hspace{38mm}(e)\hspace{78mm}(f) \par
\caption{Density ($ \rm g~cm^{-3}$) colorized with the magnetic 
field lines, at different times (ksec) of the simulation, for the case 
of $\eta_\star = 10$ (a,c,e) and $\eta_\star = 50$ (b,d,f), 
no rotation, showing the magnetic confinement and episodic outflow. 
The $x$ and $z$ axes values are in the 
scale of stellar radius $R_\star$.}
\label{fig:dynet1e1d00W0p00}
\end{figure}

The mass outflow from the surface of the star is free to flow at the 
magnetic poles and it is confined near the equatorial region by the closed
loops of the magnetic dipole field. Due to this, the wind coming from the 
either side of the equatorial region, guided  by the magnetic field lines, 
meet and cause a field lines breaking at the magnetic equator. This overall
behaviour of closed magnetic loops near the surface of the star and open
field lines as we move farther can be seen in figures 
\ref{fig:dynet1e1d00W0p00}.  

Figures \ref{fig:dynet1e1d00W0p00}a-f shows the density colorized with 
the magnetic field lines at different times of the simulation, for the
cases of magnetic field strength $\eta_\star = 10, 50$ and no rotation.
These figures show the expected dynamics of a magnetically channeled 
line driven wind. We see that each closed loop has two foot-points that 
connect to the star at similar moment arms. Owing to the radially outward 
line driving force, both foot-points have similar amounts of matter that 
is driven out from the surface of the star. 
Because Figs. \ref{fig:dynet1e1d00W0p00}a, b, c correspond to a smaller magnetic 
field than Figs 7d, e, f, we can clearly see that the former set of figures produce 
a smaller magnetosphere than the latter set of figures.

The matter driven out from the either side of the star meet at the equatorial
region, forming a dense knot. The knot is at a much higher density than the rest 
of the wind, and hence experiences a greater gravitational attraction towards
the surface of the star. However, the closed field lines just below the
knot prevents the direct fallback of matter on to the star and hence the 
path of the density knot fallback is dictated by the magnetic field lines. 
This can be seen in figures \ref{fig:dynet1e1d00W0p00}c, d. These
observations are inline with the 2D simulations of \cite{2002ApJ...576..413U}.
In addition, in this 3D simulation, it can be seen that the density knots 
at the either end of the magnetic equator fallback from the opposite sides of 
the field lines, thereby conserving the momentum of the star (figures
\ref{fig:dynet1e1d00W0p00}c-f). Also, the overall outflow is guided by the
magnetic field lines from the either sides of the pole, in an alternate fashion
and hence the subsequent density fallbacks come from the either side of
the magnetic equator. This is also inline with the observations made in 
\cite{2002ApJ...576..413U}.

\begin{figure}
    \includegraphics[scale=0.235]{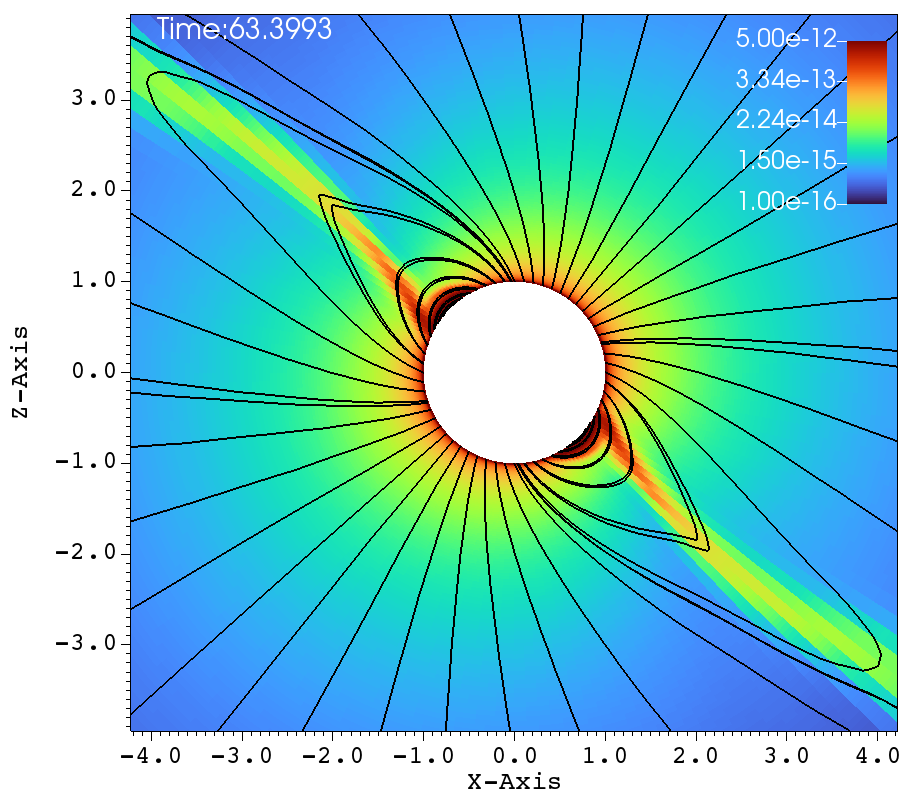}
    \includegraphics[scale=0.235]{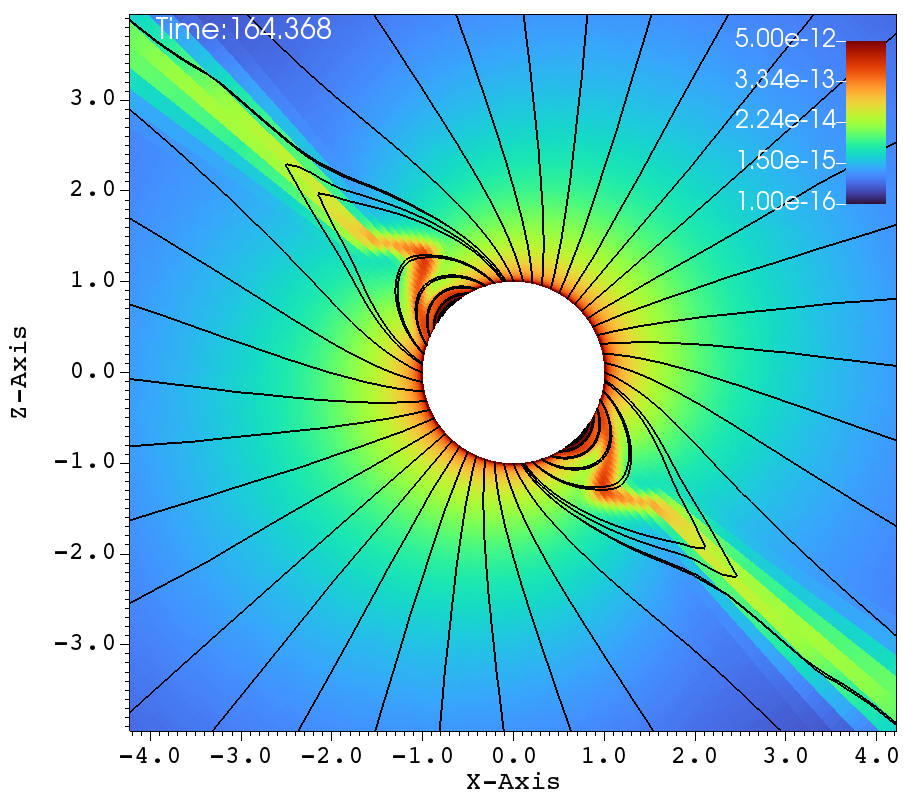}\par
    \hspace{38mm}(a)\hspace{78mm}(b) \par
    \includegraphics[scale=0.235]{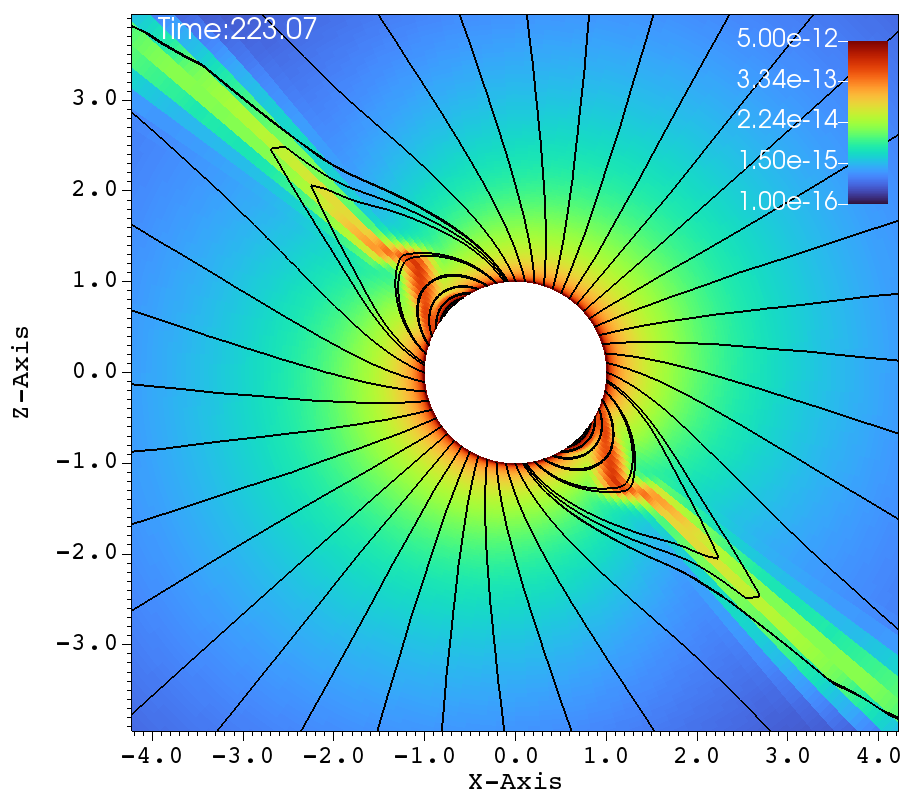}
    \includegraphics[scale=0.235]{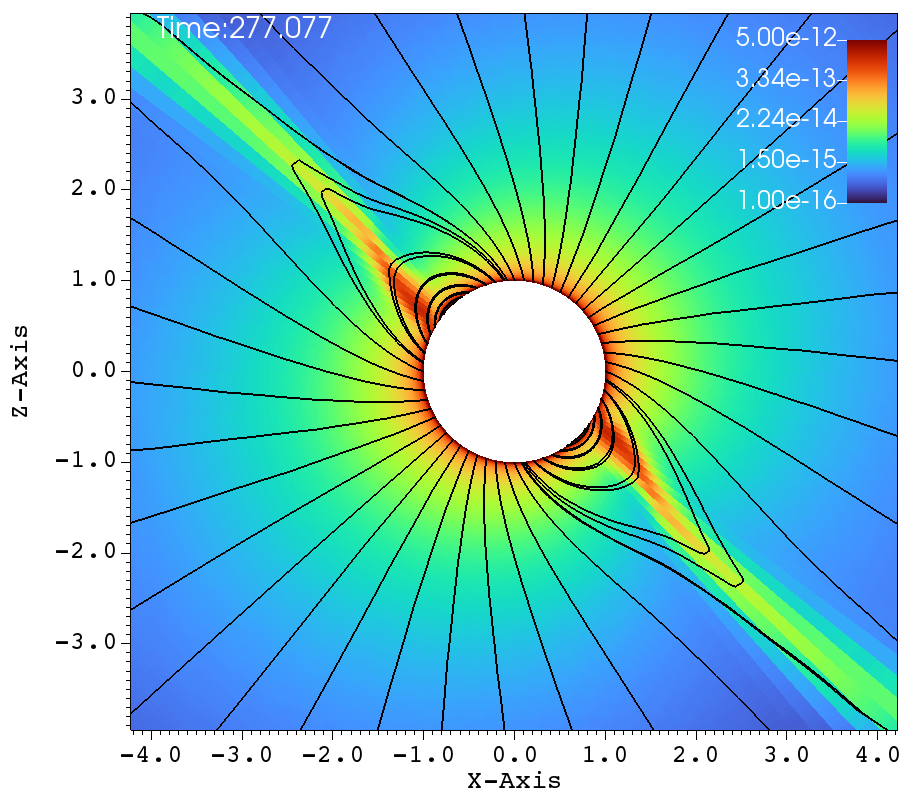}\par
    \hspace{38mm}(c)\hspace{78mm}(d) \par
    \includegraphics[scale=0.235]{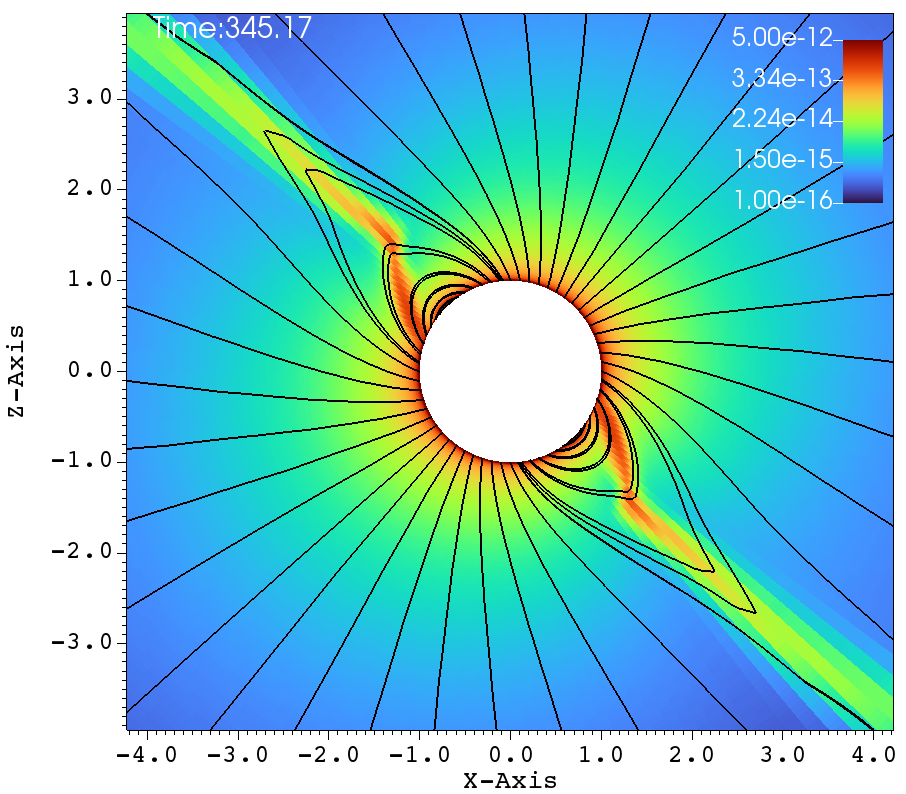}
    \includegraphics[scale=0.235]{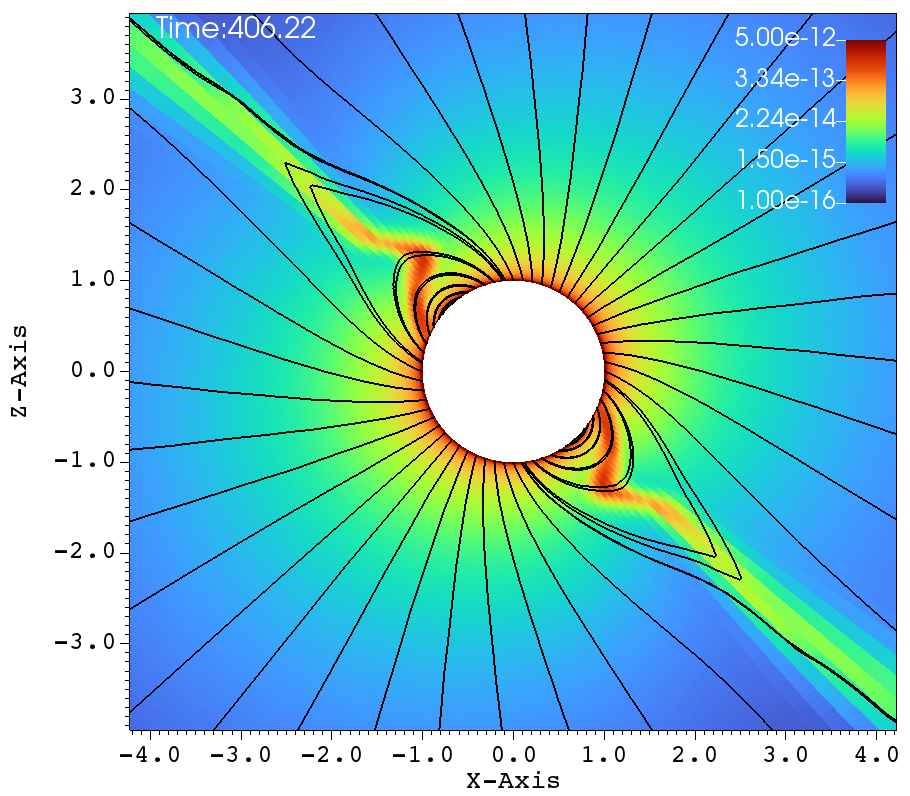}\par
    \hspace{38mm}(e)\hspace{78mm}(f) \par
\caption{Density ($\rm g~cm^{-3}$) colorized with the
magnetic field lines, at different times (ksec) of the simulation, for the case 
of $45\degr$ tilted magnetic dipole with $\eta_\star = 10$ and with the rotation
of $W=0.5$, showing the magnetic confinement and the centrifugal pull due to 
the rotation. The $x$ and $z$ axes values are in the scale of stellar 
radius $R_\star$.}
\label{fig:dynet1e1d45W0p50}
\end{figure}

The next set of figures \ref{fig:dynet1e1d45W0p50}a-f, represent the 
simulations carried out with the high rotation and an inclined magnetic
dipole. The magnetic field has the strength of $\eta_\star = 10$ and the rotation
rate is $W=0.5$, the rotation axis is the $z-$axis and the magnetic dipole
makes a $\zeta = 45\degr$ inclination with the rotation axis. 
 The dynamics are different and interesting in 
Figure~{\ref{fig:dynet1e1d45W0p50}}, when comparing to that of 
Figure~{\ref{fig:dynet1e1d00W0p00}}. 
Due to the tilt and high rotation, the streams of gas that come off from the two 
foot-points of a closed magnetic loop experiences unequal amount of 
centrifugal force. The magnetic foot-point that is closer to the
rotational  equator has more centrifugal force and is flung out faster. The other 
magnetic foot-point that is closer to the pole, does not have centrifugal 
assist when being flung outwards. Therefore the streamer that comes off 
from the foot-point that is closer to the rotational-equator dominates and 
hence that one dominant streamer is doing the flailing. 
This is in contrast to the 
case of aligned rotation, where, there are two streamers at either sides of the 
magnetic equator that go back and forth.

\begin{figure}
    \includegraphics[scale=0.22]{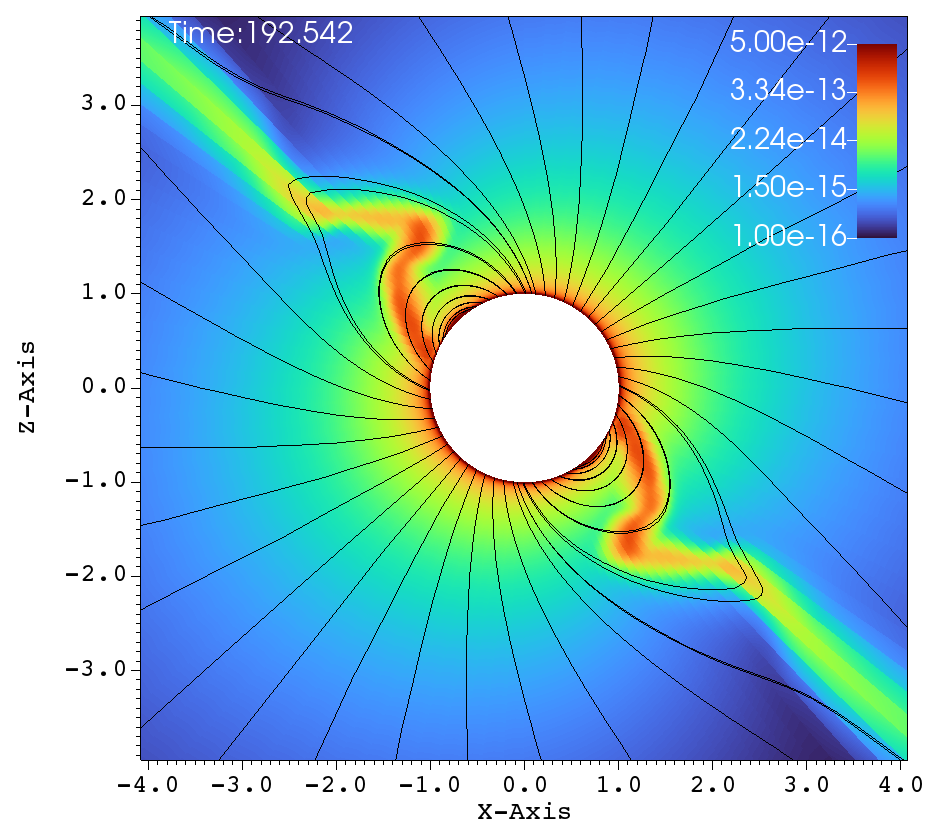}
    \includegraphics[scale=0.22]{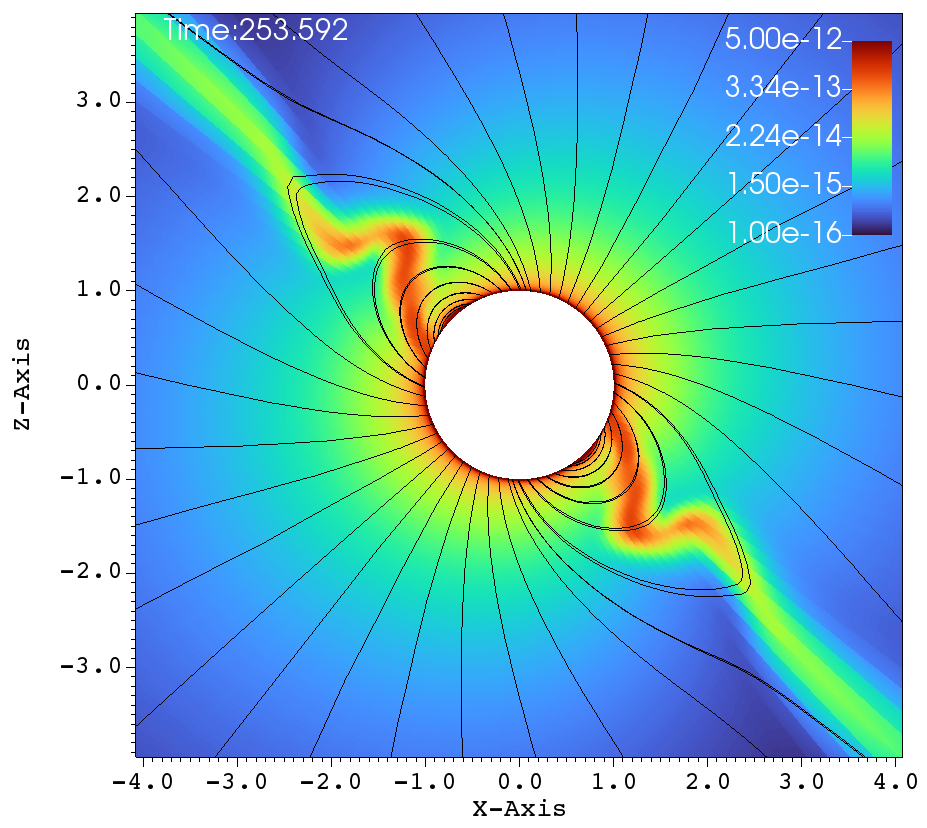}\par
    \hspace{38mm}(a)\hspace{78mm}(b) \par
    \includegraphics[scale=0.22]{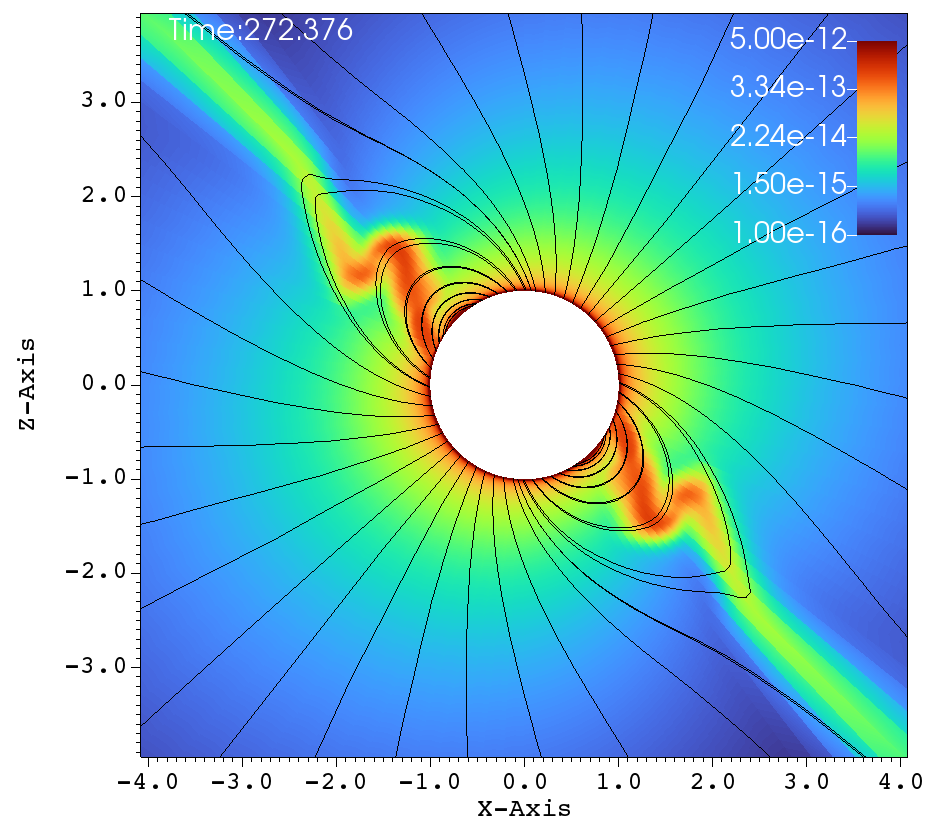}
    \includegraphics[scale=0.22]{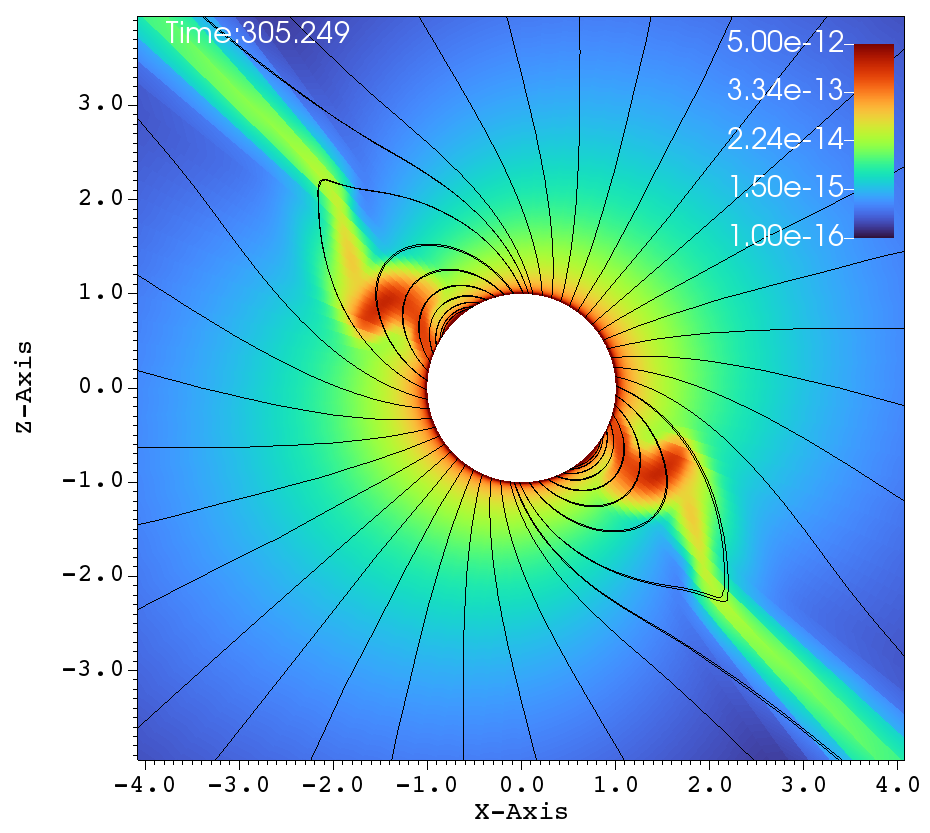}\par
    \hspace{38mm}(c)\hspace{78mm}(d) \par
    \includegraphics[scale=0.22]{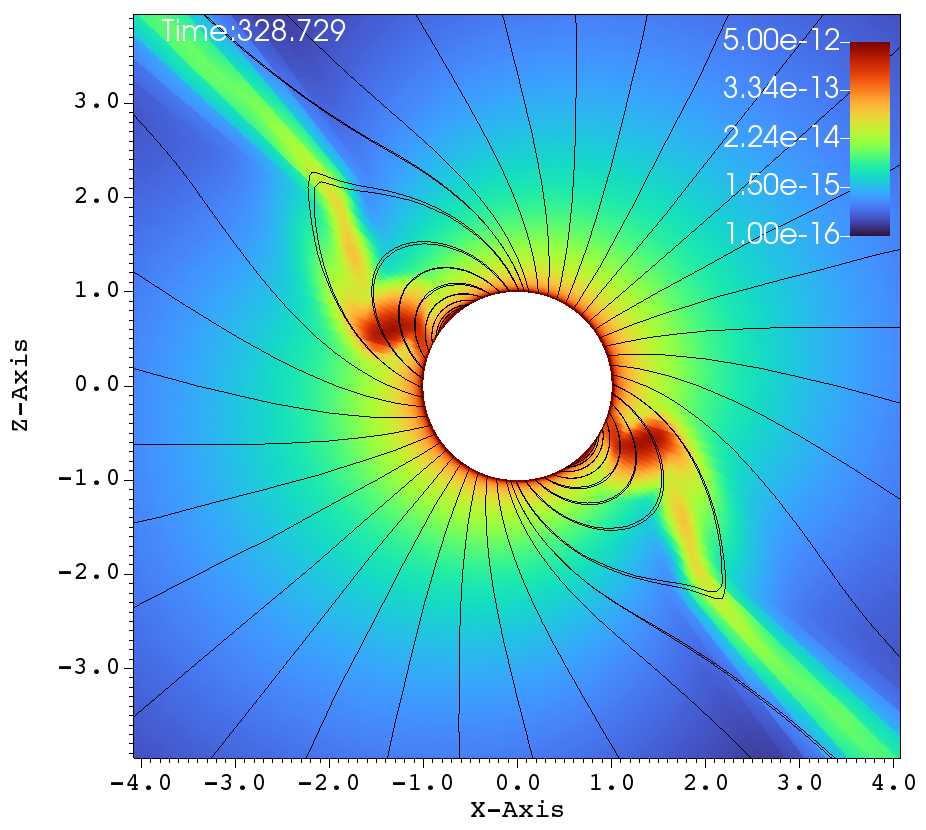}
    \includegraphics[scale=0.22]{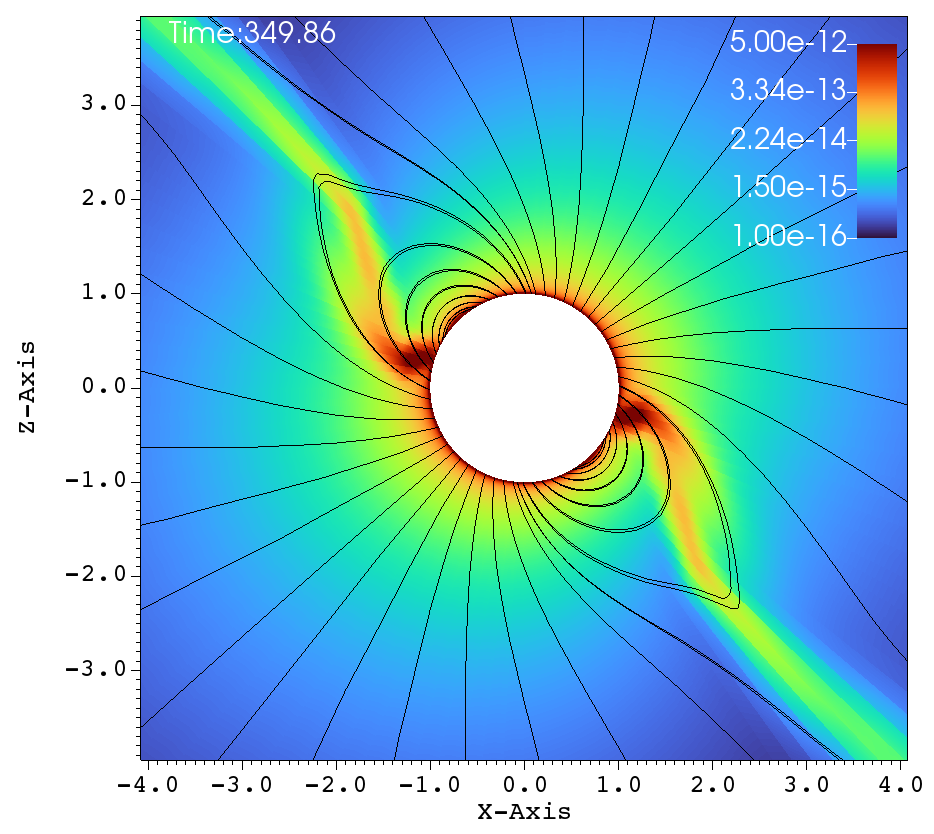}\par
    \hspace{38mm}(e)\hspace{78mm}(f) \par
\caption{Density ($\rm g~cm^{-3}$) colorized with the
magnetic field lines, at different times (ksec) of the simulation, for the case 
of $45\degr$ tilted magnetic dipole with $\eta_\star = 50$ and with the rotation
of $W=0.5$. The $x$ and $z$ axes values are in the scale of stellar 
radius $R_\star$.
While Figure~\ref{fig:dynet1e1d45W0p50} showed uniformly spread time intervals,
this figure shows a set of snapshots that are bunched in time that reveal
the filling of the magnetosphere and episodic in-fall of clumps. One entire episode
from the start of clump formation to its in-fall is shown. }
\label{fig:dynet5e1d45W0p50}
\end{figure}

Similar lines of observations can be made for the 
figures {\ref{fig:dynet5e1d45W0p50}}a-f, that corresponds 
to $\eta_\star=50$, $W=0.5$, $\zeta=45\degr$ case. 
In table {\ref{tab:mdot}}, this case has $R_A=2.9$ and $R_K=1.59$ and 
it has been classified as a centrifugal magnetosphere (CM) based on the traditional explanations that have been developed so far.
Upon observing the images at different times in 
Figure~{\ref{fig:dynet5e1d45W0p50}}a-f, it can be seen that the magnetosphere 
successively filled and with clumps falling down on the star. This 
situation is similar to that of Figure~{\ref{fig:dynet1e1d45W0p50}}a-f.
Here, one would therefore be inclined to conclude that the simulation in 
Figure~{\ref{fig:dynet5e1d45W0p50}}a-f is a 
dynamical magnetosphere (DM) even though it was classified as a CM.
But a little reflection allows us to understand that $R_A$ should be modified
by $cos \zeta$ so that $~R_A~cos \zeta \sim R_K$. Here, the tilt has brought 
the cosine-modified Alfven radius much closer to the Keplarian radius. 
This might explain two important
features in Figure~{\ref{fig:dynet5e1d45W0p50}}a-f. One, the clumps episodically 
keep filling the magnetosphere and falling back on to the star, similar to
that of a dynamical magnetosphere case. Second, the majority of the
magnetically streamlined mass outflow comes out from the magnetic foot point 
that is closer to the rotational-equator (\emph{xy}-plane).

\begin{figure}
    \includegraphics[scale=0.22]{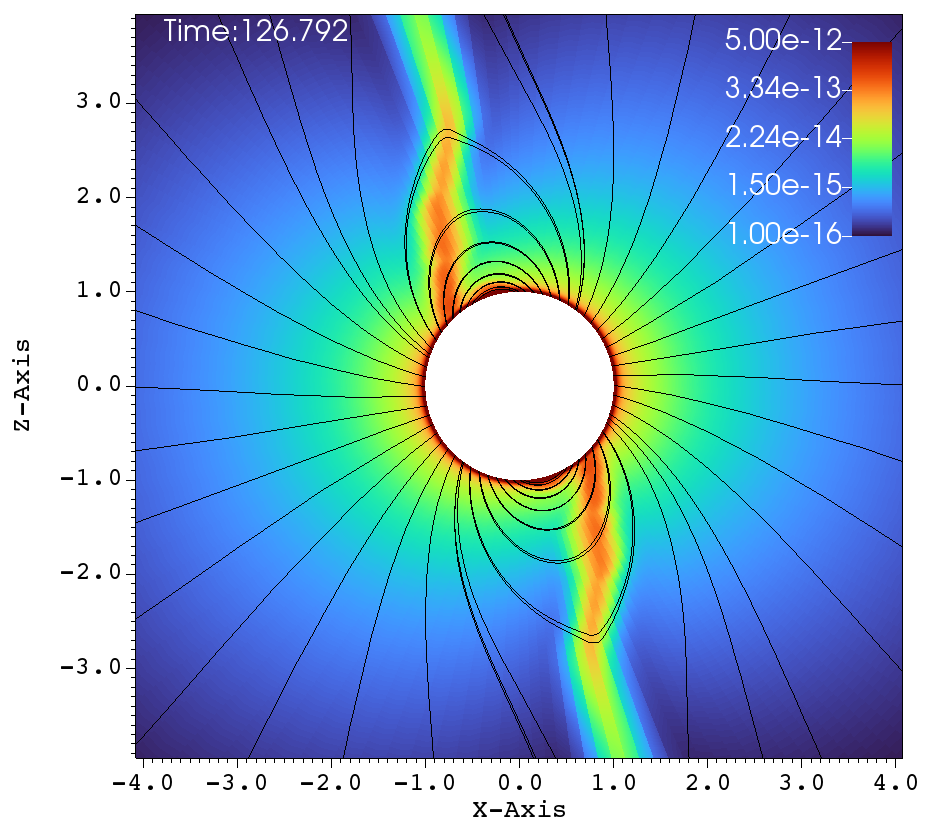}
    \includegraphics[scale=0.22]{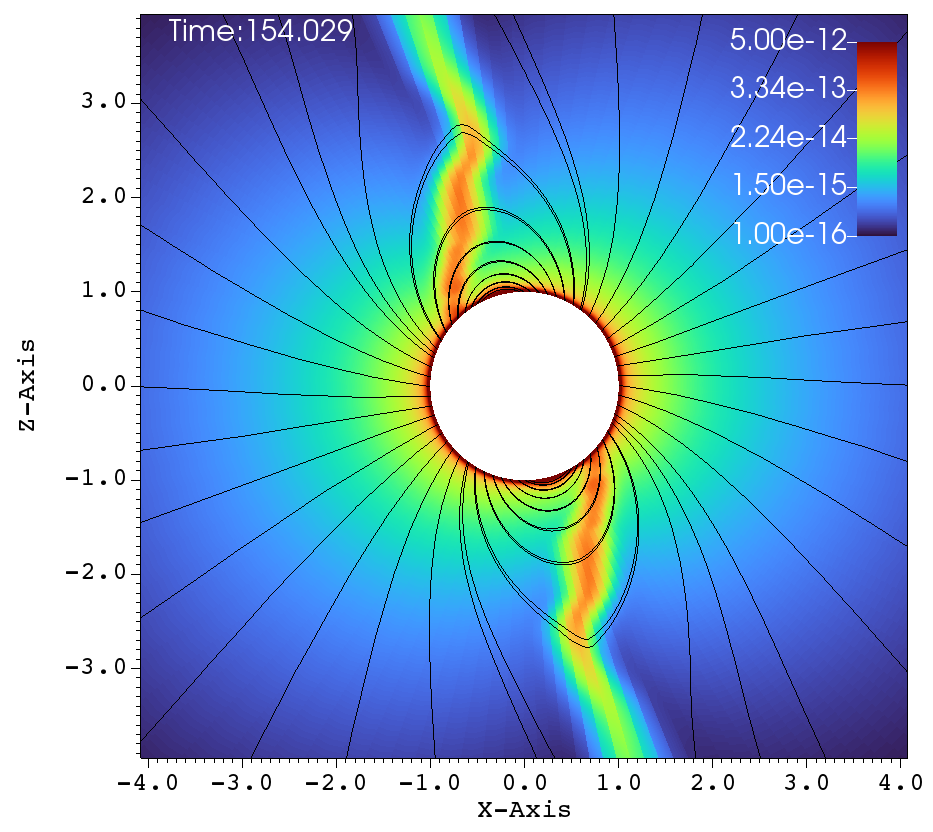}\par
    \hspace{38mm}(a)\hspace{78mm}(b) \par
    \includegraphics[scale=0.22]{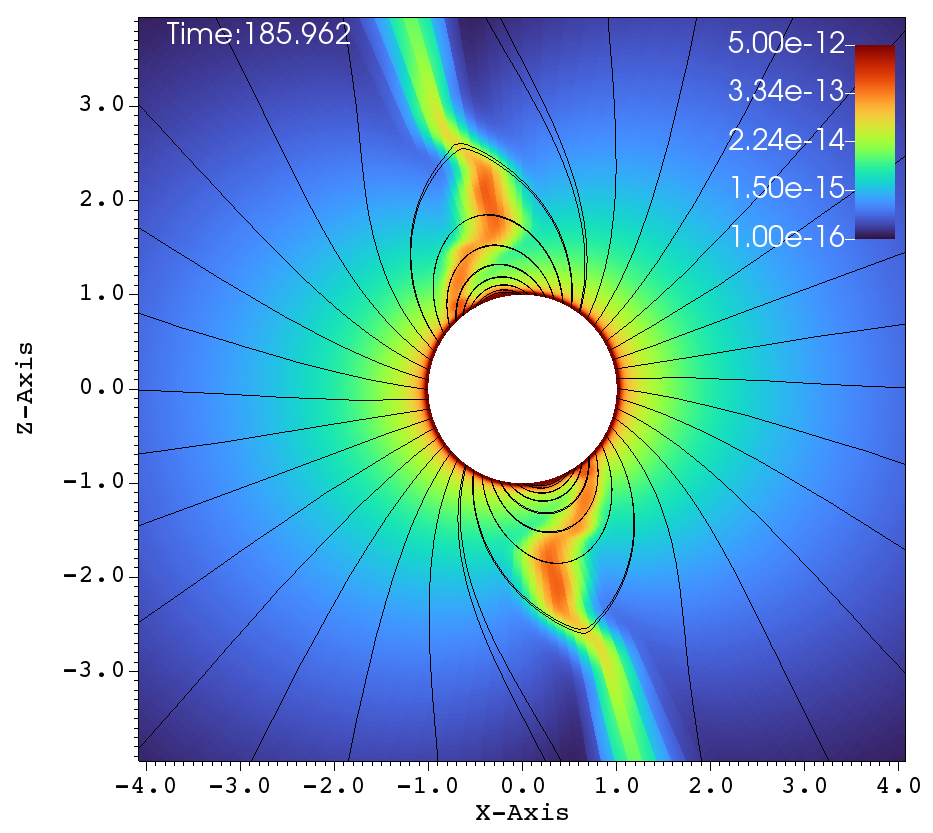}
    \includegraphics[scale=0.22]{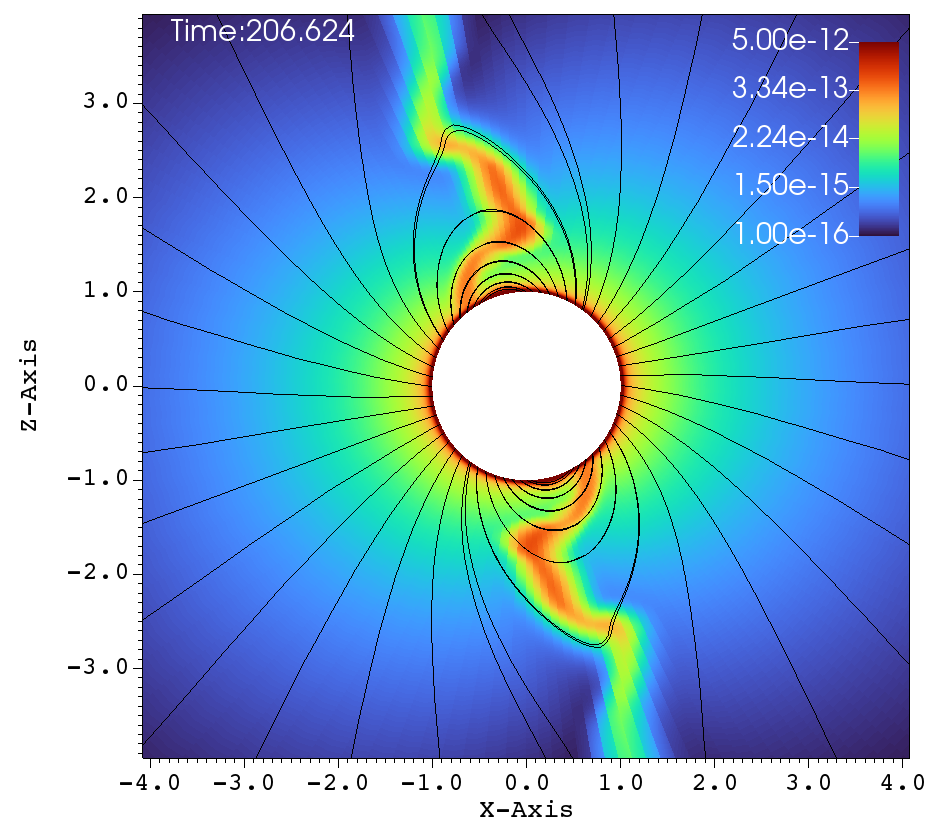}\par
    \hspace{38mm}(c)\hspace{78mm}(d) \par
    \includegraphics[scale=0.22]{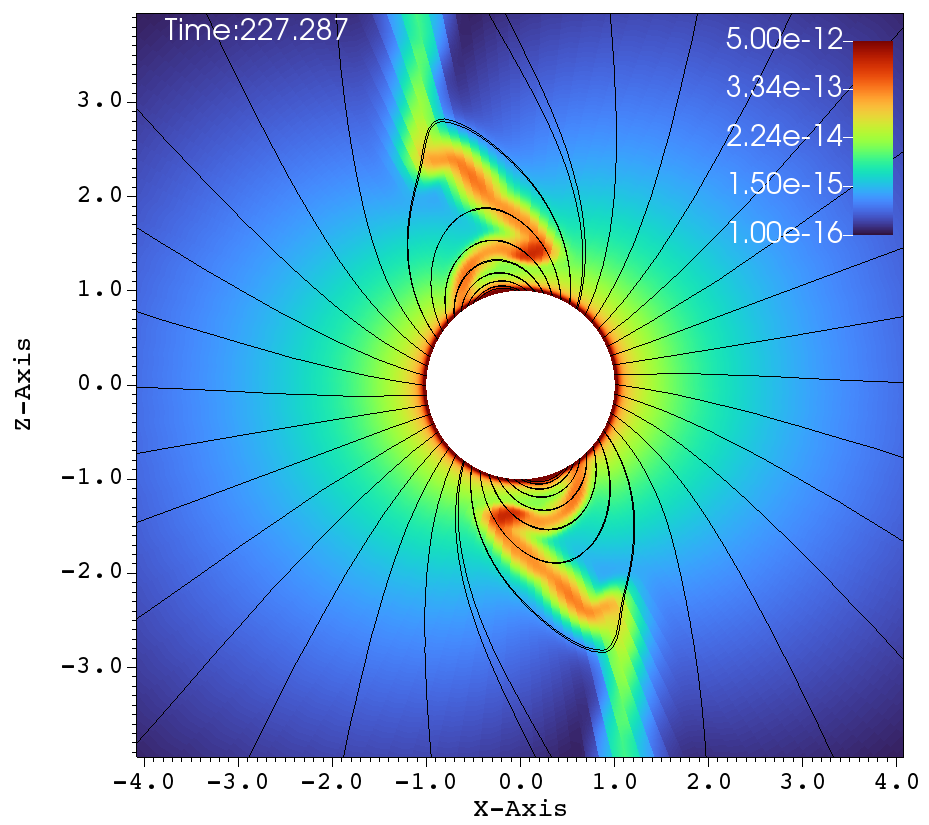}
    \includegraphics[scale=0.22]{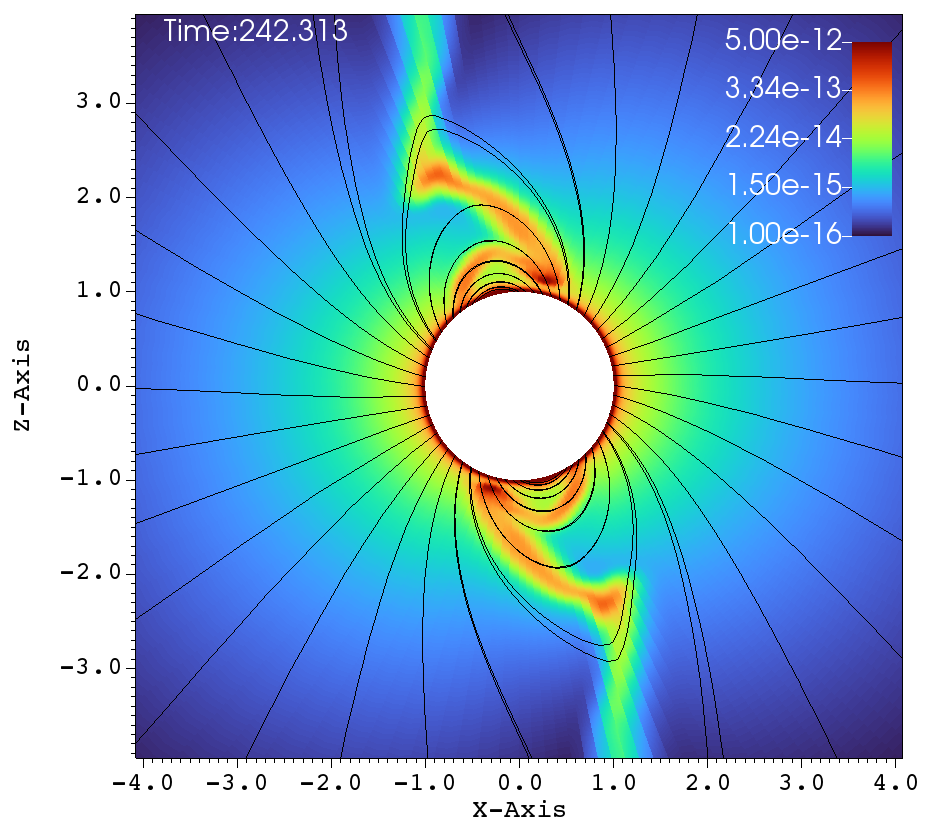}\par
    \hspace{38mm}(e)\hspace{78mm}(f) \par
\caption{Density ($\rm g~cm^{-3}$) colorized with the
magnetic field lines, at different times (ksec) of the simulation, for the case 
of $75\degr$ tilted magnetic dipole with $\eta_\star = 50$ and with the rotation
of $W=0.5$. The $x$ and $z$ axes values are in the scale of stellar 
radius $R_\star$.
While Figure~\ref{fig:dynet1e1d45W0p50} showed uniformly spread time intervals,
this figure shows a set of snapshots that are bunched in time that reveal
the filling of the magnetosphere and episodic in-fall of clumps. One entire episode
from the start of clump formation to its in-fall is shown. }
\label{fig:dynet5e1d75W0p50}
\end{figure}

Figure~\ref{fig:dynet5e1d75W0p50} again shows one episode of the $\eta_\star=50$,
$\zeta=75\degr$ and $W=0.5$ simulation, as in Figure~\ref{fig:dynet5e1d45W0p50}.
We see one clump initiated in Figure~\ref{fig:dynet5e1d75W0p50}a. Figures 
\ref{fig:dynet5e1d75W0p50}b-e show successive times when the clump is
reined in by the magnetic field and 
in-falling material fills the magnetosphere.
Figure~\ref{fig:dynet5e1d75W0p50}f shows the precise
moment when the clump impacts the star. Interestingly, at outer radii in 
Figure~\ref{fig:dynet5e1d75W0p50}f  we can see the start of another episode of 
clump initiation.
%
\begin{figure}
    \centering
    \includegraphics[scale=0.6]{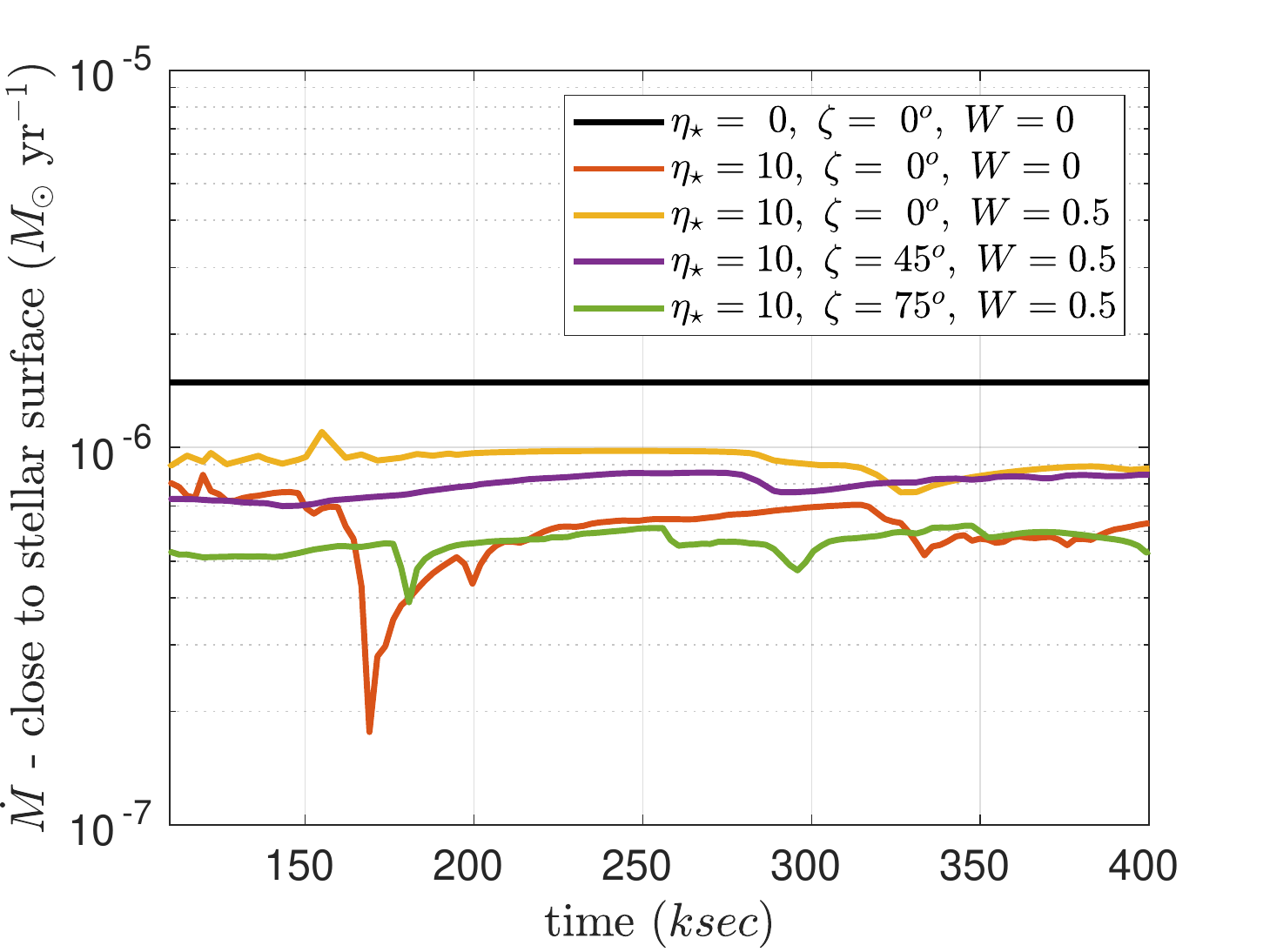}
    \caption{ Observed mass-loss rate close to the stellar surface, at 
    $R=1.1R_\star$, as a function of time (ksec), for different rotation 
    rates and tilt angles. The clump formation and successfully falling 
    back to the star happens for i) $\eta_\star=10, \zeta=0, W=0$ and 
    ii) $\eta_\star=10, \zeta=75, W=0.5$ cases. The abrupt drops in the mass 
    flow rate can be observed for the same two cases, showing the clumps 
    falling back to the star.}
\label{fig:clumpfall}
\end{figure}
The episodes of clump formation and fallback can also be seen, when observing the mass-loss rate close to the 
stellar surface.
Referring to Figure~{\ref{fig:clumpfall}}, the mass-loss rate
experiences abrupt drops for the cases, i) $\eta=10, \zeta=0, W=0$ and 
ii) $\eta=10, \zeta=75, W=0.5$, where the density knot is formed by one side of the stream and that knot proceeds to fall onto star on the other side of the magnetic loop. 
The other cases do not show such
a complete fallback and hence they do not exhibit these sharp drops
mass-loss rate close to the stellar surface.


In summary, the tilted dipole simulations shown in Figs. \ref{fig:dynet1e1d45W0p50}, \ref{fig:dynet5e1d45W0p50} and \ref{fig:dynet5e1d75W0p50}
suggest that the tilt
should be factored in when deciding whether we have a DM or CM. Larger 
simulated data sets will eventually enable us to identify the boundary between DM and CM
in cases where the magnetic dipole has a large tilt relative to the 
rotational axis. 

\begin{figure}
    \includegraphics[scale=0.165]{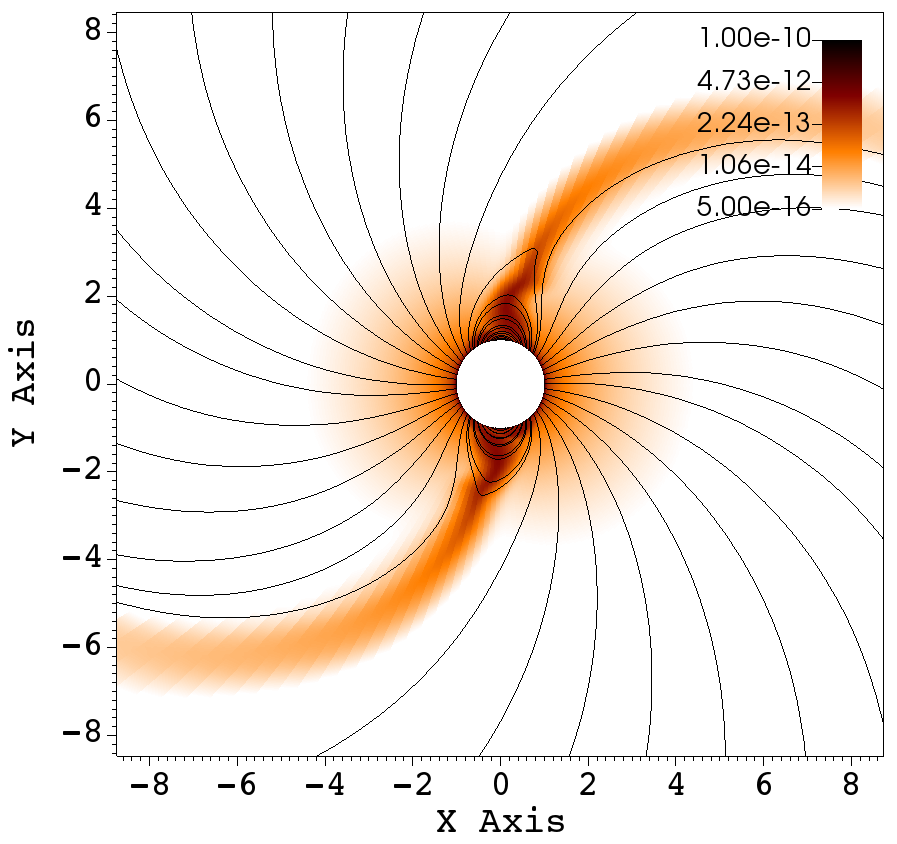}\hspace{3mm}
    \includegraphics[scale=0.165]{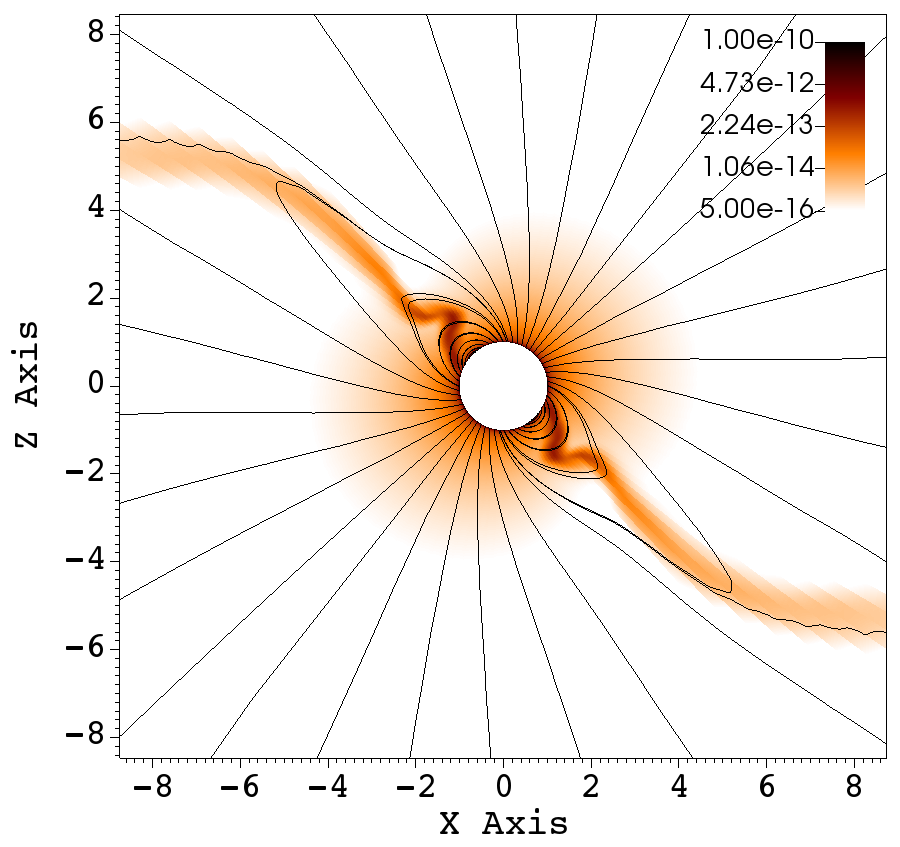}\hspace{3mm}
    \includegraphics[scale=0.165]{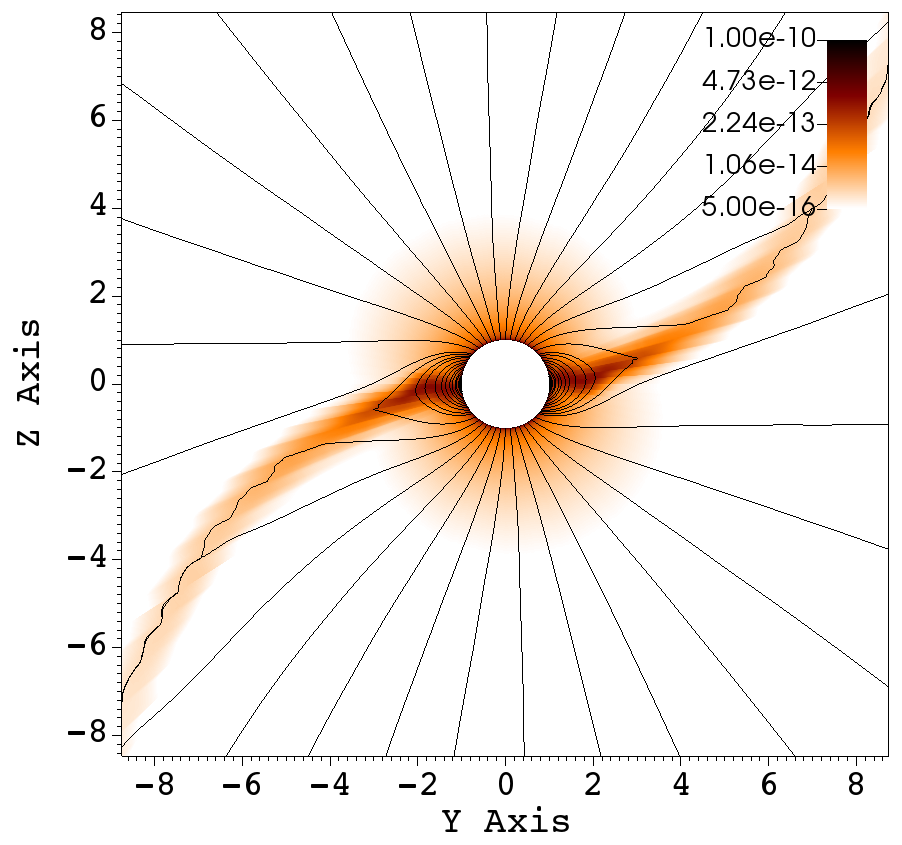}\hspace{3mm}
    \par
    \hspace{27mm}(a)\hspace{55mm}(b)
    \hspace{52mm}(c)\par
    \caption{Density ($\rm g~cm^{-3}$) colorized with the
    magnetic stream lines along the XY, XZ and YZ planes at an intermediate 
    simulation time of 240 ksec, for magnetic tilt $\zeta=45\degr$, $\eta_\star = 50$, and $W=0.5$.
    }
\label{fig:3planes}
\end{figure}
All of the simulation results presented in this work have the rotation
axis along the $z-$axis and the magnetic tilt oriented along the $xz-$plane.
Owing to this, much of the interesting phenomena are best viewed in the
$xz-$plane.
In order to illustrate the 3D nature of the simulations, 
three cross sections are presented in Figure 
{\ref{fig:3planes}}, where density and 
magnetic field lines are presented in the XY, XZ and YZ planes at an 
intermediate simulation time, for $\eta_\star = 50$, 
$W=0.5$ and $\zeta=45\degr$. 
In the next section, the angular momentum flux is described along with the
calculation of mass loss rate and spin down time of the star.



\section{Angular Momentum Loss Rate} \label{sec:jdot}


Mass loss from rotating stars carries away angular 
momentum and leads to stellar spindown. In non-magnetic massive stars with
line-driven winds, the mass-loss rate is large, but not enough to significantly
increase their rotation periods during the main sequence because the angular momentum
is only carried away by gas, and the main sequence lifetime is relatively short
\citep[e.g.,][]{2000ARA&A..38..143M}. In magnetic stars, most of the angular
momentum is lost via the Poynting stress provided by the magnetic field, and not by the 
outflowing plasma itself \citep{1967ApJ...148..217W, 2009MNRAS.392.1022U}.
\citet{1967ApJ...148..217W} modeled the angular momentum loss of the Sun
with the far-field approximation of a magnetic monopole. For stronger
dipole fields, \citet{2009MNRAS.392.1022U} showed that the angular momentum
loss rate of an aligned magnetic rotator could be approximated as:

\begin{equation}\label{eq:J_dot_dWD}
    {\dot J}_{\rm dWD} = \dfrac{2}{3} {\dot{M}}_{B=0} \Omega R_{\rm A}^2,
\end{equation}
\noindent
where ${\dot J}_{\rm dWD}$ is the total angular momentum loss rate in
the dipole-modified Weber-Davis approximation,
$R_{\rm A}$ is the Alfv\'en radius as defined in Eq. (\ref{eq:RA}),
$\Omega$ is the angular velocity, and
${\dot{M}}_{B=0}$ is the mass-loss rate in the absence of a magnetic field.
To examine the angular momentum loss in our 3D Geomesh simulations,
we define the angular momentum flux:

\begin{equation}\label{eq:djgas_dt}
    \dfrac{dj_{\rm gas}}{dt} = \rho v_r v_\phi r sin\theta,
\end{equation}

\noindent
where $v_r$ and $v_\phi$ are the radial and azimuthal velocity, respectively, and
$\theta$ is the colatitude \citep{2009MNRAS.392.1022U}.
Integrating the angular momentum flux over a spherical surface yields
the angular momentum loss rate of the outflowing gas:
\begin{equation}\label{eq:J_dot_gas}
    {\dot J}_{\rm gas} = \oint\limits_S \dfrac{dj_{\rm gas}}{dt} dS
\end{equation}

Similar to Eq. (\ref{eq:djgas_dt}), the angular momentum loss rate due to
magnetic braking is modeled with the help of the Maxwell stress tensor.
The corresponding angular momentum flux is defined as,

\begin{equation}\label{eq:djmag_dt}
    \dfrac{dj_{\rm mag}}{dt} = - \dfrac{B_r B_\phi}{4 \pi} r sin\theta,
\end{equation}

\noindent
where $B_r$ and $B_\phi$ are the radial and azimuthal components of the 
magnetic field flux density $\bf B$ \citep{2009MNRAS.392.1022U}.
The total magnetic flux density $\bf B$ is the sum of 
background magnetic flux density ${\bf B_0}$
and evolving magnetic flux density ${\bf B_1}$.
Integrating the angular momentum flux over a spherical surface yields
the magnetic angular momentum loss rate:
\begin{equation}\label{eq:J_dot_mag}
    {\dot J}_{\rm mag} = \oint\limits_S \dfrac{dj_{\rm mag}}{dt} dS,
\end{equation}
\noindent
The total angular momentum flux and total angular momentum loss rate are therefore:
\begin{equation}\label{eq:djtot_dt}
    \dfrac{dj_{\rm tot}}{dt} = \dfrac{dj_{\rm gas}}{dt} + \dfrac{dj_{\rm mag}}{dt},
\end{equation}
\begin{equation}\label{eq:J_dot_tot}
    {\dot J}_{\rm tot} = {\dot J}_{\rm gas} + {\dot J}_{\rm mag}.
\end{equation}
\begin{figure}
    \includegraphics[scale=0.22]{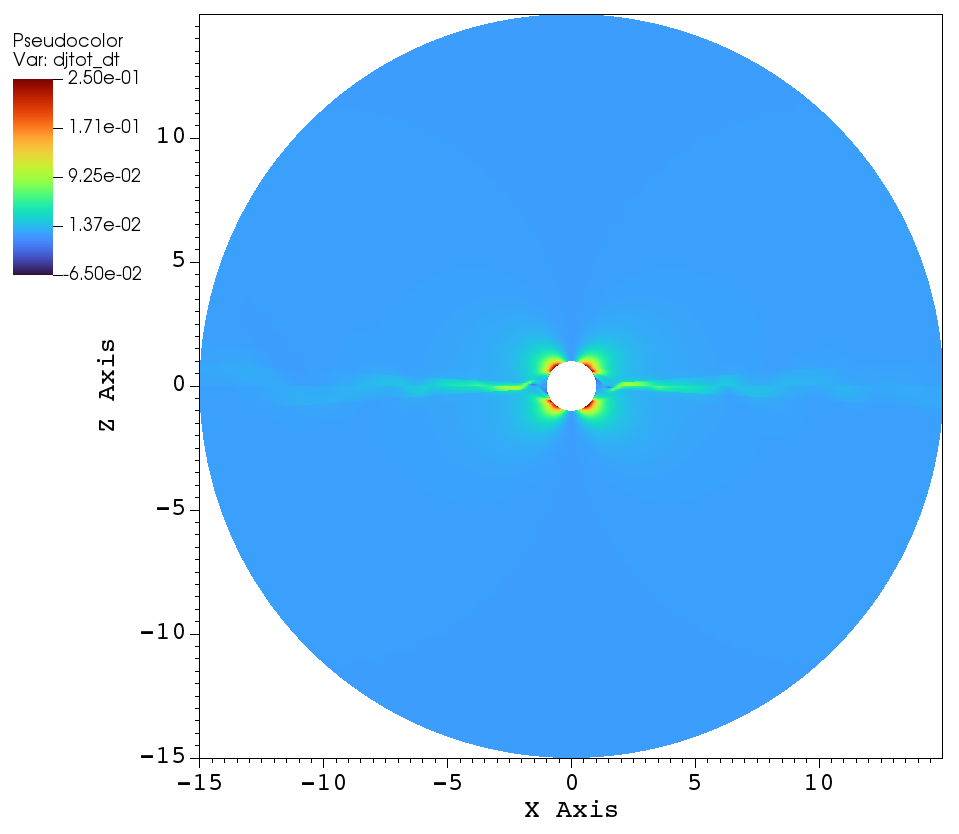}
    \includegraphics[scale=0.22]{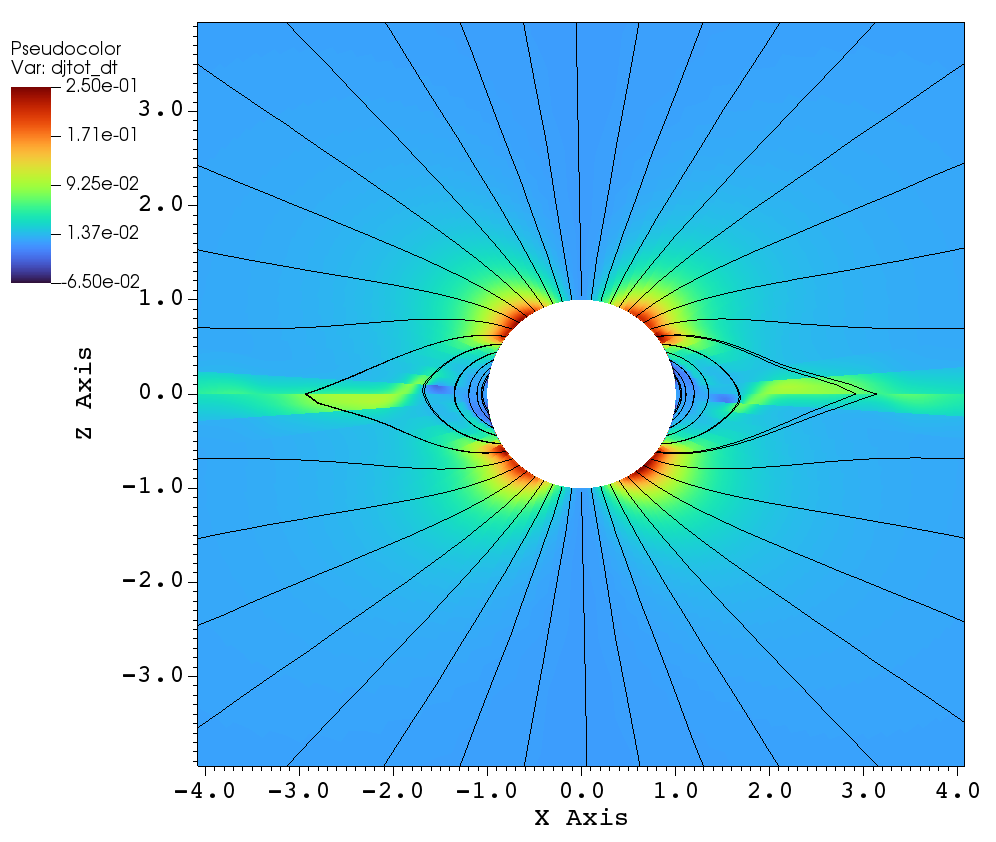}\par
    \hspace{85mm}(a)\par
    \includegraphics[scale=0.22]{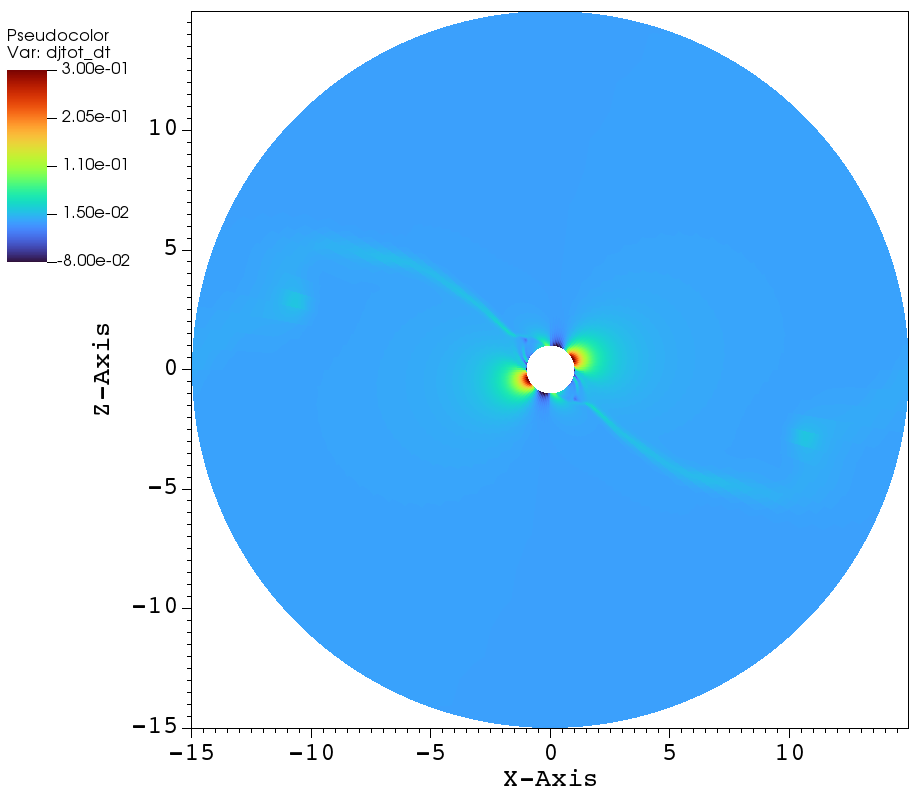}
    \includegraphics[scale=0.22]{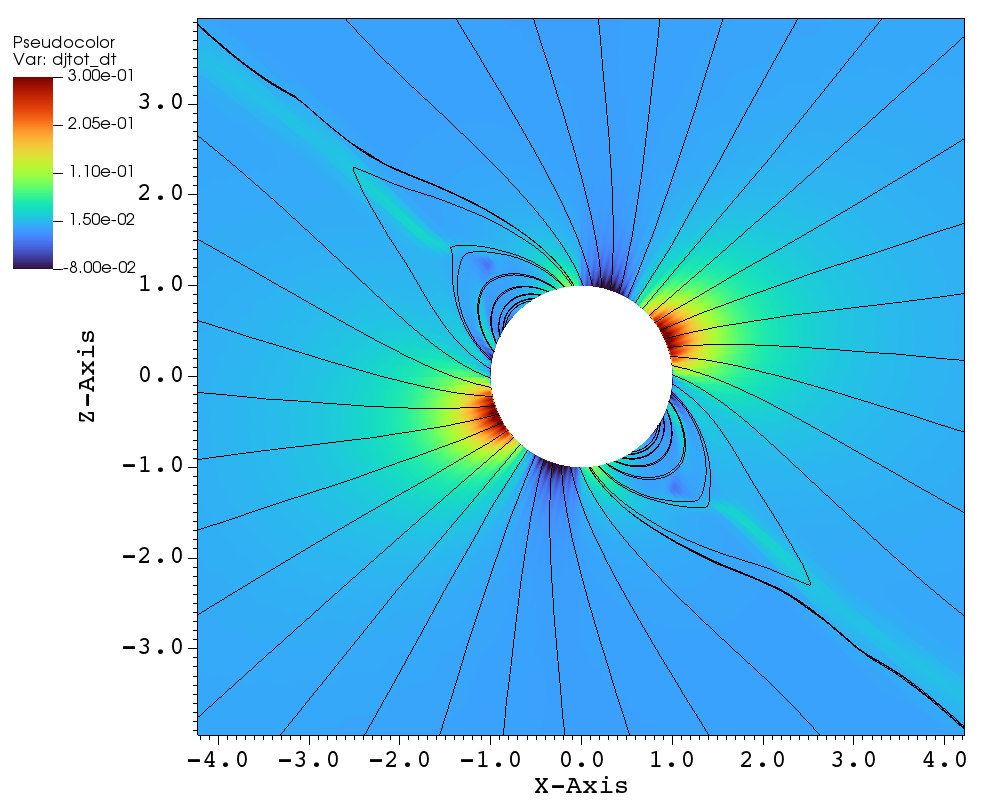}\par
    \hspace{85mm}(b)\par
    \includegraphics[scale=0.22]{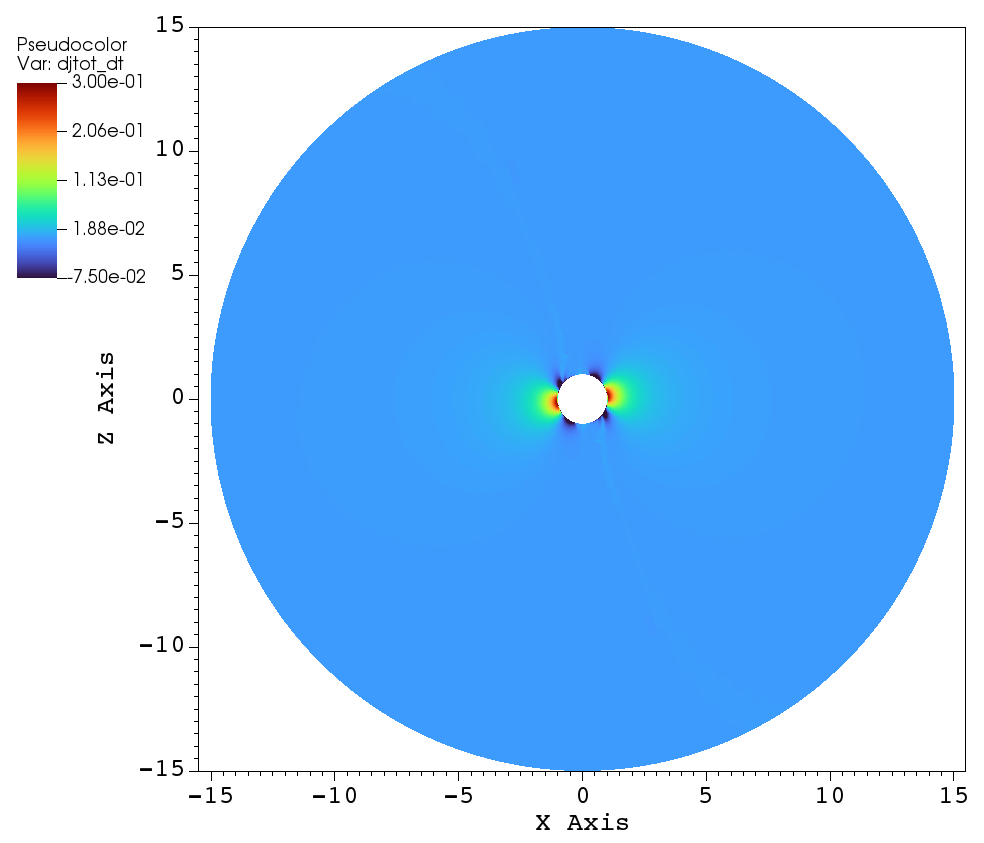}
    \includegraphics[scale=0.22]{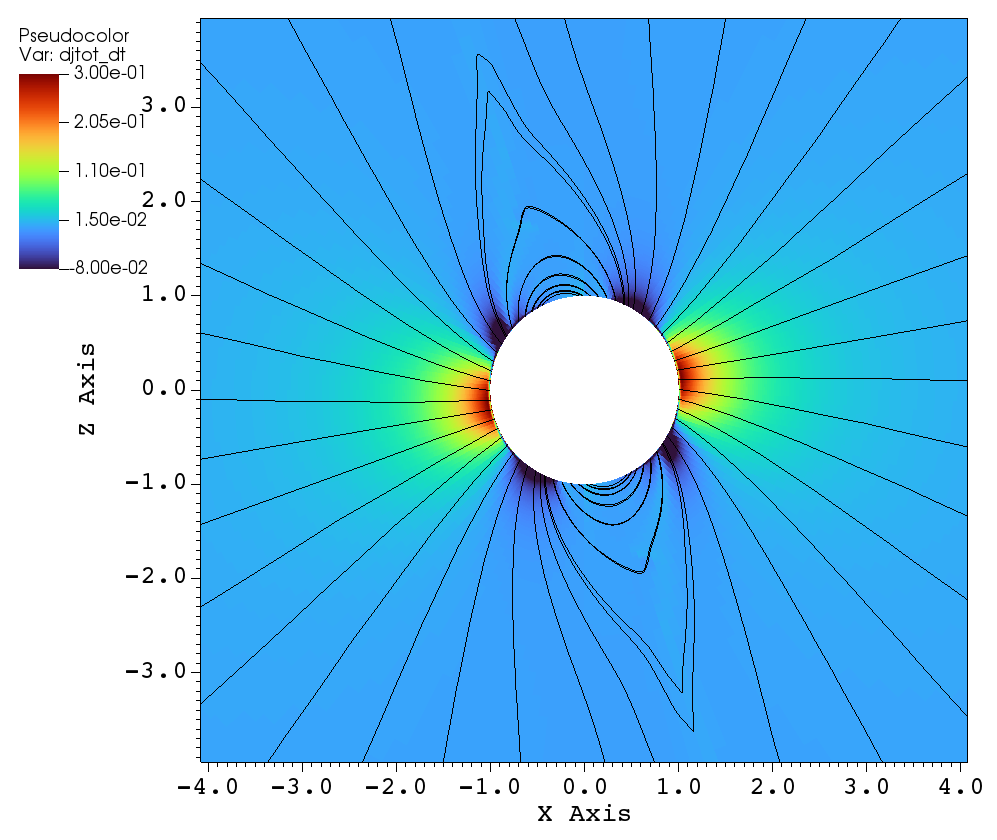}\par
    \hspace{85mm}(c)\par
\caption{Cross section of total angular momentum flux $dj_{\rm tot}/dt$ 
($\rm g\hspace{0.5mm}s^{-2})$
in the $xz$-plane for $\eta_\star=10$, $W=0.5$:
(a) no magnetic tilt ($\zeta =0\degr$), (b) tilt $\zeta =45\degr$, (c) tilt $\zeta =75\degr$.
The panels on the left show the large scale angular momentum flux in the entire xz-plane ($\pm 15 R_\star$).
The panels on the right show the zoomed in angular momentum flux in the inner xz-plane ($\pm 4 R_\star$), with magnetic field lines overlaid.}
\label{fig:Jdot_tot_sect}
\end{figure}

\begin{figure}
    \includegraphics[scale=0.23]{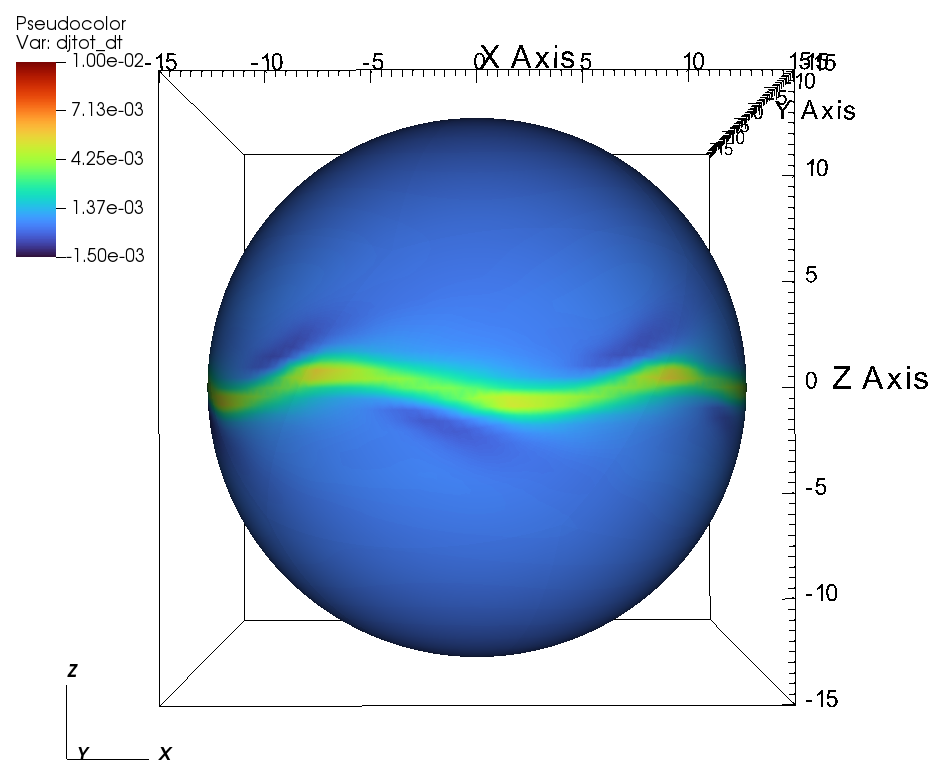}\hfill 
    \includegraphics[scale=0.23]{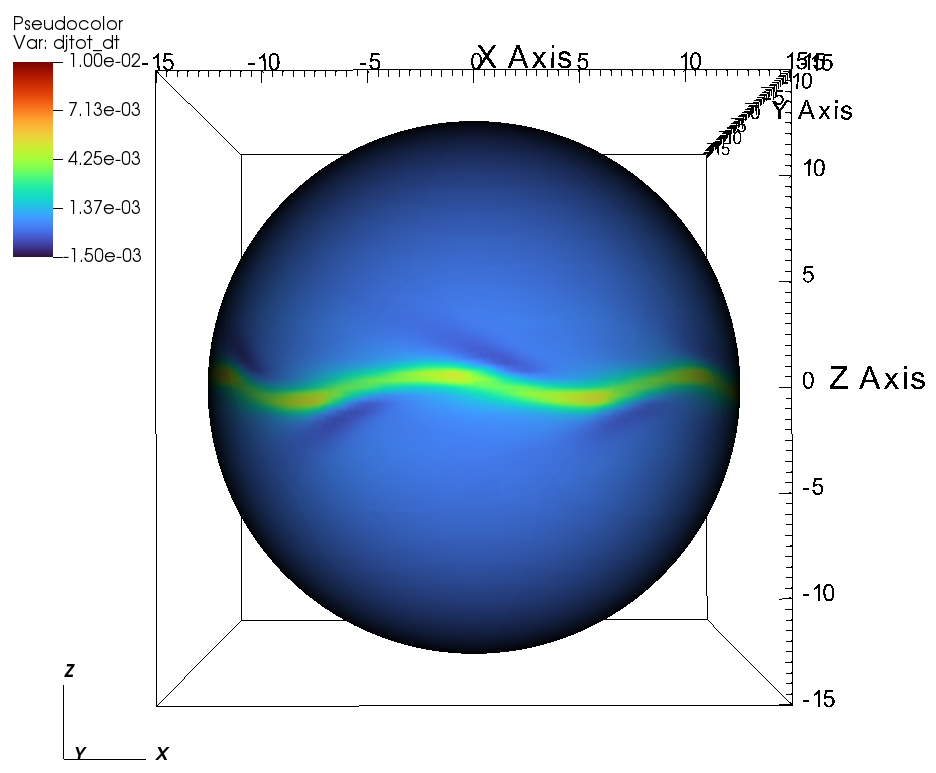}\par
    \hspace{38mm}(a)\hspace{85mm}(b) \par
    \includegraphics[scale=0.23]{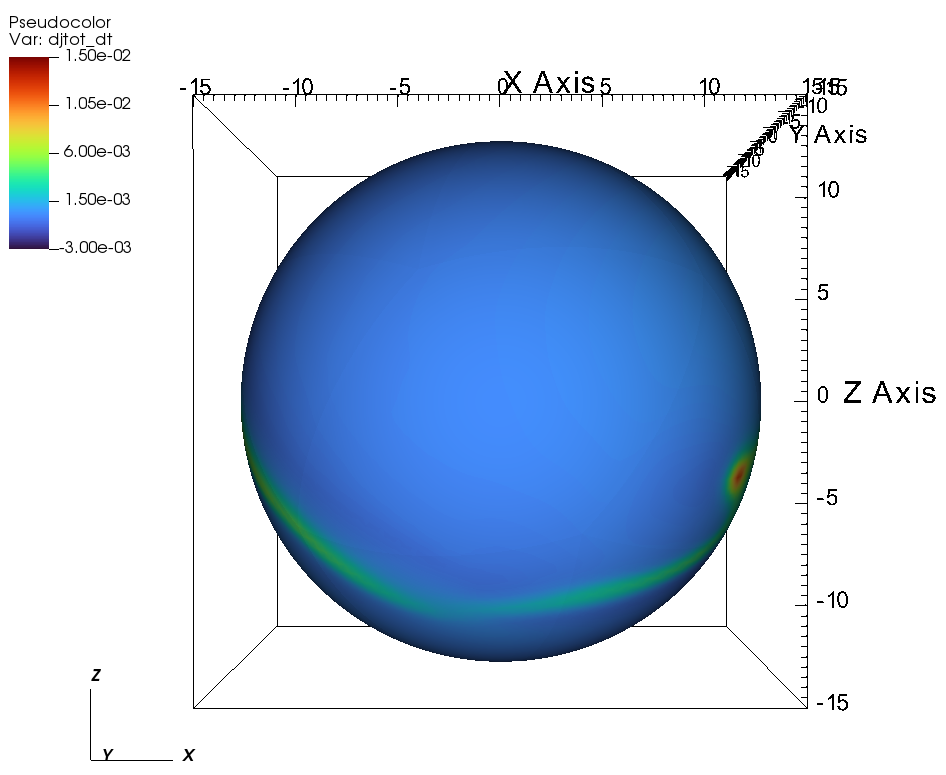}\hfill 
    \includegraphics[scale=0.23]{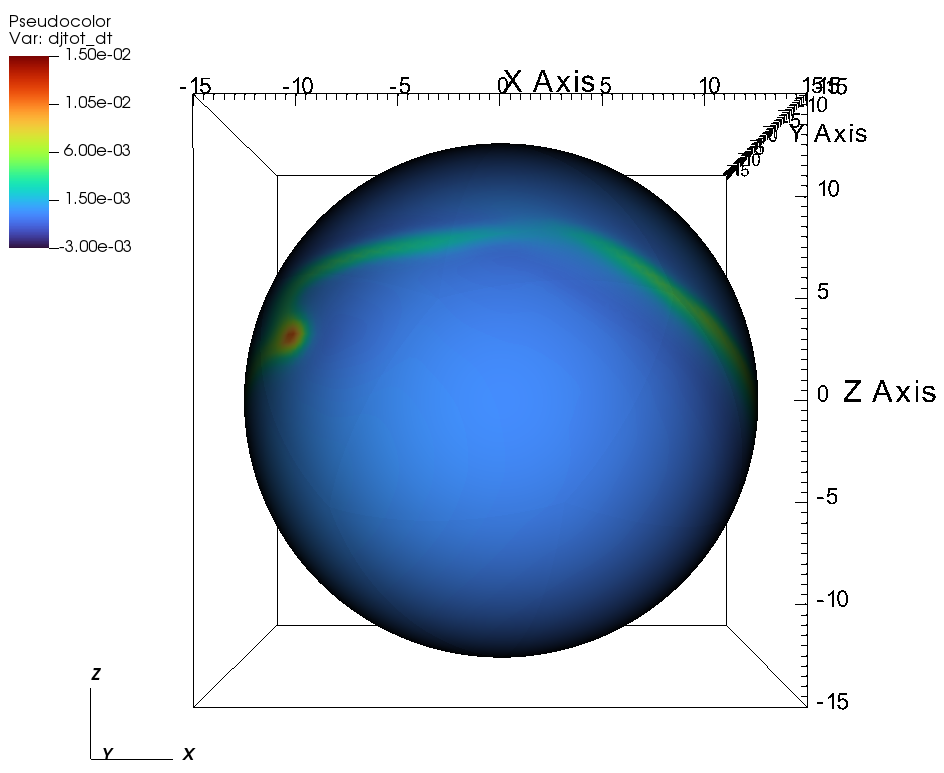}\par
    \hspace{38mm}(c)\hspace{85mm}(d) \par
    \includegraphics[scale=0.23]{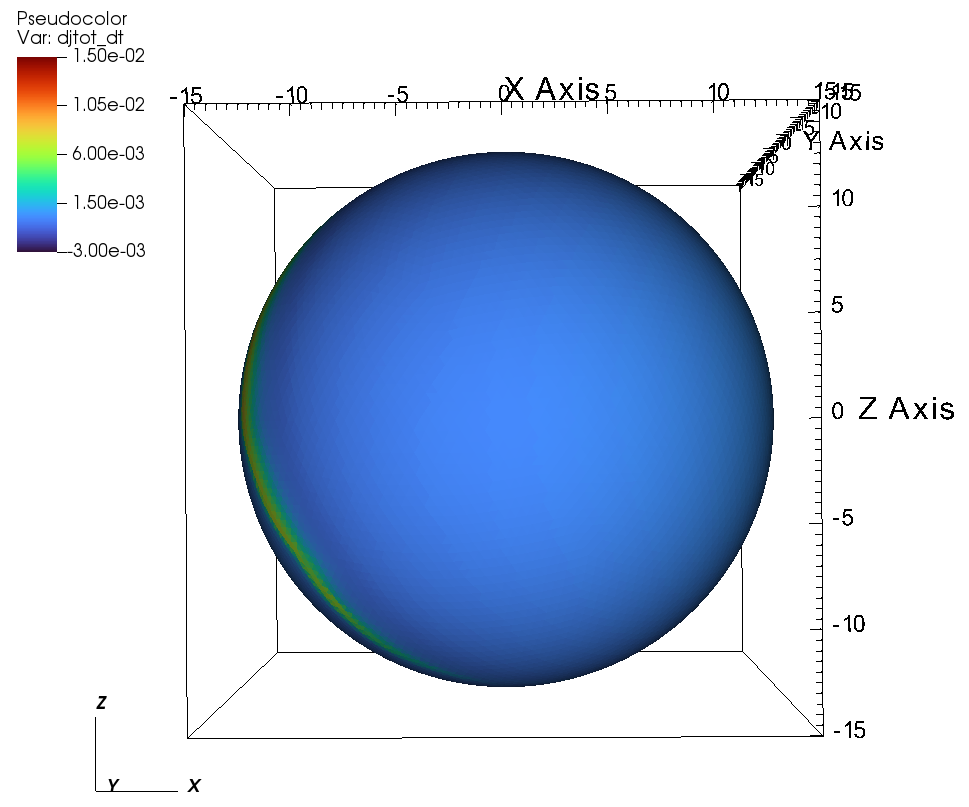}\hfill 
    \includegraphics[scale=0.23]{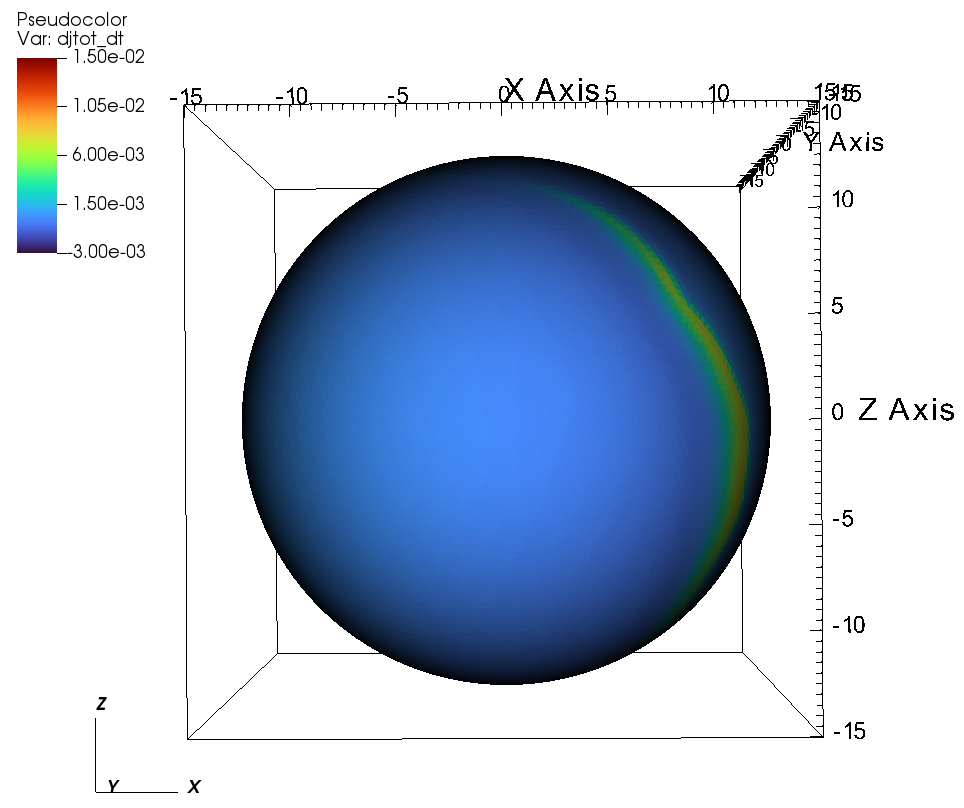}\par
    \hspace{38mm}(e)\hspace{85mm}(f) \par
\caption{Total angular momentum flux $dj_{\rm tot}/dt$ 
($\rm g\hspace{0.5mm}s^{-2})$
at the outer surface for the case of $\eta_\star=10$ and $W=0.5$ showing:
the front (a) and back (b) view for $\zeta =  0\degr$,
the front (c) and back (d) view for $\zeta = 45\degr$, and
the front (e) and back (f) view for $\zeta = 75\degr$.
}
\label{fig:Jdot_tot_surf}
\end{figure}

With this preamble, the remainder of this section describes the
angular momentum flux and angular momentum loss rate for the simulations listed in Table \ref{tab:mdot}.
Figure~\ref{fig:Jdot_tot_sect} displays the total angular momentum flux $dj_{\rm tot}/dt$ 
in the $xz-$plane for three tilt angles: 
(a) aligned rotation $\zeta=0\degr$, (b) $\zeta = 45\degr$, and (c) $\zeta=75\degr$.
Figure \ref{fig:Jdot_tot_sect}a illustrates
the total angular momentum flux for the case of aligned rotation. 
Far from the star, the angular momentum loss is expected to be maximal in the $xy-$plane
(i.e. perpendicular to the axis of rotation) and minimal along the axis of rotation (i.e. the $z-$axis).
This effect is shown in Figure~\ref{fig:Jdot_tot_sect}a (i.e. in the left panel),
where the large scale flow shows that $dj_{\rm tot}/dt$ is small along the rotation
axis and substantially larger in the plane perpendicular to the rotation axis.
However, in the in the zoomed right panel of Figure~\ref{fig:Jdot_tot_sect},
we see that the angular momentum loss perpendicular to the plane of rotation is \emph{not} maximal
because closed magnetic loops near the stellar surface inhibit mass loss.
Focusing on the right panel of Figure~\ref{fig:Jdot_tot_sect}a, we see that the maximal angular momentum flux
comes from the footpoints of the open magnetic field lines that connect to the equatorial outflow.
Once the wind propagates past $R_{\rm A}$ the angular momentum flux is channeled towards the xy-plane.
 
{
Thus the angular momentum flux close to the surface of the star is 
maximal along the open field lines (${\bf B}_{\rm open}$) in the plane of
rotation; i.e., where $|{\bf B}_{\rm open} \times {\bf \Omega}|$ is maximal. 
This result is especially evident for the case of of tilted-rotation 
with $\zeta=45\degr$, shown in Figure~{\ref{fig:Jdot_tot_sect}}b.
In the zoomed panel, we see that the angular momentum loss close to the star is low along the magnetic equator,
i.e. along the closed field lines, similar to the case of aligned rotation.
Comparing the zoomed versions of Figs. {\ref{fig:Jdot_tot_sect}}a and {\ref{fig:Jdot_tot_sect}}b,
in the $\zeta=45\degr$ case the open field lines are present on either side of the magnetic equator.
As a result of the magnetic tilt, the open field lines along the rotational equator,
leading to have larger angular momentum flux close to the star.
Conversely, the open field lines along the rotational $z-$axis have smaller angular momentum flux. 
}

Similar observations can also be made for the $\zeta=75\degr$ case (see
Figure~\ref{fig:Jdot_tot_sect}c), where the magnetic poles are nearly perpendicular to the
axis of rotation. Hence, the total angular momentum loss close to the 
stellar surface is maximum near the magnetic poles, where the open field lines
and the plane of rotation coincide.

Beyond the Alfv\'en radius, much of the angular momentum flux emerges along the magnetic equator
because the angular momentum flux is channeled by the field lines.
(see Figure \ref{fig:Jdot_tot_surf}). 
Figures \ref{fig:Jdot_tot_surf}a and \ref{fig:Jdot_tot_surf}b,
show the total angular momentum flux at the outer boundary of 
the simulation, i.e. at a radius of $R \approx 15 R_\star$.
Figures \ref{fig:Jdot_tot_surf}a and \ref{fig:Jdot_tot_surf}b 
correspond to the case of aligned rotation. We see that most of the 
angular momentum flux is channeled by the magnetic field lines.
Figures \ref{fig:Jdot_tot_surf}c-f, show the total 
angular momentum flux at the outer boundary of the simulation 
$R \approx 15 R_\star$, for the case of tilted-rotation with $\zeta=45\degr, 75\degr$. 
Here, similar to the aligned rotation, the angular momentum flux is
influenced by the magnetic channeling of the wind and rotation.
As we move farther from the star, the effect of rotation 
overpowers the strength of the magnetic field and the same can be observed 
from Figure~\ref{fig:Jdot_tot_sect}b.


\subsection{ Radial variation of angular momentum loss }

%
\begin{figure}
    \includegraphics[scale=0.58]{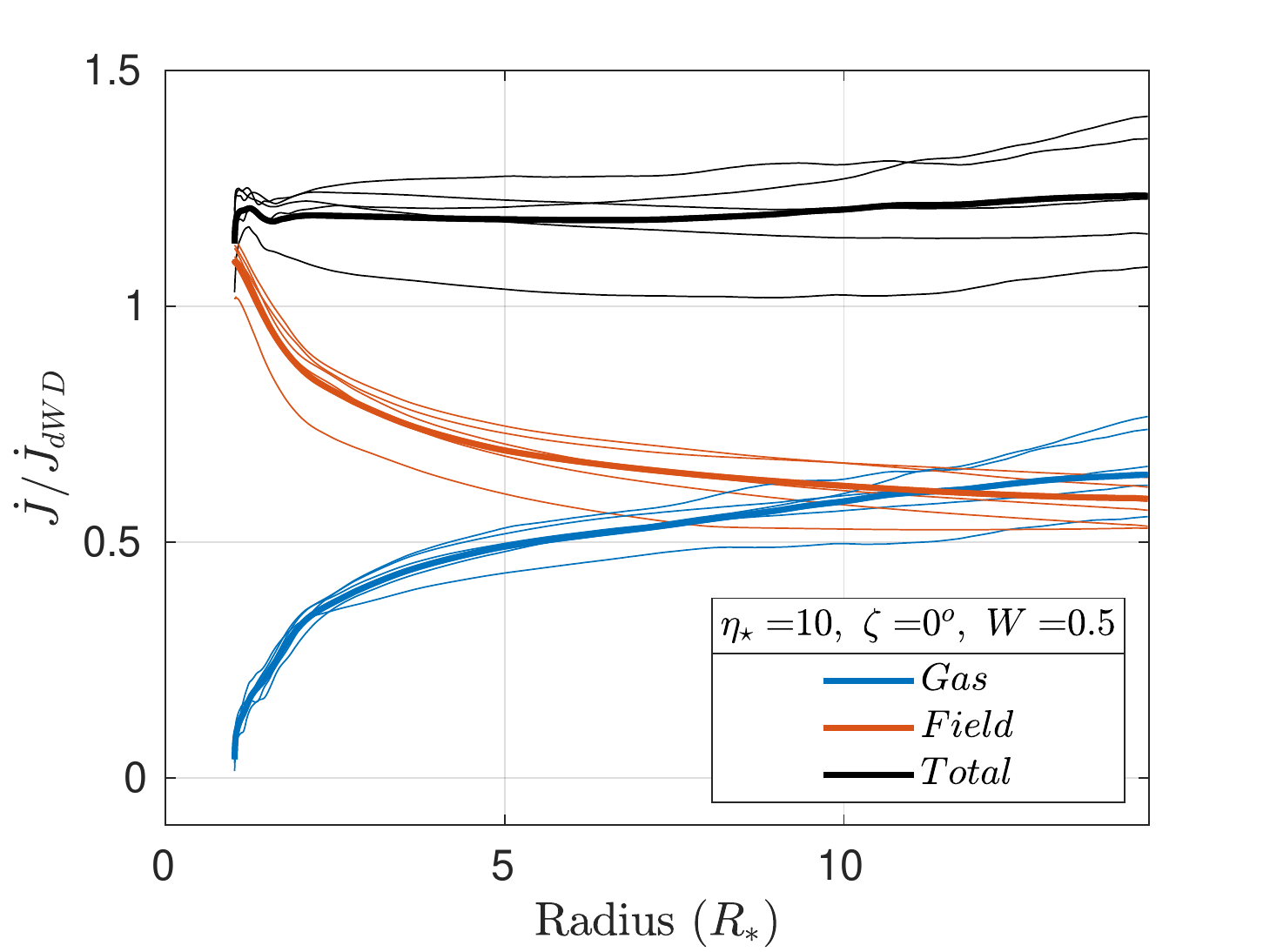}
    \includegraphics[scale=0.58]{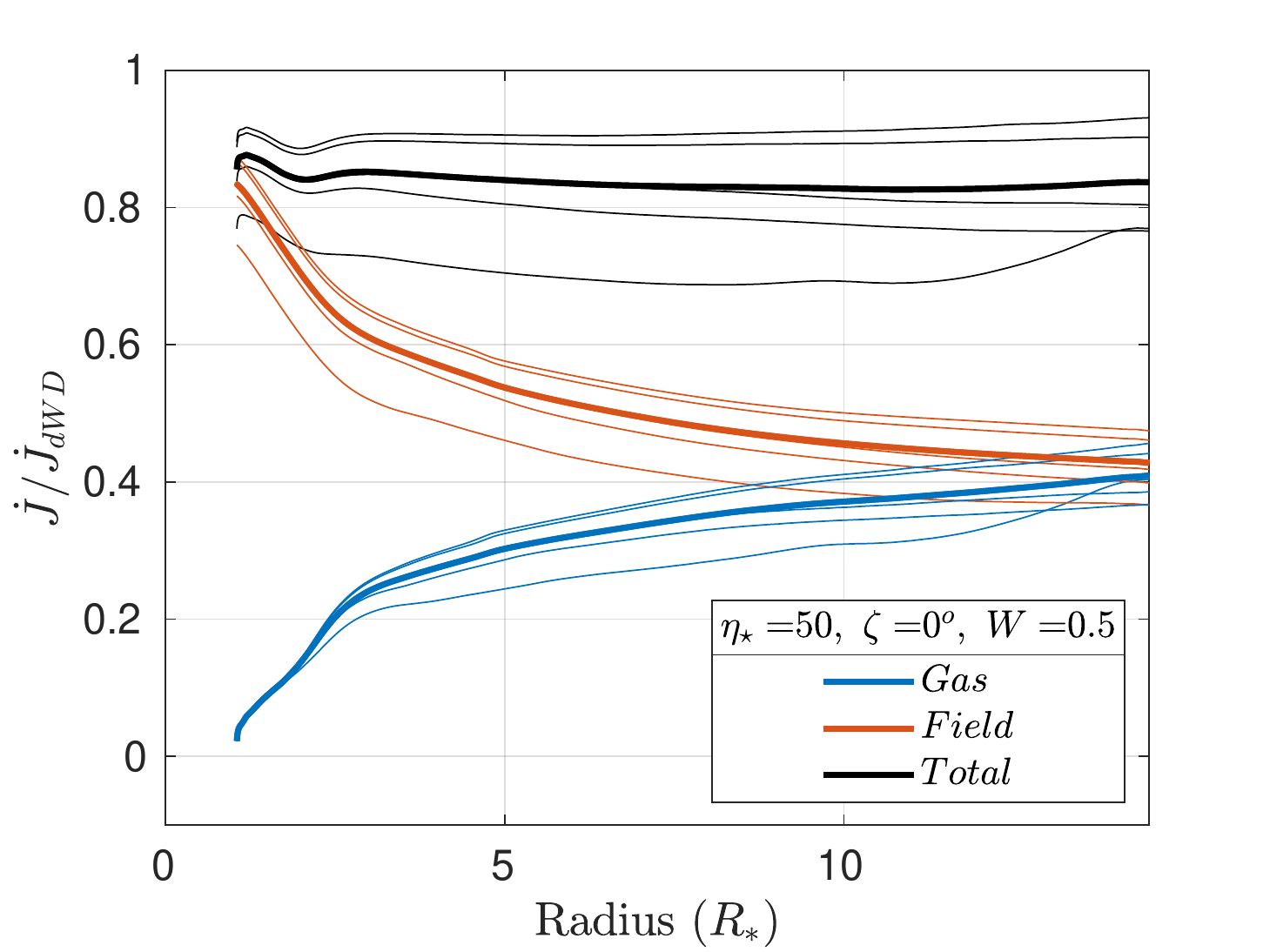}\par
    \hspace{38mm}(a)\hspace{78mm}(b) \par
    \includegraphics[scale=0.58]{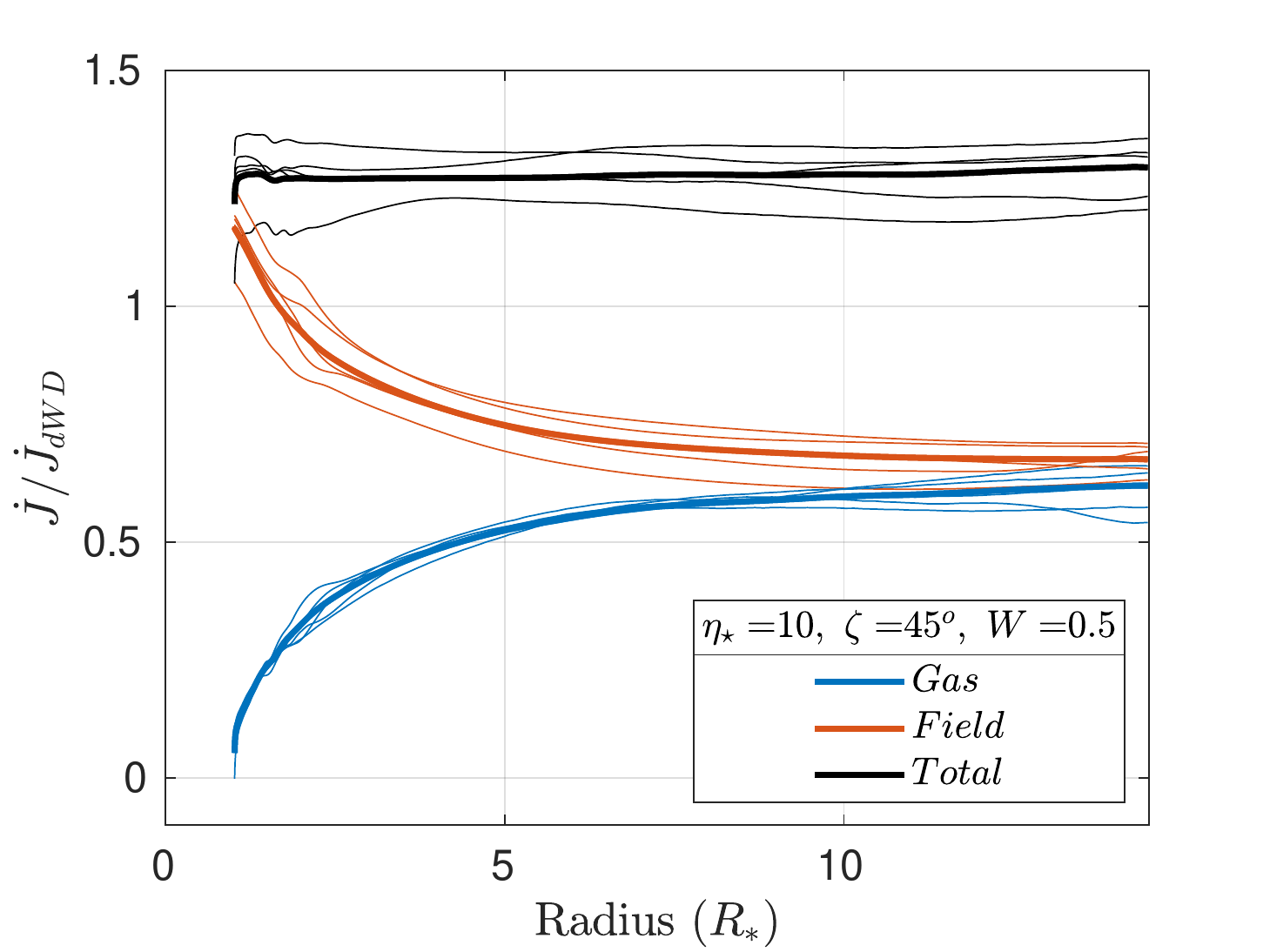}
    \includegraphics[scale=0.58]{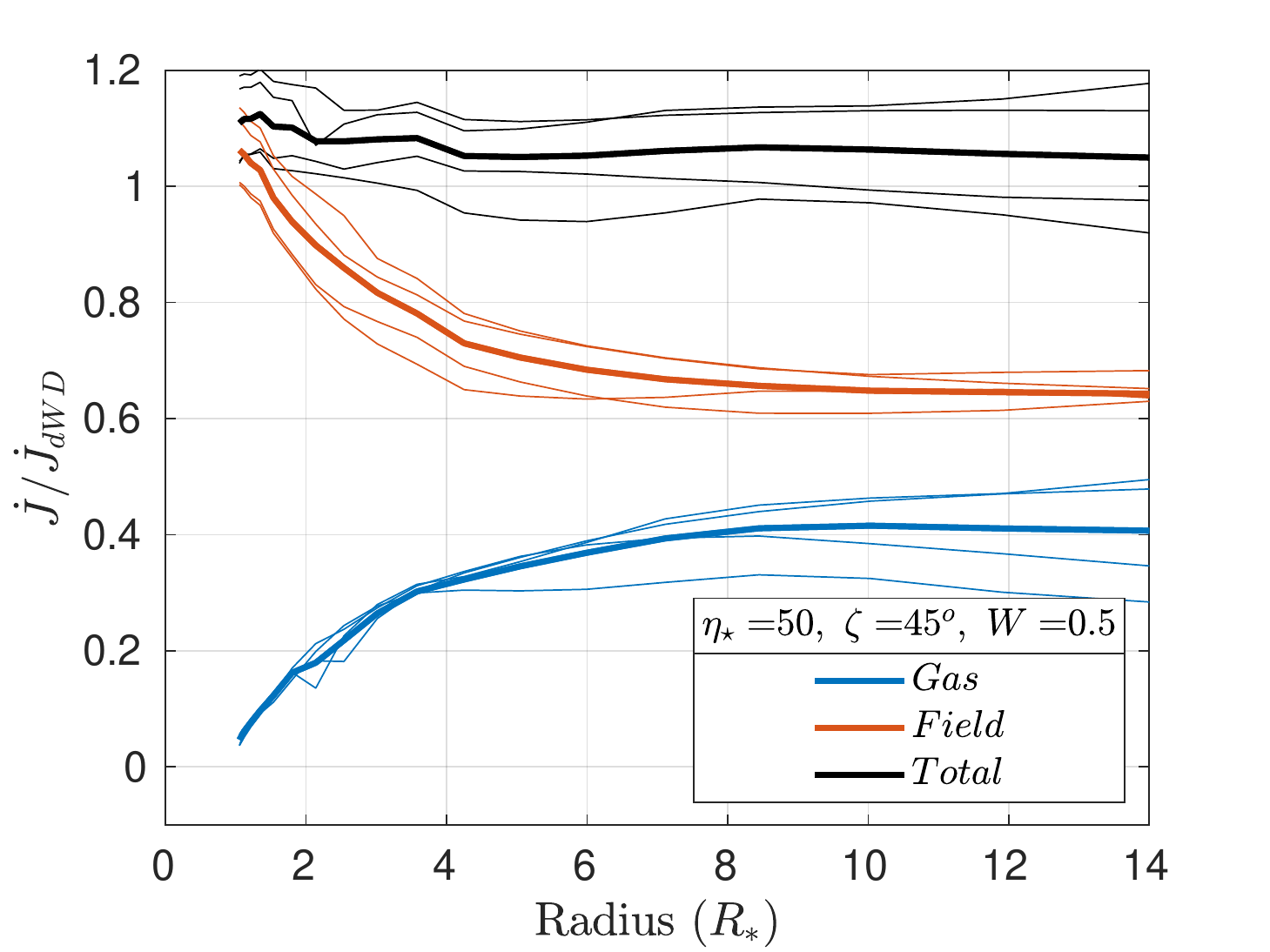}\par
    \hspace{38mm}(c)\hspace{78mm}(d) \par
    \includegraphics[scale=0.58]{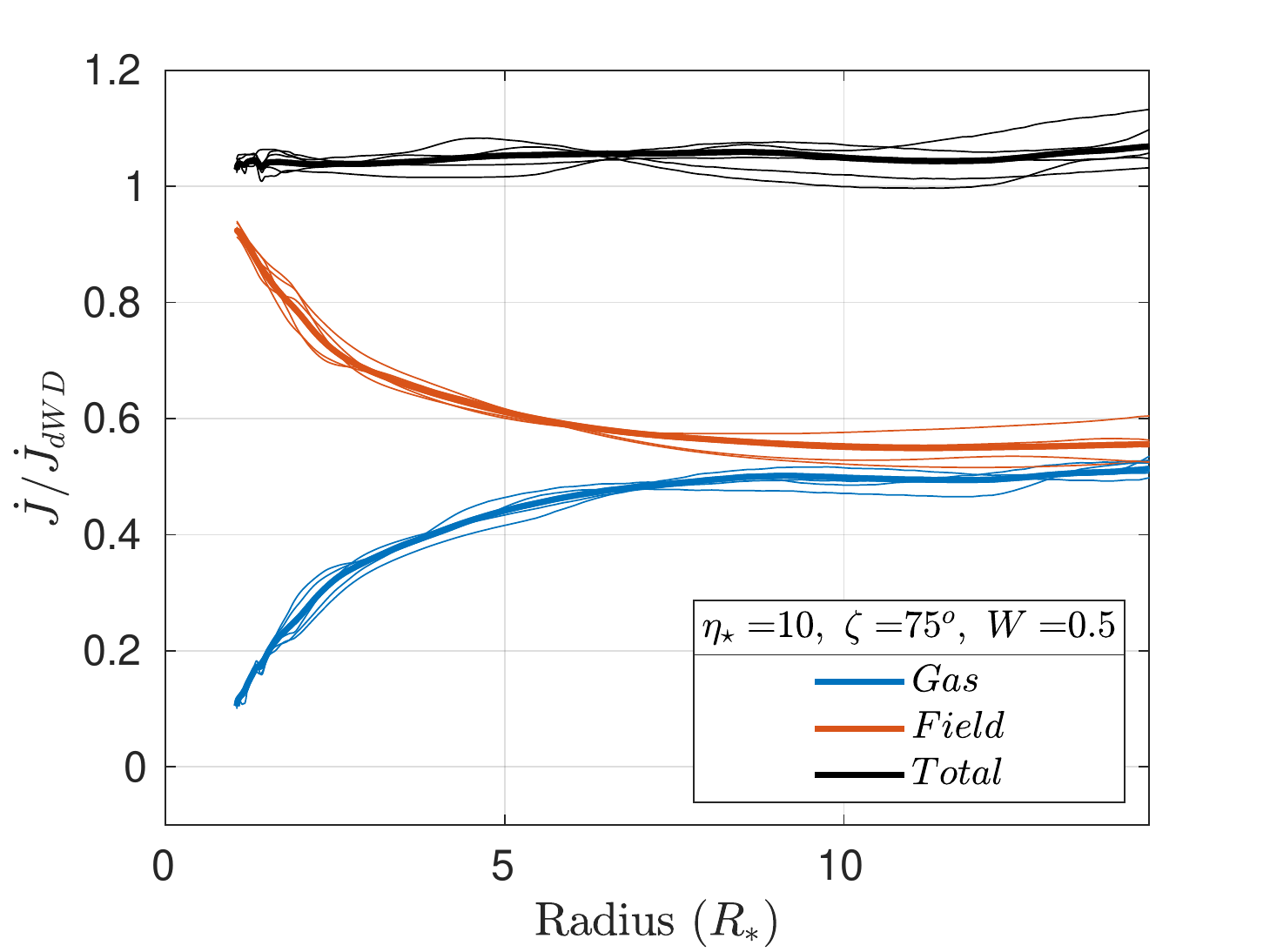}
    \includegraphics[scale=0.58]{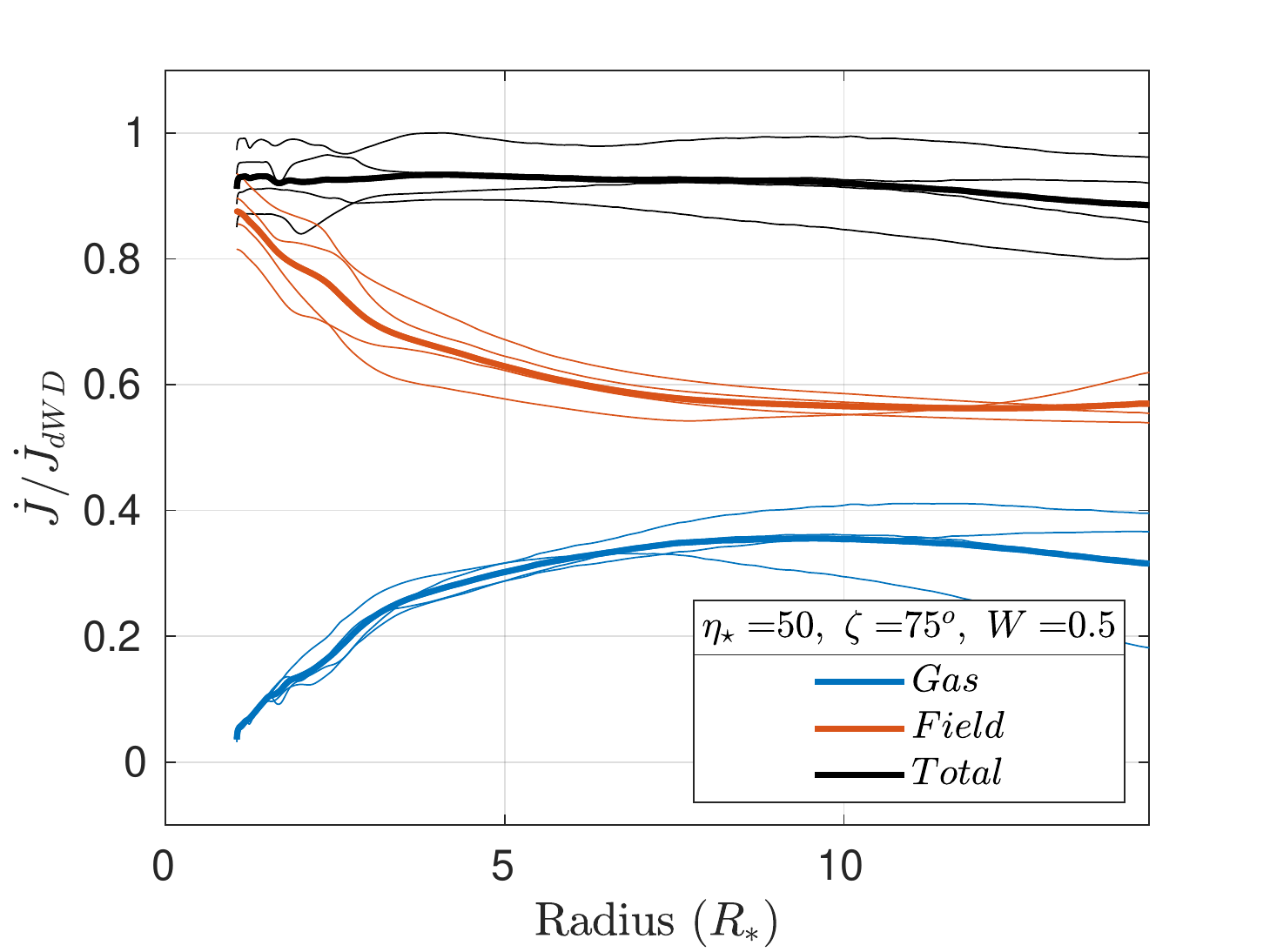}\par
    \hspace{38mm}(e)\hspace{78mm}(f) \par
\caption{Angular momentum loss rate $\dot{J}$ as a function of stellar radius
with rapid rotation ($W=0.5$)
for gas in blue, magnetic field in red, and their total in black.
$\dot{J}$ is normalized to the dipole Weber-Davis value for an aligned magnetic rotator $\dot{J}_{\rm dWD}$.
Several quasi-steady state snapshots are shown with thin lines, and their
time averages are indicated with thicker lines for the following cases:
(a) $\eta_\star=10$ and $\zeta=0\degr$,
(b) $\eta_\star=50$ and $\zeta=0\degr$,
(c) $\eta_\star=10$ and $\zeta=45\degr$,
(d) $\eta_\star=50$ and $\zeta=45\degr$,
(e) $\eta_\star=10$ and $\zeta=75\degr$, and
(f) $\eta_\star=50$ and $\zeta=75\degr$.
}
\label{fig:Jdot_vs_R}
\end{figure}
The strength of the dipole magnetic field is maximum at the photosphere,
falling steeply ($1/r^3$) as we move away from the star. Due to this steep fall off, 
the angular momentum loss rate is predominantly magnetic close to the star.
As we move away from the star, the magnetic component of the angular
momentum loss rate ${\dot J_{\rm mag}}$ decreases and the gaseous component
${\dot J_{\rm gas}}$ increases, while keeping
the total angular momentum loss rate ${\dot J_{\rm tot}}$ nearly constant.
The total, magnetic and gas-borne angular momentum loss rates are plotted versus radius in Figure~{\ref{fig:Jdot_vs_R}}
as solid black, red and blue curves, respectively, for six different parameter sets.
The $\dot{J}$ values in Figure~{\ref{fig:Jdot_vs_R}}, and the corresponding sixth column in Table~2, 
are normalized to the predicted $\dot{J}_{\rm dWD}$ in the dipole-modified Weber-Davis model
\citep{2009MNRAS.392.1022U}.
The thin lines represent individual time snapshots, and the bold black, red and blue lines
represent the average of the thin lines. As we expect from angular momentum conservation,
the black (${\dot J_{\rm tot}}$) curves are essentially flat. 
However, we also see $\approx 10\%$ snapshot-to-snapshot variations, indicating that angular momentum loss is time variable.

These results are in line with the dynamics 
of magnetically channeled winds.
2D simulations carried out for the 
case of aligned rotation also show significant time
variability in the angular momentum loss \citep{2008MNRAS.385...97U}.
In Figure~\ref{fig:Jdot_vs_R} for each $\eta_{\star}$ when the tilt angle $\zeta=0$,
the gas is flung out with maximal centrifugal force. 
Comparing Figure~\ref{fig:Jdot_vs_R}a with \ref{fig:Jdot_vs_R}c
and also comparing Figure~\ref{fig:Jdot_vs_R}b with \ref{fig:Jdot_vs_R}d, suggests that
increasing the tilt angle $\zeta$ reduces the moment arm, 
thereby reducing the gas-borne angular momentum loss. 

Further examination of the left ($\eta_{\star} = 10$) and right ($\eta_{\star} = 50$) panels 
in Figure~\ref{fig:Jdot_vs_R} suggests that the angular momentum loss due to the field
$\dot{J}_{\rm mag}$ increases with increasing $\zeta$, relative to the angular momentum
loss due to the gas $\dot{J}_{\rm gas}$.
This is especially evident in the higher-field panels on the left of Figure~\ref{fig:Jdot_vs_R}.
At low $\zeta$, the gas in the magnetic equator experiences higher centrifugal force,
and therefore contributes more to the total angular momentum loss.
At high $\zeta$, the open field lines above the magnetic poles experience
the highest centrifugal force, and therefore account for most of the total
angular momentum loss.
%
%

\subsection{Characteristic mass-loss and spindown time }

%
\begin{table}
	\centering
	\caption{Mass loss rate, characteristic mass-loss time,
	angular momentum loss rate and spin-down time.}
	\label{tab:tspin}
	\begin{tabular}{ccccccccc} 
    \hline
	$\eta_\star$ &
	$\zeta$ & $W$ &
	${\dot M}$ & $\tau_{\rm mass}$ &
    \multirow{2}{*}{\large{$\frac{{\dot J}_{\rm tot}}{{\dot J}_{\rm dWD}}$}} &
    \multirow{2}{*}{\large{$\frac{{\dot J}_{\rm tot}}{{\dot J}_{\rm dWD,B=0}}$}} &
    $\tau_{\rm spin}$ \\
    \multicolumn{3}{c}{} & ($M_\odot~{\rm yr}^{-1}$) & (Myr) &
    \multicolumn{2}{c}{} & (Myr) \\
	\hline
     0 &  $0\degr$ &  0  & 1.49 $\times 10^{-6}$ & 16.7 &  \dots & \dots 
                                                                          & \dots \\
     0 &  $0\degr$ & 0.5 & 1.70 $\times 10^{-6}$ & 14.7 &  1.04  & 1.04  
                                                                          & 2.50 \\
    \\
    10 &  $0\degr$ &  0  & 6.92 $\times 10^{-7}$ & 36.1 &  \dots & \dots 
                                                                          & \dots \\
    10 &  $0\degr$ & 0.5 & 8.37 $\times 10^{-7}$ & 29.8 &  1.29  & 5.81  
                                                                          & 0.44  \\
    10 & $45\degr$ & 0.5 & 7.96 $\times 10^{-7}$ & 31.4 &  1.35  & 6.07  
                                                                          & 0.42  \\
    10 & $75\degr$ & 0.5 & 5.88 $\times 10^{-7}$ & 42.5 &  1.05  & 4.73  
                                                                          & 0.55  \\
    \\
    50 &  $0\degr$ &  0  & 4.35 $\times 10^{-7}$ & 57.4 &  \dots & \dots 
                                                                          & \dots \\
    50 &  $0\degr$ & 0.5 & 4.58 $\times 10^{-7}$ & 54.6 &  0.95  & 7.73  
                                                                          & 0.34  \\
    50 & $45\degr$ & 0.5 & 5.84 $\times 10^{-7}$ & 42.8 &  1.24  & 10.12 
                                                                          & 0.26  \\
    50 & $75\degr$ & 0.5 & 4.41 $\times 10^{-7}$ & 56.7 &  0.96  & 7.81  
                                                                          & 0.33  \\
	\hline
	\end{tabular}
\end{table}

The characteristic mass-loss time $\tau_{\rm mass}$ and the spin-down time 
$\tau_{\rm spin}$ can be approximated from the total mass loss rate ${\dot M}$
and the total angular momentum loss rate ${\dot J_{\rm tot}}$.

\begin{equation}\label{eq:tau_mass1}
    \tau_{\rm mass} = \dfrac{M_\star}{\dot M},
\end{equation}

\begin{equation}\label{eq:tau_mass2}
    \tau_{\rm spin} = \dfrac{J}{\dot J_{\rm tot}} 
                = \dfrac{kM_\star R_\star^2  \Omega}{\dot J_{\rm tot}},
\end{equation}
where $k M_\star R_\star^2$ is the star's moment of inertia and $k \approx 0.1$
\citep{2009MNRAS.392.1022U}. 
The characteristic mass-loss time and the spin-down time are presented in 
Table \ref{tab:tspin}, along with the total angular momentum loss rate and
the mass loss rate for the different sets of simulations carried out
in this work. 
The total angular momentum loss rate presented in the sixth column of Table {\ref{tab:tspin}} is normalized to
${\dot J}_{\rm dWD}$, the predicted angular momentum loss rate from the dipole-modified Weber-Davis model.
In column~7 we normalize ${\dot J_{\rm tot}}$ by ${\dot J}_{\rm dWD,B=0}$ the prediction of the non-magnetic Weber-Davis model.
This last column highlights the expected increase in total angular momentum loss with increasing $\eta_\star$.

Comparing the non-magnetic cases, with and without rotation, the rotating case
can be seen to have an increased mass-loss rate and a reduced characteristic
mass-loss time. With the introduction of a magnetic field,
we see a significant increase in the mass-loss time due to the 
higher magnetic confinement and the reduced mass-loss rate.
In case of non-magnetic rotation, there is no spin-down due to magnetic 
braking and hence it has the longest spin-down time. This is because 
the angular momentum loss is carried by the gas. By the same token, the spin-down time is
notably reduced as we go from $\eta_\star = 10$ to $\eta_\star=50$, because
of the increase in magnetic braking torque. 


\section{Summary and Conclusions} \label{sec:summ}

The presence of a dipolar magnetic field and its tilt with respect to the rotation
axis can have notable influence on the mass outflow and the angular momentum
outflow. The mass outflows from the O and B-type stars are very significant and
thus influence the overall evolution of the star. Hence, it is inevitable 
that the presence of magnetic field and its orientation would play a
significant role in the evolution of the O and B-type stars. 
The detailed simulation and analysis of the magnetic O and B-stars are 
discussed in the literature for the case of aligned rotation
\cite{2002ApJ...576..413U, 2008MNRAS.385...97U, 2009MNRAS.392.1022U} 
with the help of 2D simulations.

For the simulation of spherical systems in full 3D, we have recently developed the {\sc Riemann Geomesh}
MHD code. The code is based on the spherical icosahedron-based  
meshing of the sphere. The surface of the sphere is mapped as uniformly as
possible, which is a significant advance compared to an $r,\theta,\phi$ mesh. This eliminates the increased discretization errors stemming from singularities at the poles
and the need for shorter time steps due to the concentration of mesh points at the poles.

In this work, we have carried out the 3D simulations of the magnetically
channeled line driven winds for a template O star. For the first time,
the simulations are done for a higher rotation rate ($0.5~\Omega_{crit}$) and 
larger magnetic field tilt angles ($0\degr, 45\degr, 75\degr$); by utilizing the 
uniform meshing and state-of-the-art MHD algorithms of the 
{\sc Riemann Geomesh} code. The simulations reach a quasi-steady 
state, as expected, and the mass-loss rate is observed to increase with 
rotation and decrease with increased magnetic field strength, due to the
magnetic confinement of the wind. For the magnetized simulations, the wind exhibits the dynamics of magnetic
channeling, as observed in previous 2D simulations. In addition, the results 
for the $45\degr$ case demonstrates the interplay between the centrifugal force,
which is maximum along the rotational equator, and the magnetic channeling, which is prominent along the
magnetic equator.

Using these insights, we proceed to catalogue the mass-loss rate and angular momentum loss rate
for different magnetic-field strengths, tilt angles and rotation rates. The spatial variations in total 
angular momentum flux ($dj_{\rm tot}/dt$) are examined for
different magnetic tilt angles, with the help of images at the 
cross-section plane, as well as, at the outer surface of the simulation
domain. Near the stellar surface, the maximal total angular momentum flux ($dj_{\rm tot}/dt$)
is observed where the  open magnetic field lines align with the plane of rotation.
Farther from the star, the maximal angular momentum flux ($dj_{\rm tot}/dt$) is aligned
with the confinement of magnetic field lines, emanating from either sides of the magnetic equator.

The gaseous, magnetic and total angular momentum loss rates ${\dot J}$ are computed as a function of radius, and plotted for different tilt angles
and magnetic field strengths. With the corresponding mass-loss rates, the characteristic mass-loss time
and spin-down time are tabulated for the simulations carried out in this work. 
The results show a longer mass-loss time with increasing magnetic
field strength, which tracks the reduced mass-loss rate. We also find
faster spin-down time with increasing magnetic field strength, showing 
the role of magnetic braking torque on the rotation of the star.

These simulations are the very first that have been reported with a  
code that can handle all possible tilt angles of the magnetic dipole with respect
to the rotation axis. 
The use of a mapped mesh makes these simulations somewhat 
more expensive, but this comes with the tremendous advantage of full
geometric generality. While we are not restricted to dipolar fields,
we have focused on dipolar fields in this study.
In order to develop a comprehensive understanding of the relations
between,  $\eta_\star,~\zeta,~W$ and ${\dot M}, {\dot J}$, a much
larger set of data with more finer variations in $\eta_\star,~\zeta,~W$
is necessary. Therefore, this is planned for the next set of study and it 
will be elaborated in a subsequent paper.

\section*{Acknowledgments}
We acknowledge the extremely valuable inputs from Stanley Owocki.
We also acknowledge the computer nodes, which are provided to us on 
  i) Compute clusters at the Notre Dame Center for Research Computing,
 ii) Bridges-2 supercomputer at the Pittsburgh Supercomputing Center, and
iii) Stampede-2 supercomputer at the Texas Advanced Computing Center.
This work used the Extreme Science and Engineering Discovery Environment (XSEDE),
which is supported by National Science Foundation grant number ACI-1548562.
Support for this work was provided by the National Aeronautics and Space Administration
through Chandra Award Number TM1-22001 issued by the Chandra X-ray Center,
which is operated by the Smithsonian Astrophysical Observatory for and on behalf
of NASA under contract NAS8-03060. AuD acknowledges support from
Pennsylvania State University Commonwealth Campuses Research Collaboration Development Program.
Data was generated through this support from the Institute of Computational and Data Sciences.
MG acknowledges summer support from a Provost's Research Grant, and a work-release 
grant from the College of Science and Mathematics at West Chester University.

\section*{Data Availability}
The data underlying this article are available in the article and in its online 
supplementary material.

\bibliography{references}
\bibliographystyle{aasjournal}

%
\appendix
%
%
\section{Description of the MHD Equations} \label{sec:App_MHD}
%
The governing MHD equations of this simulation study are presented below. The 
magnetic field ${\bf B}$ is split into a curl-free background field ${\bf B_0}$
and a time evolving field ${\bf B_1}$. The total magnetic field is defined to
be, ${\bf B} = {\bf B_0} + {\bf B_1}$. Upon imposing the vector identities,
the governing equations take the following forms 
\citep{1994arsm.rept.....P, 2016JCoPh.327..543G},
%
\\
i) conservation of mass
\begin{align}\label{eqb:c1}
\frac{\partial \rho}{\partial t} + \nabla \cdot \left(\rho {\bf v} \right) = 0
\end{align}
where, $\rho$ is the density and ${\bf v}$ is the velocity of the fluid
%
\\
ii) conservation of momentum
\begin{align}\label{eqb:c2}
\frac{\partial (\rho {\bf v})}{\partial t} + \nabla \cdot \left[ 
        \rho {\bf v}{\bf v} + P{\bf I} + 
        \frac{1}{8\pi} {B_1}^2 {\bf I}+ 
        \frac{1}{4\pi} \left( ({\bf B_0} \cdot {\bf B_1}) {\bf I}- 
                              {\bf B_1} {\bf B} - {\bf B_0} {\bf B_1} 
                      \right)
        \right]  = \rho~{\bf g_{\rm tot}}
\end{align}
where, $P$ is the pressure and ${\bf g}_{\rm tot}$ is the total net
acceleration coming from the external forces. For this simulation,
it comprises of CAK line acceleration, gravitational acceleration and
the centrifugal and Coriolis accelerations arising due to rotation.
This can be expressed as,
\[{\bf g}_{\rm tot} = g_{CAK}{\bf {\hat r}} + g_{\rm grav} {\bf {\hat r}} 
        - 2~{\bf \Omega} \times {\bf v}
        -   {\bf \Omega} \times \left ( {\bf \Omega} \times {\bf r} \right ) \]
where, ${\bf \Omega}$ is the angular velocity vector and ${\bf {\hat r}}$ is the unit vector along the radial direction.
%
\\
iii) conservation of energy
\begin{align}\label{eqb:c3}
        \frac{\partial E}{\partial t} +
        \nabla \left[ \left( E  + P + \dfrac{{B_1}^2}{8\pi} + 
               \dfrac{{{\bf B_0} \cdot {\bf B_1}}}{4\pi} \right) {\bf v} -
               \dfrac{({\bf v}   \cdot {\bf B_1}) {\bf B}}{4\pi} 
               \right]
        =
            \rho {\bf v} \cdot {\bf g}_{tot}
\end{align}
where, $E$ is the energy density. These are isothermal simulations, 
therefore the energy equation is not used however, for the sake of 
completion this is included here. The pressure $P$ is then defined
using the sound speed $v_s$ as $ P = \rho v_s^2 $.
%
\\
iv) induction equation
\begin{align}\label{eqb:c4}
\frac{\partial {\bf B_1}}{\partial t} - \nabla \times ({\bf v} \times {\bf B}) = 0
\end{align}

\end{document}